\documentclass{article}
\usepackage{graphicx}
\usepackage{amsmath}

\begin{document}

\title{Multilongitudinal mode emission in ring cavity class B lasers }
\author{Eugenio Rold\'{a}n, Germ\'{a}n J. de Valc\'{a}rcel, \\
Departament d'\`{O}ptica, Universitat de Val\`{e}ncia,\\
Dr. Moliner 50, 46100--Burjassot, Spain \and Franco Prati \\
Istituto Nazionale per la Fisica della Materia, and\\
Dipartimento di Fisica e Matematica, Universit\`{a} dell'Insubria, \\
via Valleggio 11, I--22100 Como, Italy, \and Fedor Mitschke, and Tom Voigt \\
Fachbereich Physik, Universit\"{a}t Rostock, \\
18051 Rostock, Germany}
\maketitle

\begin{abstract}
In this article we review recent theoretical and experimental
developments on multilongitudinal--mode emission in ring cavity
lasers, paying special attention to class B lasers. We consider
both homogeneously and inhomogeneously broadened amplifying media
as well as the limits of small and large cavity losses
(\textit{i.e.}, we treat cases within and outside the uniform
field limit approximation). In particular we discuss up to what
extent the experimental observations of self--mode locking in
erbium--doped fiber lasers carried out in recent years are a
manifestation of the Risken-Nummedal--Graham-Haken instability.
\end{abstract}

\section{INTRODUCTION:\newline
WHY SHOULD WE CARE ABOUT LASER INSTABILITIES?}

\label{introduction}

The very first laser -- Maiman's ruby in 1961 -- produced
extremely unstable emission, as evidenced by a figure in the
original publication \cite{maiman}. Soon thereafter, researchers
began to learn the skills of how to avoid instabilities in lasers
designed for applications. Nonetheless laser instabilities have
been around ever since both as a nuisance lurking to haunt
technically-minded people, and as an interesting nontrivial
physical phenomenon for the more fundamental-principles-minded
set. After all, lasers are nonlinear dynamical feedback systems,
and instabilities are inherent in such systems. It is certainly
warranted to gain an understanding of laser instabilities -- if
only for the minimalist purpose that they be avoided successfully
in spite of the ever increasing demands on lasers in terms of
power, speed, tunability, etc.

In lasers, typically one or several modes of the light field are subject to
resonator boundary conditions while at the same time being dynamically
coupled to the amplifying medium. Typically the coupling is highly
nonlinear, and depending on the laser type and particular circumstances, the
laser may behave in many different ways, running the gamut from smooth and
stable single-mode to irregular and unpredictable chaotic operation.

To systematically address the plethora of possibilities, one can
make a first distinction between cases in which either several
transverse modes are involved, or just a single one
(\textsf{TEM}$_{00}$). Next-neighbor longitudinal modes always
have very nearly the same frequency difference, and only a single
beat frequency, along with its overtones, can occur. In contrast,
for transverse modes there can be more beat frequencies, giving
rise to an enormous wealth of possibilities for frequency locking
and pulling phenomena. Maiman's ruby laser, for example, displayed
an instability that involved several transverse modes. The
complexity of the situation is not helped by the fact that a
mathematical description requires either an infinite set of
ordinary differential equations, or a set of fully space-dependent
partial differential equations. Nonetheless, quite some research
was devoted to multi-transverse mode dynamics (see \cite
{WeissVilaseca,Arecchi,Prati} for references, and the rest of
articles appearing in the present volume). More attention,
however, was given to single mode laser instabilities over the
last three decades. Overviews and references can be found, e.g.,
in \cite{WeissVilaseca,Arecchi,JOSA-FI85,
JOSA-FI88,Boyd86,AreHar,NarAbr88,NA88,Khanin,Mandel97,Tartwijk98}.

Here we shall consider single--transverse mode problems. In this context,
there is an important distinction whether several longitudinal cavity modes
are involved in the lasing process, or just a single one, although a
connection between singlemode and multimode instabilities can be established
\cite{Lugiato85b}. It is also of fundamental importance whether the gain
medium exhibits a homogeneously or inhomogeneously broadened lasing
transition (see, e.g., \cite{WeissVilaseca,NarAbr88}).

It was noticed by Haken in 1975 \cite{Haken75} that the
Maxwell-Bloch model for a homogeneously broadened single mode
laser is isomorphic to the Lorenz model of chaos, originally
developed for climatic instability \cite{Lorenz63}. This directly
implies that these lasers must exhibit a second threshold,
\textit{i.e.} a characteristic value of the pump power above which
a Lorenz-type instability sets on. One can show that this second
threshold is at least nine times higher than the first (ordinary)
threshold above which there is coherent oscillation (see, e.g.,
\cite{WeissVilaseca,NarAbr88,NA88,Haken75,Lorenz63}). This \lq\lq
factor--of--nine\rq\rq is of great importance to the present
paper. Later Weiss found \cite{WeissKlische} that certain
far-infrared lasers have suitable damping rates and can be pumped
hard enough to test this prediction. Indeed they displayed chaotic
behavior remarkably similar in many respects to that exhibited by
the Lorenz model \cite{WeissBrock,Weiss}. Unfortunately the level
structure of the gain atoms was much more complex than a two-level
system. The relevance of this difference for the laser dynamics
was the focus of some controversy during the eighties and the
beginning of the nineties (see \cite{WVACRVPHT,RVVCMG} for a
discussion at depth and for references). However, in the absence
of a formal proof of equivalence between the realistic laser model
and the Lorenz model, one can only cautiously conclude that some
lasers can display Lorenz chaos even when their structure does not
make it obvious that the Lorenz equations are the appropriate
model.

In contrast, multilongitudinal mode laser emission has received only
marginal attention from the viewpoint of laser instabilities. This is all
the more remarkable considering that most lasers emit in several
longitudinal modes, if only due to spectral hole burning in inhomogeneously
broadened active media (see, e.g., \cite{New83,Siegman,Svelto}).

It is a common belief that with a homogenously broadened gain line
and single-transverse mode condition, the only way to have more
than one longitudinal mode oscillating is spatial inhomogeneity in
the medium, such as the spatial hole burning occurring in linear
resonators. The spatial inhomogeneity provides the required amount
of independence of the inversion available to one mode from that
available to another mode so that the usual winner-takes-all
coupling is suspended in favor of the mild mutual coupling due the
finite overall energy balance\footnote{A good deal of theoretical
and experimental work has been done in order to correctly
understand multimode emission in Fabry--Perot cavity lasers,
starting with the well known Tang--Statz--de Mars model
\cite{TSM}. We refer the interested reader to
\cite{Mandel97,Otsuka99,Mandel00,Hill02} and references therein.
That subject, while very interesting in itself, is beyond the
scope of the present discussion.}.

This myth was shattered when in 1968 two publications, independent
of each other, discussed the situation in detail \cite{RN68,GH68}
(see also \cite{RN68(b)}). It was shown that even in a
unidirectional resonator filled with a perfectly homogeneous
medium, Rabi splitting of the lasing transition induced by the
lasing mode can provide gain for other longitudinal modes. This
means that even perfect gain homogeneity does not safeguard
against instability. While a string of four names makes a somewhat
awkward moniker, fairness dictates to call this mechanism the
\emph{Risken-Nummedal-Graham-Haken Instability}, or RNGHI for
short. Ikeda \textit{ et al.} introduced the term \emph{Resonant
Rabi Instability} for designating this instability \cite{Ikeda89}
\footnote{ Interestingly, in 1976 Graham \cite{Graham76} showed
that the multimode laser model, the RNGH model, is isomorphous to
the Lorenz model \cite {Lorenz63}, thus extending the analogy
discovered by Haken the previous year \cite{Haken75}. The
difference between the singlemode and the multimode case is that
in the latter the parameter $\sigma $ is not fixed as it depends
on the velocity of the travelling--wave solution. This would be
extended further to cover detuning in 1990 when Ning and Haken
\cite{Ning90} showed that the detuned multimode laser equations
are isomorphic to the complex Lorenz model for the baroclinic
instability \cite{Fowler82}. This isomorphism is a powerful tool
that has not been investigated enough, specially for what concerns
the multimode instability.}.

In \cite{RN68,GH68,RN68(b)} a unidirectional ring cavity with
small cavity losses filled with a homogeneously broadened
two--level active medium was assumed. The prediction was that for
RNGHI to occur the pump power must exceed a certain instability
threshold. This instability threshold is referred to as the
\emph{second threshold}, in distinction to the familiar first
threshold which defines the onset of coherent laser oscillation.
The value of the second threshold came out to be at least nine
times the lasing threshold (the same ``factor--of--nine''as in
single--mode instabilities in the Lorenz--Haken model). Another
condition was that the laser cavity length must exceed a certain
minimum. Indeed, it has to be unrealistically long for
conventional bulk lasers \cite{WeissVilaseca,NarAbr88}.

Due to these predictions and without experimental demonstration of
the opposite, interest in RNGHI abated over time: While during the
Seventies and Eighties much theoretical effort was devoted to the
understanding of the RNGHI as well as to the dynamics of the laser
above the second threshold
\cite{Milovsky70,Halford73,Haken76,Ohno76,Haken78,Gerber79,Mayr81,Zorell81},
eventually RNGHI was more or less dismissed as a merely academical
prediction rather than an actual mechanism for multimode emission.

In 1984 researchers announced that they had observed what could be
interpreted as RNGHI in a dye laser \cite{Hillman1,Hillman2}. The
dye laser seemed to be a good candidate for the observation of the
RNGHI as (i) it is a homogeneously broadened laser, and (ii) the
\textquotedblleft critical\textquotedblright cavity length (see
later for a precise definition of this term) is smaller than
typical resonator lengths \cite{WeissVilaseca}. However, it was
peculiar that in the spectrum there where only two peaks instead
of three. Finally, after some debate \cite{Lugiato85a}, Fu and
Haken showed that a suitable model for the dye (consisting of
bands rather than levels) could explain the observations without
invoking the RNGHI \cite {FuHaken1,FuHaken2,FuHaken3}. From then
on the observation in \cite{Hillman1,Hillman2} was no longer
considered as a manifestation of the RNGHI.

A few years later, a phenomenon very similar to RNGHI was
discussed in the context of optical bistability performed in the
microwave regime \cite{Segard1,Segard2}. The experimental system
consisted of a waveguide Fabry--Perot cavity with a length of
182~m filled with hydrocyanic acid vapor at low ($\approx
1\,\mathrm{mTorr}$) pressure. Driven at a frequency of 86~GHz near
a HC$^{15}$N transition, this system displayed bistability and, on
the upper branch of the bistable loop, self-oscillation at a
frequency of the order of the cavity free spectral range and of
the Rabi frequency, but could deviate by 50\% or so. Nevertheless
it is to be emphasized that this fairly unique experiment actually
displayed not the RNGHI, but the multimode instability of optical
bistability predicted in 1978 by Bonifacio and Lugiato
\cite{Bonifacio78}. Thus, for many years and after a false start,
the closest thing to the RNGH laser instability that was actually
observed was Rabi splitting in microwave optical bistability --
not very close, certainly.

The discussion on the RNGHI during the late Eighties and the
Nineties faded, although theoretical studies continued \cite
{Ikeda89,Mandel85,Mandel86,Risken86,Lugiato86a,Narducci86,Lugiato86b,
Elgin87,Gibbon88,Gibbon89,Fu89,FuHaken90,Carr94a,Carr94b,Casini97,
Tartwijk97,Jahanpanah97}, and signatures of the RNGHI were
predicted also for laser related systems such as a laser with an
intracavity parametric amplifier \cite{Pande92}, and a laser with
injected signal \cite{Castelli94}. Nonetheless, in this period it
seemed like the matter of longitudinal mode instabilities in
lasers, and the RNGHI in particular, would soon be more or less
forgotten.

The situation changed with a suggestion by Lugiato and coworkers
\cite{F95,P97}. They observed that a ring--cavity erbium--doped
fibre laser (EDFL) spontaneously mode-locked and emitted a train
of pulses with a repetition time equal to the cavity roundtrip
time. This is one signature of the RNGHI, but certainly not in
itself a sufficient criterion. However, the unusually long cavity
length of fibre lasers automatically fulfils the most difficult
prerequisite for RNGHI of sufficient resonator length. On this
basis they put forward the hypothesis that self--pulsing in these
lasers could be a manifestation of the RNGHI (see also \cite{W97}
for compatible experimental results).

Unfortunately they could not confirm the existence of a second laser
threshold, \textit{i.e.} a threshold-like onset of the instability.
Moreover, there was a grave quantitative difficulty: Instabilities were
observed at pump powers immediately above the lasing threshold \cite{P97},
in stark contrast to the prediction that the second threshold be at least
nine times higher than the first.

It was subsequently shown that this requirement is not necessarily
applicable to real-world lasers. In \cite{Roldan97} the usual two-level atom
model for Lorenz-type instability was extended to incorporate a third level
as a necessity for applying a pump source. It turned out that the
instability threshold was lowered such as to come close to the lasing
threshold. This finding was then applied to EDFLs \cite{R98}. Erbium ions,
too, behave more like three--level systems \cite{Desurvire}, rather than
two--level systems as the original theory assumed. Now the ratio of the
instability to the laser threshold was predicted to be close to unity. This
was subsequently corroborated (and expanded to include four-level cases) in
\cite{P99}.

Put simply, for EDFLs the expected \lq\lq factor--of--nine\rq\rq, became a
\lq\lq factor--of--($1+\epsilon$)\rq\rq. Thereby, the apparent contradiction
with the hypothesis in \cite{P97} is gone. This insight bestowed fresh vigor
on the debate on RNGHI and lead to dedicated experimental investigation as
will be described below.

In this paper we shall review the fundamentals of the RNGH theory
as well as the research we have carried out along the recent years
(including some inedit results), i.e., we shall not review in
detail the available literature. After this historical
introduction, the paper follows with three more sections. Sect.
\ref{homo} is devoted to the RNGHI in homogeneously broadened
media. In this section we first explain in detail modelling
issues, Sect. \ref{homo-modeling}, and then we treat the RNGHI,
first in the uniform field limit (Sect. \ref{homo-ufl}) and then
outside the limit of validity of this standard approximation
(Sect. \ref{homo-nufl}). In Sect. \ref{inhomo} we consider
multilongitudinal mode emission in inhomogeneously broadened
media, again within (Sect. \ref{inhomo-ufl}) and outside (Sect.
\ref{inhomo-nufl}) the uniform field limit. Then in Sect.
\ref{experimental} we discuss the experimental aspects of the
RNGHI concentrating our discussion on erbium--doped fibre lasers.
Finally, in Sect. \ref{conclus} we provide a general discussion of
the issue.

\section{MULTILONGITUDINAL MODE EMISSION IN HOMOGENEOUSLY
BROADENED RING LASERS: THE RNGH INSTABILITY}

\label{homo}

\subsection{Modelling}

\label{homo-modeling} In this section we introduce the standard
two--level laser theory (Sect. \ref{model-2lev}). We then discuss
how the model can be applied to three-- and four--level lasers
(Sect. \ref{model-3lev4lev}), and how it can be generalized to
treat the dependence of the laser field and of the pump field on
the radial coordinate (Sect. \ref{mode-tra}). Finally, we
determine the stationary singlemode solutions of the laser
equations (Sect. \ref{model-ss}, and we rigorously derive the
laser equations in the uniform field limit (Sect.
\ref{model-ufl}).

\subsubsection{Two--level atoms}

\label{model-2lev} We consider a collection of $\mathcal{N}$ two--level
homogeneously broadened atoms per unit volume, with transition frequency $%
\omega _{\mathrm{a}}$ interacting with a linearly polarized, plane wave,
unidirectional electric field of carrier frequency $\omega _{\mathrm{c}}$,
chosen as the longitudinal mode frequency closest to $\omega_a$. The
Maxwell--Bloch equations describing the system are \cite%
{NarAbr88,Khanin,Mandel97,Lugiato85a,Roldan03b}
\begin{eqnarray}
\partial _{z}F+v_{\mathrm{m}}^{-1}\partial _{t}F &=&\frac{a}{2}P-\frac{%
\alpha _{\mathrm{m}}}{2}F\,,  \label{eqF1} \\
\partial _{t}P &=&\gamma _{\bot }\left[ FD-(1+i\delta )P\right] \,,
\label{eqP1} \\
\partial _{t}D &=&\gamma _{\Vert }\left[ 1-D-\mathrm{Re}\left( F^{\ast
}P\right) \right] \,.  \label{eqD1}
\end{eqnarray}%
Here $F$, $P$ and $D$ are properly scaled variables representing the
electric field, medium polarization, and population difference
\begin{equation}
F=\frac{2E}{\sqrt{\gamma _{\bot }\gamma _{\Vert }}}\,,\qquad
P=-2i\frac{\rho _{12}}{d_{0}}\sqrt{\frac{\gamma _{\bot }}{\gamma
_{\Vert }}}\,,\qquad D=\frac{d}{d_{0}}\,.  \label{FPD}
\end{equation}
In these expressions $E$ is the Rabi frequency associated with the laser
field, $\rho _{12}$ is the slowly varying envelope of the coherence between
the lasing levels, and $d$ is the population difference per atom between the
upper (2) and lower (1) lasing levels; $\gamma _{\bot }$ and $\gamma _{\Vert
}$ are the decay rates of the medium polarization and population difference,
respectively, $\delta =(\omega _{\mathrm{a}}-\omega _{\mathrm{c}})/\gamma
_{\bot }$ is the detuning between the atoms and the cavity, $v_{\mathrm{m}%
}=c/n_{\mathrm{m}}$ is the light velocity in the host medium with refractive
index $n_{\mathrm{m}}$, $\alpha _{\mathrm{m}}$ is an intensity loss
coefficient per unit length, which describes (non resonant) distributed
losses inside the active medium \cite{Roldan03b}, and $d_{0}$ is the
equilibrium value towards which $d$ relaxes in the absence of an electric
field. $d_{0}$ is positive because the medium is amplifying. The unsaturated
intensity gain coefficient per unit length $a$ is proportional to the total
population difference $\mathcal{N}d_{0}$, and we can write it as
\begin{equation}
a=\frac{4\pi \mu ^{2}\omega _{\mathrm{c}}}{c\hbar \gamma _{\bot }}\mathcal{N}
d_{0}\,,  \label{a}
\end{equation}%
where $\mu $ is the dipole moment between the two levels of the laser
transition.

Equations (\ref{eqF1})--(\ref{eqD1}) must be supplied with the appropriate
boundary condition for the electric field. We assume that the active medium
fills a region of length $L_{\mathrm{m}}$ inside a ring cavity of length $L_{%
\mathrm{c}}$. Denoting by $z=0$ and $z=L_{\mathrm{m}}$ the entrance and exit
planes of the amplifying medium, the electric field obeys the boundary
condition
\begin{equation}
F(0,t)=\mathcal{R}F(L_{\mathrm{m}},t-\Delta t)\,,
\end{equation}
where ${\cal R}$ is the effective cavity reflectivity
\footnote{$\mathcal{R}$ accounts for all localized intracavity
losses occurring outside the active medium (at splices, filters,
output couplers, or other components). Namely, $\mathcal{R}^{2}=\mathcal{R}%
_{1}^{2}\cdot \mathcal{R}_{2}^{2}\ldots \mathcal{R}_{n}^{2}$, where $%
\mathcal{R}_{i}^{2}$ is the fraction of power after the $i$-th
lossy element. Distributed losses outside the amplifying medium
which damp the intensity of the electric field at a rate $\alpha
_{\mathrm{out}}$ can also be included in the effective
reflectivity $\mathcal{R}$ by multiplying it with $\exp [-\alpha
_{\mathrm{out}}(L_{\mathrm{c}}-L_{\mathrm{m}})/2]$.},
$\Delta t=(L_{\mathrm{c}}-L_{\mathrm{m}})/v_{\mathrm{c}}$, and $v_{%
\mathrm{c}}=c/n_{\mathrm{c}}$ is the speed of light within the unloaded part
of the cavity, whose refractive index is $n_{\mathrm{c}}$.

A standard procedure to make the boundary condition isochronous consists in
defining the new spatial and temporal variables \cite{Lugiato85a}
\begin{equation}
Z=\frac{z}{L_{\mathrm{m}}}\,,\qquad T=t+\frac{z}{L_{\mathrm{m}}}\Delta t\,.
\end{equation}%
This transformation amounts to ideally bend the active medium so that its
entrance and exit coincide, and the delay $\Delta t$ accumulated in the
trivial propagation outside the amplifying medium is removed. The boundary
conditions now read
\begin{equation}
F(0,T)=\mathcal{R}F(1,T)\,,
\end{equation}%
and the Maxwell--Bloch equations become
\begin{eqnarray}
\left(\partial _{T}+\frac{\alpha _{\mathrm{FSR}}}{2\pi }\partial
_{Z}\right)F &=&\kappa
\left( AP-\chi F\right) \,, \\
\partial _{T}P &=&\gamma _{\bot }\left[ FD-(1+i\delta )P\right] \,, \\
\partial _{T}D &=&\gamma _{\Vert }\left[ 1-D-\mathrm{Re}\left( F^{\ast
}P\right) \right] \,,
\end{eqnarray}%
where $\alpha _{\mathrm{FSR}}=2\pi c/\mathcal{L}_\mathrm{c}$ is the cavity
free spectral range,
\begin{equation}
\kappa =\frac{c}{2\mathcal{L}_\mathrm{c}}\left( |\ln \mathcal{R}^{2}|+\alpha
_{\mathrm{m}}L_{\mathrm{m}}\right) ,  \label{kappa}
\end{equation}%
is the cavity linewidth, $\mathcal{L}_{\mathrm{c}}=n_{\mathrm{m}}L_{\mathrm{m%
}}+n_{\mathrm{c}} (L_{\mathrm{c}}-L_{\mathrm{m}})$ is the optical length of
the cavity, and
\begin{equation}
A=\frac{aL_{\mathrm{m}}}{|\ln \mathcal{R}^{2}|+\alpha _{\mathrm{m}}L_{%
\mathrm{m}}}  \label{A}
\end{equation}%
is the adimensional pump parameter, while
\begin{equation}
\chi =\frac{\alpha _{\mathrm{m}}L_{\mathrm{m}}}{|\ln \mathcal{R}^{2}|+\alpha
_{\mathrm{m}}L_{\mathrm{m}}},  \label{ji}
\end{equation}%
$0\leq \chi \leq 1$, is an adimensional loss parameter due to distributed
losses inside the amplifying medium. In the absence of distributed loss ($%
\alpha _{\mathrm{m}}=0$), $\chi =0$, and the pump parameter becomes $A=aL_{%
\mathrm{m}}/|\ln \mathcal{R}^{2}|$, which is its usual definition.

In the RNGHI for class B lasers it is known that instability occurs when the
cavity free spectral range is a quantity of order $\sqrt{\gamma _{\Vert
}\gamma _{\bot }}$ \cite{RN68(b)} (see below). In view of this, it is
convenient to introduce the new temporal and spatial variables
\begin{equation}
\tau =\sqrt{\gamma _{\Vert }\gamma _{\bot }}\;T,\qquad \zeta =\frac{2\pi }{%
\tilde{\alpha}}\;Z,
\end{equation}%
where
\begin{equation}
\tilde{\alpha}=\frac{\alpha _{\mathrm{FSR}}}{\sqrt{\gamma _{\Vert }\gamma
_{\bot }}}=\frac{2\pi c}{\mathcal{L}_\mathrm{c}\sqrt{\gamma _{\Vert }\gamma
_{\bot } }}\,,  \label{scaledFSR}
\end{equation}%
is the scaled free spectral range. If we define also the adimensional decay
rates
\begin{equation}
\sigma =\frac{\kappa }{\sqrt{\gamma _{\Vert }\gamma _{\bot }}}\,,\qquad
\gamma =\sqrt{\frac{\gamma _{\Vert }}{\gamma _{\bot }}}\,,
\label{sigmagamma}
\end{equation}%
the Maxwell--Bloch equations read
\begin{eqnarray}
\left( \partial _{\tau }+\partial _{\zeta }\right) F &=&\sigma \left(
AP-\chi F\right) \,,  \label{eqF3} \\
\partial _{\tau }P &=&\gamma ^{-1}\left[ FD-(1+i\delta )P\right] \,,
\label{eqP3} \\
\partial _{\tau }D &=&\gamma \left[ 1-D-\mathrm{Re}\left( F^{\ast }P\right) %
\right] \,,  \label{eqD3}
\end{eqnarray}%
with the boundary condition
\begin{equation}
F(0,\tau )=\mathcal{R}F(\zeta_{\mathrm{m}},\tau )\,,  \label{bou3}
\end{equation}%
where
\begin{equation}
\zeta_{\mathrm{m}}=\frac{2\pi }{\tilde{\alpha}}  \label{Setam}
\end{equation}%
is the position of the exit plane of the amplifying medium in the
space variable $\zeta $. Note the following useful relation
\begin{equation}  \label{relatiosigmaseta}
2\sigma \zeta_{\mathrm{m}}(1-\chi)=|\ln \mathcal{R}^{2}|\,.
\end{equation}

\subsubsection{Three-- and four--level atoms}

\label{model-3lev4lev} Real lasers are not two--level lasers, even those in
which the lasing transition can be well described as a two--level system.
This is so because pump and relaxation processes connect, incoherently, the
two lasing levels to some other levels. Although under usual conditions
(essentially when the relaxation rates of the pumped transition are very
large as compared with the rest of relaxation rates) this incoherent
coupling does not have any influence on the coherent dynamics of the system,
it turns out to be essential for understanding the meaning of the laser
parameters, particularly that of the pump parameter $A$. The connection
between the two--level theory and the description of three-- and four--level
lasers (which are the usual schemes for describing real lasers) was first
treated, to the best of our knowledge, by Khanin in his book \cite{Khanin},
but he did not analyze the consequences of the parameter transformations he
derived. Later on, and independently of Khanin, these transformations were
derived again in \cite{P99} \footnote{%
In \cite{P99} equal relaxation rates for the two lasing levels in
four--level lasers were assumed. This is not a realistic approximation for
Nd:YAG lasers. In \cite{Roldan03a}, following \cite{Khanin}, it was shown
that the same model can be derived even removing that assumption}, and their
influence on the understanding of laser dynamics was clarified. The relation
between two--level and three-- and four--level lasers is important in our
context because we will apply our analysis especially to rare--earth doped
fibre lasers, where the active medium should be properly modelled as a a
collection of three-- or four--level atoms if the dopants are, respectively,
Erbium or Neodymium atoms \cite{Desurvire}.

Eqs. (\ref{eqF3}--\ref{eqD3}), describing two--level lasers, are valid also
for three-- and four--level atoms provided a dependence on the rate of
incoherent optical pumping $W_{\mathrm{opt}}$ is included in the parameters $%
\gamma _{\Vert }$ and $d_{0}$ \cite{P99,Roldan03a}. Precisely, we define the
adimensional pumping rate
\begin{equation}
W=\frac{W_{\mathrm{opt}}}{\gamma _{2}}\,,
\end{equation}%
where $\gamma _{2}$ is the total spontaneous decay rate from the upper level
of the lasing transition. Then, the parameter transformation necessary for
applying Eqs. (\ref{eqF3}--\ref{eqD3}) to three-- and four--level lasers are%
\begin{equation}
\gamma _{\Vert }=\gamma _{2}(1+W)\,,  \label{gpar}
\end{equation}%
and
\begin{equation}
d_{0}=\frac{W-\delta _{N,3}}{W+1}\,,  \label{d0}
\end{equation}%
where $\delta_{N,3}$ is the Kroenecker $\delta$ and $N$ is the number of
atomic levels. Thus, $\delta _{N,3}=1$ for three--level atoms, and $\delta
_{N,3}=0$ for four--level atoms.

In fibre lasers the gain parameter $a$ is usually defined in a way different
from Eq. (\ref{a}). If $\sigma _{\mathrm{e}}$ and $\sigma _{\mathrm{a}}$
are, respectively, the stimulated emission and absorption cross sections,
one has \cite{Desurvire}
\begin{equation}
a=\frac{\mathcal{N}}{2}\left[ \sigma _{\mathrm{e}}-\sigma _{\mathrm{a}%
}+(\sigma _{\mathrm{e}}+\sigma _{\mathrm{a}})d_{0}\right] \,.
\end{equation}%
Assuming $\sigma _{\mathrm{e}}=\sigma _{\mathrm{a}}$, the gain coefficient
turns out to be proportional to the equilibrium total population difference
per unit volume $\mathcal{N}d_{0}$, as in Eq. (\ref{a})
\begin{equation}
a=\sigma _{\mathrm{e}}\mathcal{N}d_{0}\,.  \label{a1}
\end{equation}%
Under this approximation, taking into account the definition of the pump
parameter $A$, Eq. (\ref{A}), and Eq. (\ref{d0}), the dependence of the pump
parameter on the pumping rate $W$ can be expressed in the following way
\begin{equation}
A=G\;\frac{W-\delta _{N,3}}{W+1}\,,  \label{A1}
\end{equation}%
with
\begin{equation}
G=\frac{G_{0}}{|\ln \mathcal{R}^{2}|+\alpha _{\mathrm{m}}L_{\mathrm{m}}}%
\,,\qquad G_{0}=\mathcal{N}\sigma _{\mathrm{e}}L_{\mathrm{m}}\,.  \label{G}
\end{equation}

Notice that $\gamma _{||}$ enters in the normalizations used for writing
Eqs. (\ref{eqF3}--\ref{eqD3}), and as this parameter is pump dependent in
the case of three-- and four--level lasers, Eq. (\ref{gpar}), this must be
taken into account when interpreting the results derived from Eqs. (\ref%
{eqF3}--\ref{eqD3}). The recipe is very simple: (i) replace $A$ by using Eq.
(\ref{A1}), and (ii) replace the spatial frequency $\alpha $ by $\alpha
\sqrt{1+W}$ in the final expressions (see following sections). This last
replacement is due to the normalization of the axial coordinate and,
consequently, of the spatial frequency, see Eq. (\ref{scaledFSR}).

The most relevant feature of Eq. (\ref{A1}) is that the effective
two--level pump parameter $A$ depends on the actual pump strength
$W$ in a non--linear fashion for both three-- and four--level
lasers. Although the instability threshold will be discussed
below, let us comment that for three--level lasers the
relationship between $A$ and $W$ and the fact that usually $G\gg 1
$ imply that the \lq\lq factor--of--nine\rq\rq (the ratio of $A$
at the RNGHI instability threshold and at the laser threshold)
(see Sect. \ref{homo-ufl-classb} below) becomes a \lq\lq factor of
$\left( 1+\epsilon \right)$\rq\rq for $W$, which is the actual
pump parameter that can be measured in an experiment \cite{P99}.
This is very easy to see: From Eq. (\ref{A1})
\begin{equation}
W=\frac{G+A}{G-A}\,,
\end{equation}%
and then, the instability to lasing threshold
\begin{equation}
\left( \frac{W_{\mathrm{ins}}}{W_{0}}\right) _\mathrm{3L}=\frac{G+9}{G-9}/%
\frac{G+1 }{G-1}\rightarrow 1+\frac{16}{G},  \label{ratio3L}
\end{equation}%
where the limit holds for large $G$. Contrarily, the case of four--level
lasers is very similar to that of two--level lasers as in this case%
\begin{equation*}
\left( \frac{W_{\mathrm{ins}}}{W_{0}}\right) _\mathrm{4L}=\frac{9}{G-9}/%
\frac{1}{G-1 }\rightarrow 9+\frac{72}{G},
\end{equation*}%
where the limit corresponds to large $G$ again. We see that the instability
threshold for four--level lasers is larger than that of two--level lasers,
tending to it for very large gain.

\subsubsection{Transverse effects}

\label{mode-tra}

So far we have limited our analysis to the plane wave approximation.
However, transverse effects may play an important role in real experimental
situations. In fibre lasers, for instance, both the laser and pump field
have a transverse spatial structure and the doped region has a finite
extension \cite{Desurvire}.

To include these elements in the laser equations we first notice
that the definitions of the dynamical variables $F$, $P$ and $D$,
Eq. (\ref{FPD}), are no longer appropriate, because such
definitions contain the parameters $\gamma _{\Vert }$ and $d_{0}$
which depend on the pump $W$. If $W $ is allowed to vary
spatially, this introduces an undesired spatial dependence in the
dynamical variables. The problem can be removed by introducing a
new definition of $F$, $P$ and $D$
\begin{equation}
F=\frac{2E}{\sqrt{\gamma _{\bot }\gamma _{2}}}\,,\qquad P=-2i\rho
_{12}\sqrt{\frac{\gamma _{\bot }}{\gamma _{2}}}\,,\qquad D=d\,.
\end{equation}
The dynamical variables do not depend on $W$ any longer, and they obey the
dynamical equations
\begin{eqnarray}
\left( \partial _{\tau }+\partial _{\zeta }\right) F &=&\sigma \,\left(
GP-\chi F\right) \,,  \label{eqf} \\
\partial _{\tau }P &=&\gamma ^{-1}\left[ FD-(1+i\delta )P\right] \,,
\label{eqp} \\
\partial _{\tau }D &=&\gamma \left[ W(1-D)-\delta _{N,3}-D-\mathrm{Re}\left(
F^{\ast }P\right) \right] \,,  \label{eqd}
\end{eqnarray}%
where in $\sigma $ and $\gamma $, as well as in the free spectral range $%
\tilde{\alpha}$, the decay rate $\gamma _{\Vert }$ is replaced by $\gamma
_{2}$.

Now we must include in these equations the dependence of the
electric field $F$ on the transverse coordinate. We assume that
the laser field is a Gaussian beam with $1/e$ radius equal to
$w_{0}$, which is a fairly good approximation \cite{Desurvire}.
This implies that in the Maxwell--Bloch equations all the
dynamical variables depend also on the radial coordinate $r$
through the scaled coordinate $\rho =r/w_{0}$, and in the equation
for the electric field a term describing diffraction must be
included
\begin{equation}
-\frac{i}{4\zeta_{0}}\nabla ^{2}F+\left( \partial _{\tau }+\partial _{\zeta
}\right) F=\sigma \,\left( GP-\chi F\right) \,.
\end{equation}
Here $\nabla ^{2}=\partial ^{2}/\partial \rho ^{2}+({1}/{\rho })(\partial
/\partial \rho )$ is the transverse part of the Laplacian and $\zeta_{0}$ is
the scaled Rayleigh length of the beam \textit{i.e.} the distance in the
propagation direction over which the size of the beam varies appreciably. A
Gaussian beam is described by the function \cite{Siegman,Svelto}
\begin{equation}
\psi (\rho ,\zeta )=\frac{1}{w(\zeta )}\exp \left[ -\frac{\rho ^{2}}{%
w^{2}(\zeta )}+i\frac{\rho ^{2}}{w^{2}(\zeta )}\frac{\zeta }{\zeta_{0}}%
-i\arctan \left( \frac{\zeta }{\zeta_{0}}\right) \right] \,,
\end{equation}%
with $w^{2}(\eta )=1+(\zeta /\zeta_{0})^{2}$, which is a solution of the
empty cavity equation
\begin{equation}
\frac{i}{4\zeta_{0}}\nabla ^{2}\psi =\partial _{\zeta }\psi \,.
\end{equation}%
We can write the electric field as $F(\rho ,\zeta ,\tau )=\psi (\rho ,\zeta
)f(\zeta ,\tau )$ and, taking into account the normalization property of $%
\psi $
\begin{equation}
\int_{0}^{\infty }\!\!\!d\rho \,\rho \,|\psi (\zeta ,\rho )|^{2}=\frac{1}{4}%
\,,
\end{equation}%
valid for any $\zeta $, we can project the equation for the total field $F$
on the mode amplitude $f$ by multiplying each term by $\psi ^{\ast }$ and
integrating over $\rho $ from 0 to $\infty $
\begin{equation}
\left( \partial _{\tau }+\partial _{\zeta }\right) f=\sigma \,\left(
4G\int_{0}^{\infty }\!\!\!d\rho \,\rho \,\mathrm{e}^{-\rho ^{2}}\,P-\chi
f\right) \,.
\end{equation}%
Here we have also assumed $\zeta \ll \zeta _{0}$, so that the dependence on $%
\zeta $ of $\psi $ can be neglected, and $\psi (\rho )=\exp (-\rho ^{2})$.
We make the same assumption for the pump field, and we denote its maximum
amplitude by $\beta^{1/2}$ and its $1/e$ radius by $w_\mathrm{p}$.
Introducing the parameter $\eta =(w_{0}/w_\mathrm{p})^{2}$, the pump term,
which is proportional to the intensity of the pump field, can be written as $%
W(\rho )=\beta \exp (-2\eta \rho ^{2})$. Finally, we replace $f$ again with $%
F$ and introduce the new radial coordinate $u=2\rho ^{2}$. The
Maxwell--Bloch equations suitable to describe transverse effects read
\begin{eqnarray}
\left( \partial_{\tau }+\partial_{\zeta }\right) F &=& \sigma \, \left( G
\int_{0}^{u_{\mathrm{m}}} \! \! \! du \, \mathrm{e}^{-u/2} \, P- \chi F
\right) \,,  \label{tra-f} \\
\partial _{\tau }P &=&\gamma ^{-1}\left[ \mathrm{e}^{-u/2}FD-(1+i\delta )P%
\right] \,,  \label{tra-p} \\
\partial _{\tau }D &=&\gamma \left[ \beta \mathrm{e}^{-\eta u}(1-D)-\delta
_{N,3}-D-\mathrm{e}^{-u/2}\mathrm{Re}\left( F^{\ast }P\right) \right] \,,
\label{tra-d}
\end{eqnarray}%
with the boundary condition
\begin{equation}
F(0,u,\tau )=\mathcal{R}F(\zeta_{\mathrm{m}},u,\tau
)\,,\qquad\zeta_{\mathrm{m}}=2\pi/\tilde\alpha\,.
\end{equation}
We have fixed the upper limit in the integral over the radial coordinate
equal to $u_{\mathrm{m}}=2(r_{\mathrm{m}}/w_{0})^{2}$, where $r_{\mathrm{m}}$
is the radius of the amplifying medium. In the case of a fibre $r_{\mathrm{m}%
}$ coincides with the radius of the doped region.

\subsubsection{Stationary singlemode solutions}

\label{model-ss}

We now return to the Maxwell--Bloch equations (\ref{eqF3})--(\ref{eqD3}) in
the plane wave approximation and look for their single frequency
(singlemode) solutions, which can be found by setting
\begin{equation*}
F\left( \zeta ,\tau \right) =F_\mathrm{s}\left( \zeta \right) e^{-i\omega
\tau }\,,\qquad P\left( \zeta ,\tau \right) =P_\mathrm{s}\left( \zeta\right)
e^{-i\omega \tau }\,,\qquad D\left( \zeta ,\tau\right) =D_\mathrm{s}\left(
\zeta \right)\,.
\end{equation*}
The stationary atomic variables are
\begin{eqnarray}
P_\mathrm{s} &=&\frac{1-i\Delta }{1+\Delta^2+\left\vert F_\mathrm{s}%
\right\vert ^{2}}F_\mathrm{s}\,, \\
D_\mathrm{s} &=&\frac{1+\Delta^{2}}{1+\Delta^2+\left\vert F_\mathrm{s}%
\right\vert ^{2}}\,,
\end{eqnarray}%
where $\Delta \equiv \delta -\gamma \omega $ and $F_\mathrm{s}$ verifies the
differential equation
\begin{equation*}
\frac{\mathrm{d}F_s}{\mathrm{d}\zeta }=\sigma \left( AP_\mathrm{s}-\chi F_%
\mathrm{s}\right)+i\omega F_\mathrm{s}.
\end{equation*}
Using the polar decomposition $F_\mathrm{s}=I_\mathrm{s}^{1/2}e^{i\phi _%
\mathrm{s}}$ and splitting the above equation into its real and imaginary
parts, one obtains
\begin{eqnarray}
\frac{\mathrm{d}I_\mathrm{s}}{\mathrm{d}\zeta } &=&2\sigma \left( \frac{A}{%
1+\Delta^2+I_\mathrm{s}}-\chi \right) I_\mathrm{s}\,, \label{dFdz} \\
\frac{\mathrm{d}\phi _\mathrm{s}}{\mathrm{d}\zeta } &=&\omega
-\frac{\sigma \Delta A }{1+\Delta^2+I_\mathrm{s}}\,, \label{dfidz}
\end{eqnarray}
with the boundary conditions
\begin{eqnarray}
I_\mathrm{s}(0)&=&{\cal
R}^2I_\mathrm{s}\left(\zeta_\mathrm{m}\right)\,,\label{bouI}\\
\phi_\mathrm{s}\left(\zeta_\mathrm{m}\right)-\phi_\mathrm{s}(0)&=&2\pi
n\,,\qquad\quad n\; \mathrm{integer}\,.\label{bouphi}
\end{eqnarray}
Integration of Eq. (\ref{dFdz}) yields
\begin{equation}
\left( 1+\Delta ^{2}\right) \ln \frac{I_\mathrm{s}\left( \zeta \right) }{ I_%
\mathrm{s}\left( 0\right) }-\frac{A}{\chi }\ln \frac{A-\chi \left[%
1+\Delta^2+I_\mathrm{s}\left( \zeta \right)\right] }{A-\chi\left[
1+\Delta^2+I_\mathrm{s}\left( 0\right)\right] }=\left[A-\chi\left(1+\Delta^2%
\right)\right]2\sigma\zeta\,.  \label{ecIs}
\end{equation}
Particularizing this equation to $\zeta =\zeta_{\mathrm{m}}$ and
taking into account Eqs. (\ref{relatiosigmaseta}) and
(\ref{bouI}), we find that the stationary intensity at the exit of
the active medium is
\begin{equation}  \label{ecIsetam}
I_\mathrm{s}\left( \zeta_{\mathrm{m}}\right) =\left( \frac{A}{\chi }%
-1-\Delta^2\right) \frac{1-\exp \left[ -\left\vert \ln \mathcal{R}%
^{2}\right\vert \frac{\chi }{1-\chi }\frac{A-1-\Delta^2}{A}\right] }{1-%
\mathcal{R}^{2}\exp \left[ -\left\vert \ln \mathcal{R}^{2}\right\vert \frac{%
\chi }{1-\chi }\frac{A-1-\Delta^2}{A}\right] }\,.
\end{equation}
It can be easily seen that the lasing threshold, \textit{i.e} the value of $%
A $ for which $I_s(\zeta_\mathrm{m})=0$, is $1+\Delta^{2}$. In the limit of
no distributed losses ($\chi\rightarrow 0$) Eq. (\ref{ecIsetam}) reduces to
\begin{equation}\label{ischizero}
I_\mathrm{s}\left( \zeta_{\mathrm{m}}\right) =\frac{\left\vert \ln \mathcal{R%
} ^{2}\right\vert }{1-\mathcal{R}^{2}}\left[ A-\left( 1+\Delta ^{2}\right)%
\right]\,,\qquad\qquad (\chi =0)\,.
\end{equation}
The frequency $\omega $ of the lasing solution has still to be fixed. To do
that we must determine how the stationary phase $\phi_s$ varies along $\zeta$%
. If we insert Eq. (\ref{dFdz}) into Eq. (\ref{dfidz}), the latter can be
written as
\begin{equation}
\frac{d\phi_s}{d\zeta}=\omega-\sigma\Delta\chi-\frac{\Delta}{2I_s}\frac{dI_s%
}{d\zeta}\,.
\end{equation}
from which it follows that $\phi_s$ depends on $\zeta$ in a nonlinear way as
\begin{equation}  \label{phizeta}
\phi _\mathrm{s}\left( \zeta\right) -\phi _\mathrm{s}\left( 0\right)
=\left(\omega-\sigma\Delta\chi\right)\zeta -\frac{\Delta}{2}\ln\frac{%
I_s(\zeta)}{I_s(0)}\,.
\end{equation}
The total phase shift experienced by the field in a roundtrip from 0 to $%
\zeta_{\mathrm{m}}$ is then
\begin{equation}
\phi _\mathrm{s}\left( \zeta_\mathrm{m} \right) -\phi
_\mathrm{s}\left( 0\right)
=\left(\omega-\sigma\Delta\right)\zeta_\mathrm{m}
=\left(\omega-\sigma\delta+\sigma\gamma\omega\right)\frac{2\pi}{\tilde\alpha}\,,
\end{equation}
where we have used Eqs. (\ref{relatiosigmaseta}) and (\ref{Setam})
and the fact that $\Delta=\delta-\gamma\omega$. Imposing the
boundary condition (\ref{bouphi}), we obtain for the $n$--th
longitudinal mode
\begin{equation}  \label{omegadeltan}
\omega_{n}=\frac{n\tilde{\alpha}+\sigma\delta }{1+\gamma \sigma}\,,
\qquad\Delta_{n}=\frac{\delta -\gamma n\tilde{\alpha}}{1+\gamma \sigma}\,.
\end{equation}
The singlemode solution associated with the $n$--th mode has threshold $%
1+\Delta_n^2$. Recalling that $\delta =\left(\omega_{\mathrm{a}}-\omega_{%
\mathrm{c}}\right)/\gamma _{\bot }$, and that $\left\vert \omega _{\mathrm{a}%
}-\omega _{\mathrm{c}}\right\vert \leq \alpha _{\mathrm{FSR}}/2$ by
definition ($\alpha _{\mathrm{FSR}}$ is the cavity free spectral range), we
have $\left\vert \delta \right\vert \leq \tfrac{1}{2}\gamma \tilde{\alpha}$.
Hence, the mode with the lowest threshold is the one with $n=0$, according
to our initial choice of the reference frequency. For this mode
\begin{equation}
\omega_0=\frac{\sigma\delta }{1+\gamma \sigma}=\frac{\kappa\omega_\mathrm{a}%
+\gamma_\bot\omega_\mathrm{c}}{\kappa+\gamma_\bot}\,, \qquad \Delta_0=\frac{%
\delta}{1+\gamma\sigma}=\frac{\omega _{\mathrm{a}}-\omega _{\mathrm{c}}}{%
\kappa+\gamma _{\bot }}\,.
\end{equation}
The expression for $\omega_0$ corresponds to the well-known mode pulling
formula. From now on $\Delta=\Delta_0$ will be our laser--atoms detuning
parameter, and the condition of perfectly resonant laser will be equivalent
to $\Delta=0$.

\subsubsection{The uniform field limit}

\label{model-ufl}

A widely adopted approximation in the study of laser dynamics is the
so--called Uniform Field Limit (UFL), which is based on the assumption that
the steady state of the electric field is nearly constant along the
propagation direction inside the amplifying medium. In other words, in the
uniform field limit the electric field obeys a periodic boundary condition,
as if the cavity mirrors were perfectly reflecting. From the analytic point
of view this represents a great simplification of the problem, because it
allows for an expansion of the electric field in terms of the Fourier modes $%
\exp {(i\alpha _{n}\zeta )}$ of the empty cavity, where $\alpha _{n}=n\tilde{%
\alpha}$ is the frequency of the $n$--th mode and the singlemode
stationary solution associated with mode $n=0$ is spatially
homogeneous. From the numerical point of view, the benefits of the
UFL are even more relevant, because the expansion of electric
field in Fourier modes allows to transform the original set of
Maxwell-Bloch equations from partial differential equations to
ordinary differential equations, which can be numerically
integrated much easier and faster. Although our work on the RNGHI
shows that the above mentioned features of the UFL persist even
outside that limit for class B lasers (see Sect. \ref{homo-nufl}),
in the most complex situations -- for instance when transverse
effects or inhomogeneous broadening are included in the model --
we still limit our analysis to the UFL for the sake of algebraic
simplicity.

In \cite{deValcarcel03a} we introduced a new technique for dealing with the
UFL. In comparison to the standard technique \cite{Lugiato86a}, the new
technique is advantageous because it allows to clearly demonstrate that the
only requirement for applying the UFL is that the resonator reflectivity
must be close enough to unity (the single-pass gain does not have to verify
any constraint). In \cite{deValcarcel03a} the simplest case of exact
resonance and absence of distributed loss was considered. In this section we
show, following \cite{deValcarcel03a}, how the Maxwell-Bloch equations in
the UFL can be derived for homogeneously broadened lasers in the plane wave
approximation in the presence of both detuning and distributed loss. The
generalization to other situations (including, e.g., transverse effects or
inhomogeneous broadening) is straightforward.

To derive the dynamical equations in the UFL we first observe that in the
limit $\mathcal{R}^{2}\rightarrow 1$ the solution of Eq. (\ref{ecIs}) for
the $n$-th mode varies slowly along $\zeta$ according to the equation
\begin{equation}
I_\mathrm{s}\left( \zeta \right) =\left[ A-\left( 1+\Delta_n ^{2}\right) %
\right] \left[ 1+\mathcal{T}\left( \frac{\zeta }{\zeta_\mathrm{m}}-\frac{1}{2%
}\right) \right] +\mathcal{O}\left( \mathcal{T}^{2}\right) ,
\label{IsalmostUFL}
\end{equation}
where $\mathcal{T}\equiv 1-\mathcal{R}^{2}$ is a small parameter. We thus
see that, independently of the value of all other laser parameters, the
laser intensity is almost uniform along the active medium in the limit $%
\mathcal{R}^{2}\rightarrow 1$, hence the name UFL. If we insert the above
expression for $I_s(\zeta)$ in Eq. (\ref{phizeta}), and take into account
Eqs. (\ref{relatiosigmaseta}) and (\ref{omegadeltan}), we find that the
stationary phase also varies linearly along $\zeta$ according to the simple
expression
\begin{equation}  \label{pshisalmostUFL}
\phi_\mathrm{s}\left(\zeta\right)-\phi_\mathrm{s}\left(0
\right)=n\tilde\alpha\zeta+\mathcal{O}\left( \mathcal{T}^{2}\right)\,,
\end{equation}
which shows that the space frequency associated with the $n$--th
solution is $n\tilde\alpha$. We focus on the $n=0$ solution for
which the stationary phase is constant along $\zeta$. Thus, in the
UFL, this stationary solution can be written as
\begin{equation}  \label{ssufl}
\begin{array}{cc}
  I_\mathrm{s}^\mathrm{UFL}=A-1-\Delta^2\,,
  & F_\mathrm{s}^\mathrm{UFL}=\sqrt{I_\mathrm{s}^\mathrm{UFL}}\,\mathrm{e}^{i\phi}\,, \\
  P_\mathrm{s}^\mathrm{UFL}=\frac{1-i\Delta}{A}F_\mathrm{s}^\mathrm{UFL}\,,
  & D_\mathrm{s}^\mathrm{ UFL}=\frac{1+\Delta^2}{A}\,. \\
\end{array}
\end{equation}
Now we introduce new dynamical variables obtained from the old
ones dividing them by the stationary solution $n=0$ outside the
UFL and multiplying by the same solution in the UFL
\begin{eqnarray}
F^{\prime}(\zeta,\tau)&=&\frac{F(\zeta,\tau)}{F_\mathrm{s}(\zeta)
\mathrm{e}^{-i\omega\tau}}F_\mathrm{s}^\mathrm{UFL}\,, \\
P^{\prime}(\zeta,\tau)&=&\frac{P(\zeta,\tau)}{P_\mathrm{s}(\zeta)
\mathrm{e}^{-i\omega\tau}}P_\mathrm{s}^\mathrm{UFL}=\frac{F_\mathrm{s}^\mathrm{UFL}}{\eta(\zeta)}\frac{P(\zeta,\tau)}
{F_\mathrm{s}(\zeta)\mathrm{e}^{-i\omega\tau}}\,, \\
D^{\prime}(\zeta,\tau)&=&\frac{D(\zeta,\tau)}{D_\mathrm{s}(\zeta)}D_\mathrm{s}^\mathrm{UFL}
=\frac{D(\zeta,\tau)}{\eta(\zeta)}\,,
\end{eqnarray}
with
\begin{equation}  \label{eta}
\eta (\zeta
)=\frac{1+\Delta^2+I_\mathrm{s}^\mathrm{UFL}}{1+\Delta^2+I_\mathrm{s}(\zeta
)}=\frac{A}{1+\Delta^2+I_\mathrm{s}(\zeta )}\,.
\end{equation}
Since $F_\mathrm{s}(0)=\mathcal{R}F_\mathrm{s}(\zeta_\mathrm{m})$, the new
electric field $F^{\prime}(\zeta,\tau)$ obeys a \textit{periodic} boundary
condition
\begin{equation}\label{bouFprime}
F^{\prime}(0,\tau)=F^{\prime}(\zeta_\mathrm{m},\tau)\,.
\end{equation}
The Maxwell-Bloch equations (\ref{eqF3}--\ref{eqD3}) for the primed
variables take the form
\begin{eqnarray}
(\partial _{\tau }+\partial _{\zeta })F^{\prime}&=&\sigma\eta(\zeta) \left[
AP^{\prime}-\left( 1-i\Delta \right) F^{\prime}\right] \,,  \label{numint-f1}
\\
\partial _{\tau }P^{\prime}&=&\gamma ^{-1}\left[ F^{\prime}D^{\prime}-\left(
1+i\Delta \right) P^{\prime}\right] \,,  \label{numint-p1} \\
\partial _{\tau }D^{\prime}&=&\gamma \left[\frac{1}{\eta(\zeta)}-D^{\prime}-%
\frac{I_s(\zeta)}{I_\mathrm{s}^\mathrm{UFL}}\mathrm{Re}\left(
{F^{\prime}}^{\ast }P^{\prime}\right)\right] \,.
\label{numint-d1}
\end{eqnarray}
Because of the periodicity condition (\ref{bouFprime}) these
equations are particularly suitable for numerical integration, as
we shall discuss in Sect. \ref{pulse-numres}.\\
Obviously, Eqs. (\ref{numint-f1}--\ref{numint-d1}) admit as a
stationary solution the UFL stationary solution (\ref{ssufl}).
Yet, they are still equivalent to the original Maxwell--Bloch
equations, and this fact is reflected in the dependence on $\zeta$
of $\eta$ and $I_s$. But from Eqs. (\ref{IsalmostUFL}),
(\ref{ssufl}), and (\ref{eta}) we see that in the UFL
\begin{eqnarray}
\frac{I_\mathrm{s}(\zeta)}{I_\mathrm{s}^\mathrm{UFL}}&=&
1+\mathcal{T}\left( \frac{\zeta
}{\zeta_\mathrm{m}}-\frac{1}{2}\right) +\mathcal{O}\left( \mathcal{T}%
^{2}\right)\,.  \label{approx1} \\
\eta(\zeta)&=&1+\frac{A-1-\Delta^2}{A} \mathcal{T}\left( \frac{\zeta }{\zeta_%
\mathrm{m}}-\frac{1}{2}\right) +\mathcal{O}\left( \mathcal{T}^{2}\right)\,,
\label{approx2}
\end{eqnarray}
Thus, in the very limit we can set both terms equal to unity, and, dropping
the primes, the laser equations in the UFL take the well--known form
\begin{eqnarray}
(\partial _{\tau }+\partial _{\zeta })F &=&\sigma\left[ AP-\left( 1-i\Delta
\right) F\right] \,,  \label{ufl-f1} \\
\partial _{\tau }P &=&\gamma ^{-1}\left[ FD-\left( 1+i\Delta \right) P\right]
\,,  \label{ufl-p1} \\
\partial _{\tau }D &=&\gamma \left[1-D -\mathrm{Re}\left( F^{\ast }P\right)%
\right] \,,  \label{ufl-d1}
\end{eqnarray}
with the boundary condition
\begin{equation}
F(0,\tau )=F\left(\zeta_\mathrm{m},\tau
\right)\,,\qquad\zeta_\mathrm{m}=2\pi/\tilde\alpha\,.
\label{bouufl}
\end{equation}
These equations differ from the general ones (Eqs.
(\ref{eqF3})--(\ref{bou3})) only in one respect: The (localized)
cavity losses, represented by the mirror reflectivity
$\mathcal{R}$, have disappeared from the boundary condition (which
now is periodic) and have appeared in the dynamical equation for
the electric field through the term $-\sigma F$, which contains
\textit{both localized and distributed losses}. The appearance of
a detuning in the equation for $F$ as well as the replacement of
$\delta$ with $\Delta$ in the equation for $P$ are only apparent
changes, due to the fact that in Eqs. (\ref{ufl-f1}--\ref{ufl-d1})
the reference frequency is the lasing frequency $\omega_0$, while
in Eqs. (\ref{eqF3})--(\ref{eqD3}) the reference frequency is the
empty cavity frequency $\omega_c$.

Let us stress that to derive the Maxwell--Bloch equations in the
UFL we had \textit{only} to assume that
$\mathcal{R}^{2}\rightarrow 1$. Eqs. (\ref
{approx1}--\ref{approx2}) show that this assumption suffices to
approximate $ I_\mathrm{s}(\zeta)/I_\mathrm{s}^\mathrm{UFL}$ and
$\eta(\zeta)$ with unity, for \textit{any} value of the pump
parameter $A$ greater than one, no matter how large it is.
Therefore, the condition of small single pass gain $aL_{\mathrm{
m}}\rightarrow 0$, which invariably accompanies the condition
$\mathcal{R}^2\rightarrow 1$ in the literature about the UFL, is
actually completely superfluous. Notice also that the amount of
distributed loss does not influence the validity of the UFL.

\subsection{The RNGH instability in the uniform field limit}

\label{homo-ufl}

The uniform field limit (UFL), rigorously introduced in Sect.
\ref{model-ufl}, is the case in which the RNGHI can be analyzed in
the simplest way, as well as the limit where the effects of
additional features not considered in the original analyzes by
Risken and Nummedal \cite{RN68} and by Graham and Haken
\cite{GH68} can be studied most easily. Still another reason for
the interest in the UFL is that it still captures the very basic
signatures of the multimode instability of ring lasers found in
more complex models.

In this section, we shall first derive again the classical results
of RNGH (Sect. \ref{homo-ufl-rnghi}). Then we shall consider the
relevant case when the population inversion is an extremely slow
variable (class B laser). This
applies, in particular, to fiber lasers (Secs. \ref{homo-ufl-classb} and \ref%
{homo-ufl-expansion}). Next we will show that detuning plays no
role in class B lasers (Sect. \ref{homo-ufl-detuning}). Finally,
we shall analyze the spatial (transverse) effects due to the modal
structure of fiber lasers (Sect. \ref{homo-ufl-tra}).

\subsubsection{The RNGH instability}

\label{homo-ufl-rnghi} The original studies by RNGH
\cite{RN68,GH68}\ considered a perfectly resonant two--level ring
laser in the UFL. The Maxwell--Bloch equations that describe such
laser are given by Eqs. (\ref{ufl-f1})--(\ref{ufl-d1}) introduced
in Sect. \ref{model-ufl} with $\Delta =0$
\begin{eqnarray}
(\partial _{\tau }+\partial _{\zeta })F &=&\sigma (AP-F)\,,  \label{fg1} \\
\partial _{\tau }P &=&\gamma ^{-1}(FD-P)\,,  \label{pg1} \\
\partial _{\tau }D &=&\gamma \left[ 1-D-\mathrm{Re}\left( F^{\ast }P\right) %
\right] \,,  \label{dg1}
\end{eqnarray}
supplemented by the \textit{periodic} boundary condition
\begin{equation}
F(0,\tau )=F\left(\zeta_\mathrm{m},\tau
\right)\,,\qquad\zeta_\mathrm{m}=2\pi/\tilde\alpha\,.
\label{boug1}
\end{equation}
The steady, \textit{spatially uniform}, lasing solution is given
by Eq. (\ref{ssufl}) with $\Delta=0$
\begin{equation}
F_\mathrm{s}=\sqrt{A-1}\,\mathrm{e}^{i\phi}\,,\qquad P_\mathrm{s}=\frac{F_%
\mathrm{s}}{A}\,,\qquad D_\mathrm{s}=\frac{1}{A}\,,  \label{ss-ufl}
\end{equation}
where $\phi $ is an arbitrary phase, and it exists for $A>1$ ($A=1$
corresponds to the first laser threshold). This solution is the basic,
singlemode lasing solution. Other singlemode solutions exist (in fact an
infinite countable set) that can be obtained from Eqs. (\ref{fg1})--(\ref%
{boug1}) by setting $F\left( \zeta ,\tau \right)
=F_{\mathrm{s},n}\exp \left[ i\left( n\tilde\alpha\zeta -\omega_n
\tau \right) \right]$, $P\left( \zeta ,\tau \right)
=P_{\mathrm{s},n}\exp \left[ i\left(n\tilde\alpha\zeta
-\omega_n\tau \right) \right]$, $D\left( \zeta ,\tau \right)
=D_{\mathrm{s},n}$. These solutions were computed in Sect.
\ref{model-ufl} outside the UFL and correspond to different values
of the longitudinal index $n$. As we showed in that section, all
these additional singlemode solutions have a larger oscillation
threshold and hence solution Eq.~(\ref{ss-ufl}) is the first
lasing mode.

Stability of this singlemode solution is analyzed by perturbing the state of
the system as%
\begin{eqnarray}
F\left( \zeta ,\tau \right) &=&F_\mathrm{s}+e^{i\phi }\delta F\left( \zeta
,\tau \right) , \\
P\left( \zeta ,\tau \right) &=&P_\mathrm{s}+e^{i\phi }\delta P\left( \zeta
,\tau \right) , \\
D\left( \zeta ,\tau \right) &=&D_\mathrm{s}+\delta D\left( \zeta ,\tau
\right) ,
\end{eqnarray}%
and linearizing the dynamical equations for the perturbations. The study is
facilitated by expressing any perturbation $\delta X$ in terms of its real ($%
\delta X_{\mathrm{R}}$) and imaginary ($\delta X_{\mathrm{I}}$) parts ---
note that $\delta D$ is real by definition. The system of equations can be
expressed in vector form as%
\begin{equation}
\partial _{\tau }\overrightarrow{\delta X}=L\cdot \overrightarrow{\delta X},
\label{LSA}
\end{equation}%
where $\overrightarrow{\delta X}=\mathrm{col}\left( \delta F_{\mathrm{R}
},\delta P_{\mathrm{R}},\delta D,\delta F_{\mathrm{I}},\delta P_{\mathrm{I}
}\right) $ and
\begin{equation}
L=%
\begin{bmatrix}
-\sigma -\partial _{\zeta } & \sigma A & 0 & 0 & 0 \\
\gamma ^{-1}D_\mathrm{s} & -\gamma ^{-1} & \gamma ^{-1}\left\vert F_\mathrm{s%
}\right\vert & 0 & 0 \\
-\gamma \left\vert P_\mathrm{s}\right\vert & -\gamma \left\vert F_\mathrm{s}%
\right\vert & -\gamma & 0 & 0 \\
0 & 0 & 0 & -\sigma -\partial _{\zeta } & \sigma A \\
0 & 0 & 0 & \gamma ^{-1}D_\mathrm{s} & -\gamma ^{-1}%
\end{bmatrix}%
,  \label{matrixUFL}
\end{equation}%
with $F_\mathrm{s}$, $P_\mathrm{s}$, and $D_\mathrm{s}$ given by
Eq. (\ref{ss-ufl}).

Because of the linear nature of the system (\ref{LSA}), any solution $%
\overrightarrow{\delta X}$ can be calculated as $\overrightarrow{\delta X}%
\left( \zeta ,\tau \right) =\sum\nolimits_{\lambda }\overrightarrow{\delta X}%
_{\lambda }\left( \zeta \right) \mathrm{e}^{\lambda \tau }$. On the other
hand, given the dependence of $L$ on the spatial coordinate $\zeta $ only
through the gradient $\partial _{\zeta }$, any $\overrightarrow{\delta X}%
_{\lambda }\left( \zeta \right) $ can be written as $\overrightarrow{\delta X%
}_{\lambda }\left( \zeta \right) =\sum\nolimits_{\alpha }\overrightarrow{%
\delta X}_{\lambda ,\alpha }\exp \left( i\alpha \zeta \right) $. (Outside
the UFL, this ansatz is not valid; see below.) Hence any solution $%
\overrightarrow{\delta X}$ can be finally calculated as%
\begin{equation*}
\overrightarrow{\delta X}\left( \zeta ,\tau \right) =\sum\nolimits_{\lambda
,\alpha }\overrightarrow{\delta X}_{\lambda ,\alpha }\mathrm{e}^{\lambda
\tau +i\alpha \zeta }.
\end{equation*}%
Thus the dynamical problem (\ref{LSA}) is transformed into the following
eigenvalue problem%
\begin{equation*}
\lambda \overrightarrow{\delta X}_{\lambda ,\alpha }=L_{\alpha }\cdot
\overrightarrow{\delta X}_{\lambda ,\alpha }
\end{equation*}%
where $L_{\alpha }$ is given by matrix $L$ with $\partial _{\zeta }$
substituted by $i\alpha $. Given the block-diagonal form of $L_{\alpha }$
the characteristic polynomial $\mathcal{P}\left( \lambda ;\alpha \right) $
that determines the Lyapunov exponents $\lambda $ is factorized as%
\begin{eqnarray}
\mathcal{P}\left( \lambda ;\alpha \right) &=&\mathcal{P}_{\mathrm{I}}\left(
\lambda ;\alpha \right) \mathcal{P}_{\mathrm{R}}\left( \lambda ;\alpha
\right)\,,  \label{polRNGH} \\
\mathcal{P}_{\mathrm{I}}\left( \lambda ;\alpha \right) &=&\lambda
^{2}+\left( \gamma ^{-1}+\sigma +i\alpha \right) \lambda +i\gamma
^{-1}\alpha\,,  \label{polI} \\
\mathcal{P}_{\mathrm{R}}\left( \lambda ;\alpha \right) &=&\lambda
^{3}+\left(\gamma ^{-1}+\gamma +\sigma +i\alpha\right)\lambda ^{2}+  \notag
\\
&& \left[A+\gamma \sigma +i\alpha \left( \gamma ^{-1}+\gamma \right)\right]%
\lambda +2\sigma \left(A-1\right) +iA\alpha\,.  \label{polR}
\end{eqnarray}
These polynomials, equated to zero, allow to determine the dependence $%
\lambda \left( \alpha \right) $ of the eigenvalues on the spatial frequency $%
\alpha $ of the sidemode perturbation. The boundaries of the unstable domain
for the multimode instability correspond to cases where $\mathrm{Re}\lambda
\left( \alpha \neq 0\right) =0$ (for $\alpha =0$ one is considering the well
known singlemode, or Lorenz--Haken, laser instability, which is not treated
here). The singlemode solution will become unstable against sidemodes of
spatial frequency $\alpha $ if $\mathrm{Re}\lambda \left( \alpha \right) >0$.

It is easy to show that $\mathcal{P}_{\mathrm{I}}$ has no roots with $%
\mathrm{\ Re}\lambda \left( \alpha \right) >0$ for any $\alpha $. Only
marginally $\lambda \left( \alpha =0\right) =0\,$ is always a root. This
solution does not entail an instability as it never gets positive real part;
in fact that root is merely reflecting the phase arbitrariness of the lasing
solution. As $\mathcal{P}_{\mathrm{I}}$ is associated with the subspace
formed by $\mathrm{col}\left( \delta F_{\mathrm{I}},\delta P_{\mathrm{I}%
}\right) $, see Eqs. (\ref{LSA}) and (\ref{matrixUFL} ), and that
subspace controls the possible growth of perturbations in
phase-quadrature with the lasing mode, one concludes that the
RNGHI is not a phase instability but an amplitude instability
\cite{RN68,GH68}. This result remains valid outside the uniform
field limit, as well as when other factors (inhomogeneous
broadening, etc) are included, as far as the resonance condition
is maintained. As we discuss below (Sect. 2.2.4), detuning does
not have any influence in the stability of class--B lasers, which
are our main interest. Nevertheless, in class--A and class--C
lasers, detuning has a
large influence on the stability properties \cite%
{Gerber79,Zorell81,Narducci86}.

\begin{figure}[t]
\begin{center}
\scalebox{0.5}{\includegraphics{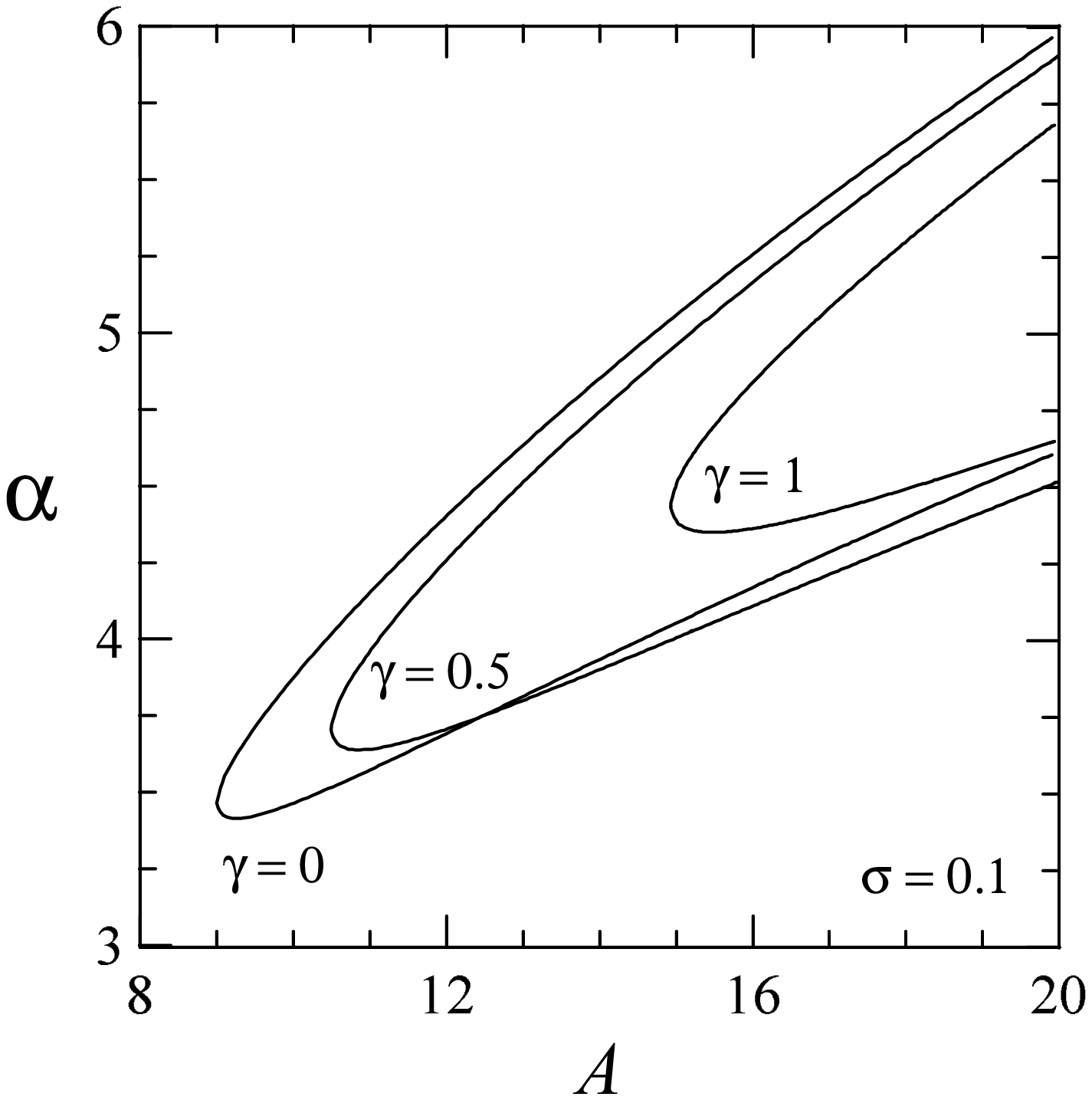}}
\end{center}
\caption{{}RNGHI threshold in the $\left\langle r,\protect\alpha %
\right\rangle $ plane for the three values of $\protect\gamma $ marked in
the figure.}
\label{fig:1}
\end{figure}

Regarding the polynomial $\mathcal{P}_{\mathrm{R}}$ --- which, \textit{\
mutatis mutandi}, governs amplitude instabilities --- the boundaries of the
multimode instability are found by setting $\lambda =-i\omega $. Upon
splitting the thus obtained polynomial (now in $\omega $) into its real and
imaginary parts and by equating them to zero one can solve for $\alpha $ and
$\omega $ as%
\begin{eqnarray}
\alpha _{\pm } &=&\omega _{\pm }\left( 1+\frac{\gamma \sigma }{A-\omega
_{\pm }^{2}}\right) ,  \label{tonguea} \\
\omega _{\pm }^{2} &=&\tfrac{1}{2}\left[ 3\left( A-1\right) -\gamma ^{2}\pm
\sqrt{R}\right] ,  \label{tonguew} \\
R &=&\left( A-1\right) \left( A-9\right) -6\gamma ^{2}\left( A-1\right)
+\gamma ^{4}.  \label{tongueR}
\end{eqnarray}%
The homogeneous solution turns out to be unstable for values of $\alpha $
verifying $\left\vert \alpha _{-}\right\vert \leq \left\vert \alpha
\right\vert \leq \left\vert \alpha _{+}\right\vert $. This unstable domain
has the shape of a tongue in the plane $<A,\alpha >$, and the two branches $%
\alpha _{\pm }$ merge at the critical point $\left( A_{\mathrm{c}},\alpha _{%
\mathrm{c}}\right) $ defined by the condition $R=0$. The critical pump
represents the minimum value of $A$ for which the instability can exist, and
is given by \cite{RN68,GH68}%
\begin{equation}
A_{\mathrm{c}}=5+3\gamma ^{2}+4\sqrt{1+\tfrac{1}{2}\gamma ^{2}\left(
3+\gamma ^{2}\right) }.  \label{Ac}
\end{equation}%
The value of $\alpha _{\mathrm{c}}$ is obtained from Eqs. (\ref{tonguea})
and (\ref{tonguew}) by setting $R=0$ and $A=A_{\mathrm{c}}$. The results of
this linear stability analysis are summarized in Fig. \ref{fig:1}.

\subsubsection{The RNGH instability in class B lasers}

\label{homo-ufl-classb}Class B lasers are defined by the inequalities $%
\gamma _{\Vert }\ll \kappa \ll \gamma _{\bot }$ concerning the
decay rates of the population inversion, intracavity field
amplitude, and medium polarization, respectively. With the used
normalizations those inequalities transform into $\gamma \ll
\sigma \ll \gamma ^{-1}$. This limit is especially interesting as
fiber lasers --- and erbium-doped fiber lasers in particular ---
belong to this class. On the other hand that limit is interesting
also from the theoretical viewpoint as it allows to obtain a
wealth of analytical information. The latter does not show up in
the \lq\lq classical\rq\rq version of the RNGHI we have just
considered, as the linear stability analysis of the singlemode
solution against multimode perturbations is completely analytical.
In this section we apply the class B limit to the general
expressions already obtained above. In the following section we
show how the same results can be derived using asymptotic
techniques that employ $\gamma $ as an expansion parameter.

The expressions for the instability boundaries in the class B limit are
retrieved from Eqs. (\ref{tonguea})--(\ref{tongueR}) by setting $\gamma =0$%
\begin{eqnarray}
\alpha _{\pm }^{2} &=&\tfrac{1}{2}\left[ 3\left( A-1\right) \pm \sqrt{\left(
A-1\right) \left( A-9\right) }\right] ,  \label{tongueclassB} \\
\omega _{\pm } &=&\alpha _{\pm }.  \label{omegaclassB}
\end{eqnarray}%
The critical pump $A_\mathrm{c}$ is obtained from Eq. (\ref{Ac}) by making $%
\gamma =0$ and reads%
\begin{equation}
A_\mathrm{c}=9,
\end{equation}%
which is in fact the minimum value $A_\mathrm{c}$ can attain. As
the first (lasing) threshold occurs at a pump $A=1$, we conclude
that in order to have a RNGHI the pump parameter $A$ should be at
least a factor of nine greater than its value at lasing threshold.
This is the famous \lq\lq factor--of--nine\rq\rq.

As for the sidemode critical frequency $\alpha _\mathrm{c}$, we obtain%
\begin{equation}
\alpha _\mathrm{c}=\sqrt{12}.  \label{alphac}
\end{equation}

So far we have assumed implicitly a continuum of longitudinal modes labelled
by their wavenumber offset $\alpha $. But it must be remembered that the
periodic boundary condition (\ref{boug1}) imposes, in particular, that the
perturbation wavenumber $\alpha $ must be an integer multiple of the scaled
cavity free spectral range $\tilde{\alpha}$, Eq. (\ref{scaledFSR})%
\begin{equation}
\alpha =\alpha _{n}=n\tilde{\alpha}=n\frac{2\pi c}{\mathcal{L}_\mathrm{c}%
\sqrt{ \gamma _{\Vert }\gamma _{\bot }}},\;n\text{ integer.}  \label{nalpha}
\end{equation}%
Thus, in order to have an instability at the lowest possible pump $A=A_%
\mathrm{c}$ the critical frequency $\alpha _\mathrm{c}$ must verify (\ref%
{nalpha}); alternatively, the cavity (optical) length $\mathcal{L}_\mathrm{c}
$ must verify
\begin{equation}
\mathcal{L}_\mathrm{c}=n\frac{\pi c}{\sqrt{3\gamma _{\Vert }\gamma
_{\bot }}},\;n \text{ integer.}  \label{Lccrit}
\end{equation}%
Hence the shortest cavity length that allows the RNGHI at $A=A_
\mathrm{c} $ is
\begin{equation}
\mathcal{L}_\mathrm{c,\min }=\frac{\pi c}{\sqrt{3\gamma _{\Vert }\gamma
_{\bot }}},  \label{Lcmin}
\end{equation}
which is the \lq\lq critical cavity length\rq\rq which we referred to in the
introduction.

Next we analyze the influence of the actual value of the cavity length (or,
alternatively, the consequences on the instability of discrete nature of the
cavity modes) on the instability. The analysis will clarify the meaning of $%
\mathcal{L}_{\mathrm{c,\min }}$ , which we anticipate is not a true minimum
(critical) value for the cavity length, but only the minimum value of $%
\mathcal{L}_{\mathrm{c}}$ for the instability to occur at the lowest pump.
\begin{figure}[t]
\begin{center}
\scalebox{0.5}{\includegraphics{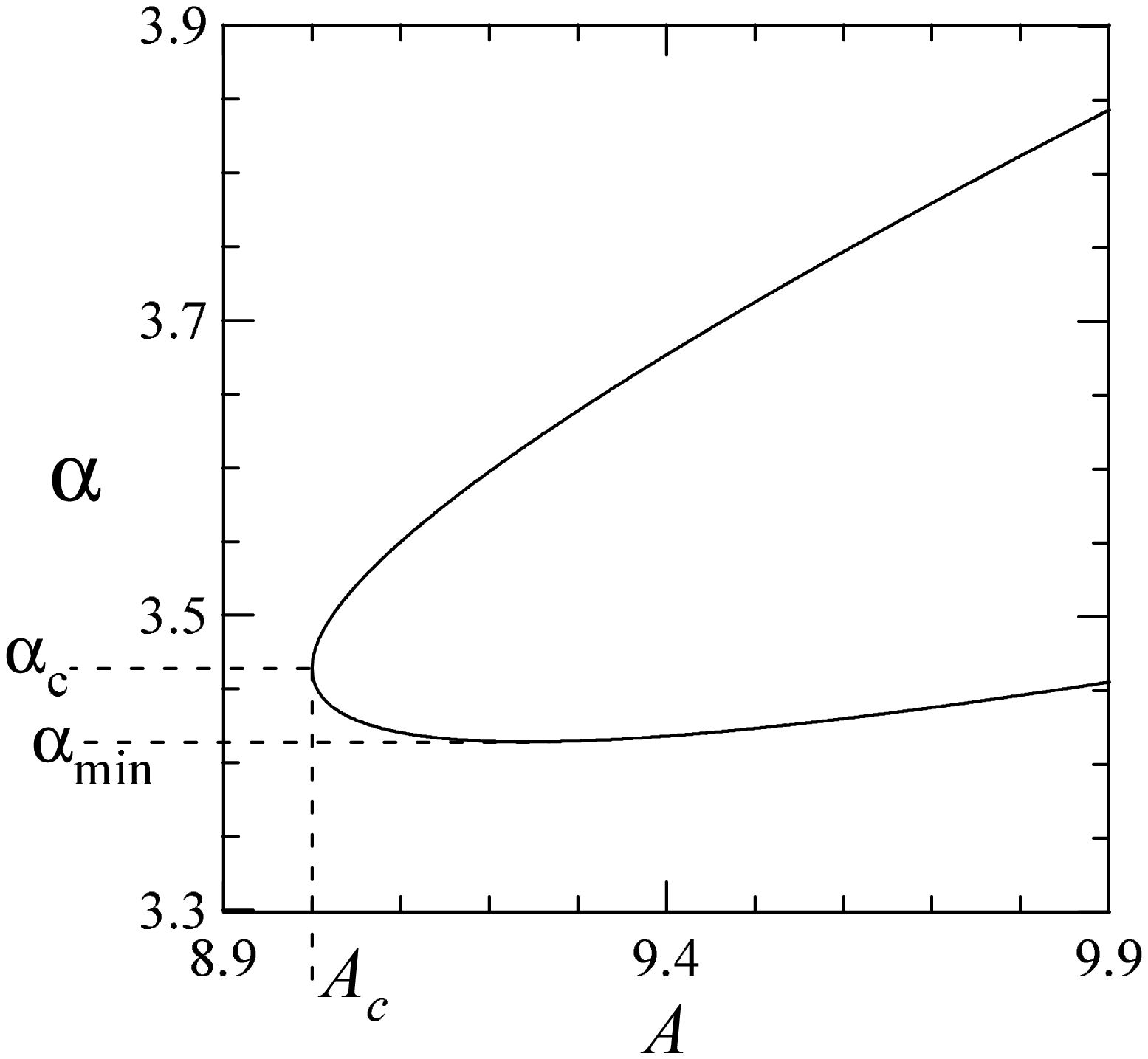}}
\end{center}
\caption{{}Enlargement of Fig. 1 for $\protect\gamma =0$. }
\label{fig:2}
\end{figure}
Let us consider the RNGHI boundary, Eq. (\ref{tongueclassB}),
shown in Fig. \ref{fig:2}. Besides the critical point $\left( A_{\mathrm{c}%
},\alpha _{\mathrm{c}}\right) $ another relevant point is the minimum of $%
\alpha _{-}$ vs. $A$, which is easily determined from Eq. (\ref{tongueclassB}%
) and evaluates to $\alpha _{\min }=2+\sqrt{2}$. This quantity is
relevant as it settles the minimum frequency of a sideband that
can be linearly amplified after the RNGHI. If the cavity is short
in the sense that its free spectral range $\tilde{\alpha}$ is
greater than $\alpha _{\min }$ (\textit{i.e.} if
$\mathcal{L}_{\mathrm{c}}<\mathcal{L}_{\star }\equiv \frac{2\pi
}{2+\sqrt{2}}\frac{c}{\sqrt{\gamma _{\Vert }\gamma _{\bot }}}$)
then the sideband with lowest threshold (which we denote by $A_{\mathrm{thr}%
} $) will be the first one (the one with $\alpha =\tilde{\alpha}$). This
threshold value is obtained from Eq. (\ref{tongueclassB}) and reads%
\begin{equation}
A_{\mathrm{thr}}=\frac{1}{4}\left( 2+3\alpha ^{2}-\sqrt{4-12\alpha
^{2}+\alpha ^{4}}\right) ,  \label{Athr}
\end{equation}%
where $\alpha $ must be substituted by $\tilde{\alpha}$ and the expression
is only valid for $\alpha \geq \alpha _{\min }=2+\sqrt{2}$ \footnote{%
We note that this equation corrects some typos appearing in Eq.
(45) of \cite{P99}.}. As by decreasing $\mathcal{L}_{\mathrm{c}}$,
$\tilde{\alpha}$ increases, the RNGHI is produced at increasing
pump values according to Eq. (\ref{Athr}) so that
$A_{\mathrm{thr}}\rightarrow \infty $ for
$\mathcal{L}_{\mathrm{c}}\rightarrow 0$. This reasoning evidences
that $\mathcal{L}_{\mathrm{c,\min }}$ is not a true minimum value
for the cavity length in order to observe the RNGHI, contrarily to
what is commonly believed. Clearly, by lowering
$\mathcal{L}_{\mathrm{c}}$ $A_{\mathrm{thr}}$ can become so huge
that, in practice, the RNGHI can be ruled out. But we stress that
this is a practical, not a fundamental, limitation. On the other
hand when $\mathcal{L}_{\mathrm{c}}>\mathcal{L} _{\star }$ then
$\tilde{\alpha}<\alpha _{\min }$ and the instability is produced
at a higher order sideband. The identification of the sideband
with lowest threshold is now a more involved task (see \cite{P99}
for a discussion). In general, for each value of
$\mathcal{L}_{\mathrm{c}}$, $A_{ \mathrm{thr}}$ must be computed
from Eq. (\ref{Athr}) by taking $\alpha =n \tilde{\alpha}$ with
$n=1,2,3,\ldots $ and the lowest value of $A_{\mathrm{\ thr}}$
corresponds to the actual instability threshold value. Fig. \ref
{fig:3} displays the actual RNGHI threshold as a function of the
cavity length. We note that the graph has periodically recurring
minima. This is due to the fact that whenever Eq. (\ref{Lccrit})
is verified, $A_{ \mathrm{thr}}=A_{\mathrm{c}}=9$.
\begin{figure}[t]
\begin{center}
\scalebox{0.5}{\includegraphics{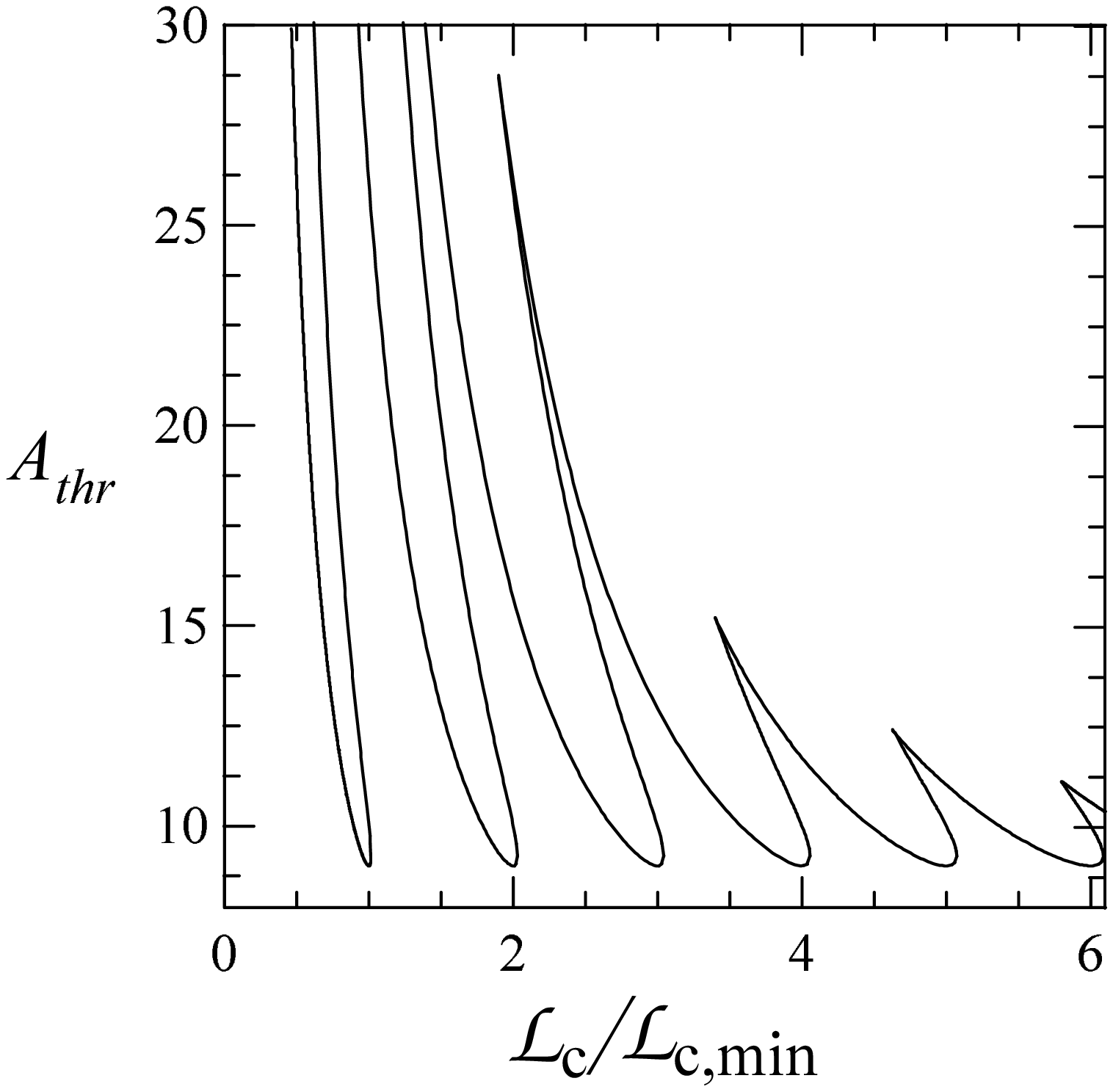}}
\end{center}
\caption{RNGHI threshold as a function of the cavity length for $\protect%
\gamma =0$.}
\label{fig:3}
\end{figure}
All the above analysis can be applied to three-- and four--level
lasers\ making use of the transformations defined in Sect.
\ref{model-3lev4lev}: (i) replace $A$ by using Eq. (\ref{A1}), and
(ii) replace the spatial frequency $ \alpha $ by $\alpha
\sqrt{1+W}$ in the final expressions (remember that $W$ is the
actual pumping strength). As already advanced in Sect. \ref
{model-3lev4lev}, for four--level lasers the situation is quite
similar to that of two--level lasers, but for three--level lasers
the scenario changes dramatically \cite{P99} as we discuss next.
On the one hand the critical
value for two--level lasers $\alpha _{\mathrm{c}}=\sqrt{12}$, Eq. (\ref%
{alphac}), must be replaced by $\sqrt{12\left( 1+W_{\mathrm{c}}\right) }$
where $W_{\mathrm{c}}$ is the minimum (critical) instability threshold
referred to the actual optical pumping $W$, which is given by
\begin{equation}
W_{\mathrm{c}}=\frac{G+9}{G-9}.
\end{equation}%
Here $G$ is the gain coefficient for three--level lasers defined
in Eq. (\ref {G}). For $G\gg 1$, as is typical in erbium-doped
fiber lasers \cite{P99}, $W_{\mathrm{c}}\rightarrow 1$ and then
$\alpha _{\mathrm{c}}^{\left( 3L\right) }=\sqrt{12\left(
1+W_{\mathrm{c}}\right) }\rightarrow \sqrt{24}$, which is a factor
$\sqrt{2}$ larger than for two--level lasers. With respect to
$\mathcal{L}_{\mathrm{c,\min }}$
\begin{equation}
\mathcal{L}_{\mathrm{c,\min }}^{\left( 3L\right) }=\frac{\pi c}{\sqrt{%
6\gamma _{\Vert }\gamma _{\bot }}}\sqrt{\frac{G-9}{G}}\simeq \frac{\pi c}{%
\sqrt{6\gamma _{\Vert }\gamma _{\bot }}}
\end{equation}%
This value of $\mathcal{L}_{\mathrm{c,\min }}$ is $\sqrt{2}$ times
smaller than the corresponding two--level value (\ref{Lcmin}).
However the most fascinating difference with two--level lasers is
the extremely low instability threshold obtained in three--level
lasers. Denoting by $W_{\mathrm{on}}$ the lasing threshold
referred to $W$, one has $W_{c}/W_{\mathrm{on}}\simeq 1+16/G$ when
$G\gg 1$, as already discussed in Sect. \ref {model-3lev4lev} (see
Eq. (\ref{ratio3L})). Thus in erbium-doped fiber lasers the RNGHI
is predicted to occur just above threshold \cite{R98,P99}.
Finally, the variation of the instability threshold with the
cavity length in three--level lasers is much less pronounced than
in two--level lasers and, in fact, the cavity can become
substantially shorter than $\mathcal{L}_{\mathrm{c,\min }}$ while
the instability threshold keeps moderate values; see Fig. 6 in
\cite{P99}.

\subsubsection{Asymptotic expansions for class B lasers}

\label{homo-ufl-expansion}The results discussed in the previous
section have been obtained by applying the class B limit $\gamma
\rightarrow 0$ to the general expressions obtained in Sect.
\ref{homo-ufl-rnghi}, which are valid for arbitrary values of the
laser parameters. In other instances (e.g., outside the uniform
field limit) this strategy is not possible as general expressions
are not available. In such cases one can still obtain a lot of
analytical information if the study is done, \textit{ab initio},
for class B lasers. Those treatments rely on asymptotic expansions
of the problem that use $\gamma $ as a smallness parameter.
Although in the UFL this analysis is not necessary, we prefer to
introduce the technique at this point as here the explanations are
more transparent and straightforward than in more complex cases
that will be considered below.

The starting point of the analysis is the characteristic
polynomial governing the RNGHI, which must be equated to zero in
order to obtain the eigenvalues $\lambda $ that govern the
stability of the singlemode lasing solution. In the UFL it is
given by Eq. (\ref{polR}), which we recall for convenience
\begin{eqnarray}
\mathcal{P}_{\mathrm{R}}\left( \lambda ;\alpha \right) &=&\lambda
^{3}+\left( \gamma ^{-1}+\gamma +\sigma +i\alpha \right) \lambda ^{2}+
\notag \\
&&\left[ A+\gamma \sigma +i\alpha \left( \gamma ^{-1}+\gamma
\right) \right] \lambda +2\sigma \left( A-1\right) +iA\alpha
\,.\notag
\end{eqnarray}%
(We ignore the polynomial $\mathcal{P}_{\mathrm{I}}$ associated with the
phase as we showed it does not contain any instability.) Now, as $0<\gamma
\ll 1$ defines class B lasers (together with $\sigma \sim \gamma ^{0}$), we
assume the following ansatz%
\begin{equation}
\lambda =\lambda _{0}+\gamma \lambda _{1}+\gamma ^{2}\lambda _{2}+\cdots ,
\end{equation}%
for the eigenvalues. The ansatz is substituted into
$\mathcal{P}_{\mathrm{R}}\left( \lambda ;\alpha \right)$ and the
resulting polynomial is expanded in series of $\gamma $ as%
\begin{equation}
\mathcal{P}_{\mathrm{R}}=\sum\limits_{n=N}^{\infty }\gamma ^{n}p_{n}\left(
\left\{ \lambda _{i}\right\} ;\alpha \right) ,  \label{Pexpand}
\end{equation}%
and, in our case, the expansion (\ref{Pexpand}) starts at $N=-1$.

Now $\mathcal{P}_{\mathrm{R}}$ must be equated to zero. As the expansion is
assumed to be uniformly valid for any $\gamma $, we impose that each of the $%
p_{n}$ be null. Starting at the leading order $n=N\left( =-1\right) $, we
have%
\begin{equation*}
p_{-1}=\lambda _{0}\left( \lambda _{0}+i\alpha \right) ,
\end{equation*}%
which has the roots
\begin{equation}
\lambda_{0}=0 \quad \mbox{and} \quad \lambda_{0}=-i\alpha .  \label{lambda0}
\end{equation}%
The next order ($n=0$) reads%
\begin{equation}
p_{0}=\lambda _{0}\lambda _{1}+\left( \lambda _{0}+i\alpha \right) \left(
\lambda _{1}+A+\lambda _{0}^{2}\right) +2\sigma \left( A-1\right) +\sigma
\lambda _{0}^{2}.  \label{p0}
\end{equation}%
Making use of the first root $\lambda _{0}=0$ and equating $p_{0}$ to zero
we obtain $\lambda _{1}=-A+2i\sigma \left( A-1\right) /\alpha $. As $\mathrm{%
Re}\lambda _{1}<0$ the roots corresponding to $\lambda _{0}=0$ do not entail
an instability and can be discarded. Then we must seek instabilities
associated with $\lambda _{0}=-i\alpha $, Eq. (\ref{lambda0}). Substituting
this root into Eq. (\ref{p0}) and making $p_{0}=0$, we obtain
\begin{equation}
\lambda _{1}=i\sigma\alpha \left[1
-\frac{2}{\alpha^2}\left(A-1\right)\right] . \label{lambda1}
\end{equation}%
As $\mathrm{Re}\lambda _{1}=0$ we must continue the analysis. At the next
order ($n=1$), once Eqs. (\ref{lambda0}) and (\ref{lambda1}) have been used,
and after making $p_{1}=0$ we obtain%
\begin{equation}
\lambda _{2}=-\frac{\sigma }{\alpha ^{2}}\left[ \alpha
^{4}-3\alpha^2\left( A-1\right) +2A\left( A-1\right) \right]
-i\sigma ^{2}\alpha\left[1-\frac{4}{\alpha^4}\left( A-1\right)
^{2}\right] .  \label{lambda2}
\end{equation}%
The analysis can stop here as $\mathrm{Re}\lambda _{2}$ is not identically
zero. The instability boundary $\mathrm{Re}\lambda =0$ reads in this case $%
\mathrm{Re}\lambda _{2}=0$, which exactly reduces to the instability
boundary for class B lasers Eq. (\ref{tongueclassB}), or Eq. (\ref{Athr}),
as can be checked easily. Regarding the oscillation frequency at the
instability threshold, which we denoted by $\omega $ in the previous
sections, it corresponds to $\mathrm{Im}\lambda $. To the leading order $%
\mathrm{Im}\lambda = \mathrm{Im}\lambda _{0}=-\alpha $, Eq. (\ref{lambda0}).
Thus it is predicted that $\omega =\alpha $, in agreement with Eq. (\ref%
{omegaclassB}).

This kind of analysis will be used later in order to treat more involved
problems which, usually, do not admit general analytical expressions to
which the class B limit $\gamma \rightarrow 0$ can be applied.

To conclude, we note that the above expansion assumes implicitly that all
parameters are of order $\gamma ^{0}$. Obviously, additional scalings can be
incorporated if needed, as it occurs in the next section. For instance, in
order to fully understand the bifurcation in the presence of inhomogeneous
broadening, the scaling $\alpha $ $=\gamma ^{-1}\alpha _{-1}$ must be
studied separately \cite{Roldan01a}.

\subsubsection{Role of detuning in class B lasers}

\label{homo-ufl-detuning} Throughout this paper we shall be
dealing with resonant models, as we have just done in the previous
sections. Here we prove that detuning between the cavity and the
gain line, which is almost unavoidable in real experiments, can be
ignored in class B lasers. This is important as it demonstrates
that the predictions of resonant models remain valid, to the
leading order, even when cavity detuning is present. The analysis
will be done in the UFL for the sake of simplicity. Thus the
starting point of the analysis are Eqs.
(\ref{ufl-f1})--(\ref{ufl-d1})
\begin{eqnarray}
(\partial _{\tau }+\partial _{\zeta })F &=&\sigma \left[ AP-\left( 1-i\Delta
\right) F\right] \,,  \notag \\
\partial _{\tau }P &=&\gamma ^{-1}\left[ FD-\left( 1+i\Delta \right) P\right]
\,,  \notag \\
\partial _{\tau }D &=&\gamma \left[ 1-D-\mathrm{Re}(F^{\ast }P)\right]
\,,\notag
\end{eqnarray}
with the boundary condition (\ref{bouufl})
\begin{equation}
F(0,\tau )=F\left(\zeta_\mathrm{m},\tau
\right)\,,\qquad\zeta_\mathrm{m}=2\pi/\tilde\alpha\,. \notag
\end{equation}
where $\tilde{\alpha}$ is the scaled cavity free spectral range,
Eq. (\ref {scaledFSR}), and
\begin{equation}
\Delta =\frac{\delta}{1+\gamma\sigma}\,.
\end{equation}%
As pointed out in the derivation of the equations in the UFL, by definition $%
|\delta|\le\gamma\tilde\alpha/2$. In the class B limit $\gamma \ll 1$ and $%
\sigma =\mathcal{O}\left( \gamma ^{0}\right) $, hence $|\Delta|\leq \tfrac{1%
}{2}\gamma \tilde{\alpha}$ to the leading order. Next we determine the
maximum admissible order of magnitude of $\Delta $. For that we note, based
on the results of the resonant model discussed in the previous sections,
that $\tilde{\alpha}$ must be at most a quantity of order $\gamma ^{0}$ for
the instability to occur at physically realizable pumping levels. Hence $%
\Delta $ is, at most, of order $\gamma $. This is shown by the
following argument: As the case of small $\tilde{\alpha}$ (say
$\tilde{\alpha}\sim \gamma ^{k}$ with $k=1,2,\ldots $) poses no
problem (in this case $\Delta $ would be still smaller) we
concentrate on the large free spectral range limit
$\tilde{\alpha}\gg 1$. In this case, as discussed in Sect. \ref
{homo-ufl-classb}, the first sideband ($\alpha =\tilde{\alpha}$)
is the one with lowest instability threshold, whose value is given
by Eq. (\ref{Athr}). The asymptotic form of such equation in the
limit $\tilde{\alpha}\gg 1$ reads
\begin{equation*}
A_{\mathrm{thr}}\overset{\tilde{\alpha}\gg 1}{\longrightarrow }\tfrac{1}{2}%
\tilde{\alpha}^{2}.
\end{equation*}%
Then, if $\tilde{\alpha}\sim \gamma ^{-1}$, $A_{\mathrm{thr}}\sim \gamma
^{-2}\sim 10^{10}$ (for typical fiber lasers $\gamma \sim 10^{-5}$), and
this is nonsense\footnote{%
Even for a moderate value as $\tilde{\alpha}=20$ (which is still of order $%
\gamma ^{0}$), the actual instability threshold is $A_{\mathrm{thr}}=202.01$%
, which is clearly a huge pumping ratio.}. The conclusion is that, unless
completely unrealistic pumping values are considered, $\tilde{\alpha}$ must
be a quantity of order $\gamma ^{0}$ (at most) and then an appropriate
scaling for class B lasers is $\Delta =\gamma \Delta _{1}$, with $\Delta
_{1} $ of order $\gamma ^{0}$.

The singlemode solution with lowest threshold is the stationary solution
given by Eq. (\ref{ssufl}), that we recall here for convenience
\begin{equation}
F_{\mathrm{s}}=\sqrt{A-1-\Delta ^{2}}\,\mathrm{e}^{i\phi
}\,,\qquad P_{\mathrm{s}}= \frac{F_{\mathrm{s}}}{A}\left(
1-i\Delta \right) \,,\qquad D_{\mathrm{s}}= \frac{1+\Delta
^{2}}{A}\,.\notag
\end{equation}%
For $\Delta =0$ this solution reduces to the resonant singlemode
solution (\ref{ss-ufl}) analyzed in the previous sections. The
lasing threshold $1+\Delta ^{2}$ differ from that of a perfectly
resonant laser only by a term of order $\gamma ^{2}$ as $\Delta $
is of order $\gamma $.

The stability of the stationary solution can be studied as in
Sect. \ref {homo-ufl-rnghi}. In this case however, the presence of
the atomic detuning $\Delta$ introduces an entanglement between
the real and imaginary parts of the fluctuations, and the fifth
order characteristic polynomial cannot be any longer factorized as
in the resonant case, Eq. (\ref{polRNGH}). The analysis of the
eigenvalues is much more complicated and it is convenient to apply
the asymptotic expansion presented in Sect.
\ref{homo-ufl-expansion}. We do not give the details of the
derivation, which is straightforward but lengthy. We just want to
emphasize that the relevant eigenvalue obtained in this way reads
$\lambda =-i\alpha +\gamma \lambda _{1}+\gamma ^{2}\lambda
_{2}+O\left( \gamma ^{3}\right) $, where $\lambda _{1}$ and
$\lambda _{2}$ are the very same we obtained in the resonant case
analyzed in Sect. \ref{homo-ufl-expansion}. In particular this
means that the detuning $\Delta =\gamma \Delta _{1}$ does not have
any influence to the leading order (probably it appears at
corrections of order $\gamma ^{3}$ or smaller, but these do not
control the instability). The conclusion is then clear: In class B
lasers the influence of the cavity detuning does not show up to
the leading order. Then one can safely ignore that detuning and
the results obtained in resonant models can be considered as an
excellent approximation to detuned models.

\subsubsection{Spatial effects}

\label{homo-ufl-tra}

When the plane--wave approximation is abandoned and one assumes, more
realistically, that the laser operates on the fundamental Gaussian mode (%
\textsf{TEM}$_{00}$), the RNGHI (as well as the Lorenz--Haken instability)
turns out to be strongly influenced by the relative size of the beam and the
amplifying medium. If $r_{\mathrm{m}}$ is the radius of the amplifying
medium and $w_{0}$ the beam waist, the relevant parameter is $u_{\mathrm{m}
}=2(r_{\mathrm{m}}/w_{0})^{2}$. In two--level lasers it was demonstrated
that the instability disappears in the limit $u_{\mathrm{m}}\gg 1$ \cite%
{LugiatoMilani83,Stuut84,LugiatoMilani85}, but it is recovered as soon as $%
u_{\mathrm{m}}$ becomes of order unity \cite{SmithDykstra,Urchueguia98}.
This result can be easily understood considering that in the limit $u_{
\mathrm{m}}\ll 1$ of very narrow amplifying medium the model equations for a
Gaussian beam reduce to those of the plane--wave approximation, because the
wavefront can be assumed to be plane inside the amplifying medium. The
results of \cite{Urchueguia98} apply well to fibre lasers, where the dopant
is usually confined inside the fibre core, whose radius $r_{\mathrm{core}}$
is typically smaller than the waist of the laser beam. Yet, we know that
when considering fibre lasers one must distinguish between three-- and
four--level atoms, because of the different dependence on the pump of the
equilibrium population difference $d_{0}$.

The Maxwell-Bloch equations for a Gaussian beam in three-- and
four--level lasers are Eqs. (\ref{tra-f}--\ref{tra-d}), derived in
Sect. \ref{mode-tra}. Here we limit our analysis to the UFL.
Following the same steps as in Sect. \ref{model-ufl}, it can be
easily shown that the Maxwell-Bloch equations in the resonant case
in the UFL are \cite{Roldan03a,Urchueguia00}
\begin{eqnarray}
\left( \partial _{\tau }+\partial _{\zeta }\right) F &=&\sigma \left[
G\int_{0}^{u_{\mathrm{m}}}\!\!\!du\,\mathrm{e}^{-u/2}\,P-F\right] \,.
\label{tra-fufl} \\
\partial _{\tau }P &=&\gamma ^{-1}\left( \mathrm{e}^{-u/2}FD-P\right) \,,
\label{tra-pufl} \\
\partial _{\tau }D &=&\gamma \left[ \beta \mathrm{e}^{-\eta u}(1-D)-\delta
_{N,3}-D-\mathrm{e}^{-u/2}FP \right] \,. \label{tra-dufl}
\end{eqnarray}%
with the periodic boundary condition
\begin{equation}
F(0,u,\tau )=F(\zeta_{\mathrm{m}},u,\tau
)\,,\qquad\zeta_{\mathrm{m}}=2\pi/\tilde\alpha\,.
\end{equation}%
Here we assumed without loss of generality that both $F$ and $P$ are real.
We recall that $\beta $ is the intensity of the pump field, and the
parameter $\eta $ is defined as $\eta =(w_{0}/w_\mathrm{p})^{2}$, where $w_%
\mathrm{p}$ is the waist of the pump beam. The parameter $\delta _{N,3}$,
which is equal to 1 for three--level lasers and to 0 for four--level lasers,
makes the difference between the two situations.

Transverse effects in Erbium--doped (three--level) fibre lasers were
investigated in \cite{Urchueguia00}. It was assumed that the dopant is
confined inside the core of the fibre, \textit{i.e.} $r_{\mathrm{m}}\leq r_{%
\mathrm{core}}$. The ratio
\begin{equation}
x_{\mathrm{dop}}=\frac{r_{\mathrm{m}}}{r_{\mathrm{core}}}\,
\end{equation}%
was used as a free parameter, ranging 0 to 1. For fibre lasers of the type
used in \cite{P97} adequate parameters are $w_\mathrm{p}=2.086\,\mu $m and $%
w_{0}=2.828\,\mu $m, which implies $\eta =1.839$, and $r_{\mathrm{core}
}=1.875\,\mu $m \cite{P97,Desurvire}. Since $u_{\mathrm{m}}$ can be written
as $u_{\mathrm{m}}=2(r_{\mathrm{core}}/w_{0})^{2}x_{\mathrm{dop}}^{2}$, with
our parameters we have $u_{\mathrm{m}}=0.879\,x_{\mathrm{dop}}^{2}$.
\begin{figure}[t]
\begin{center}
\scalebox{0.5}{\includegraphics{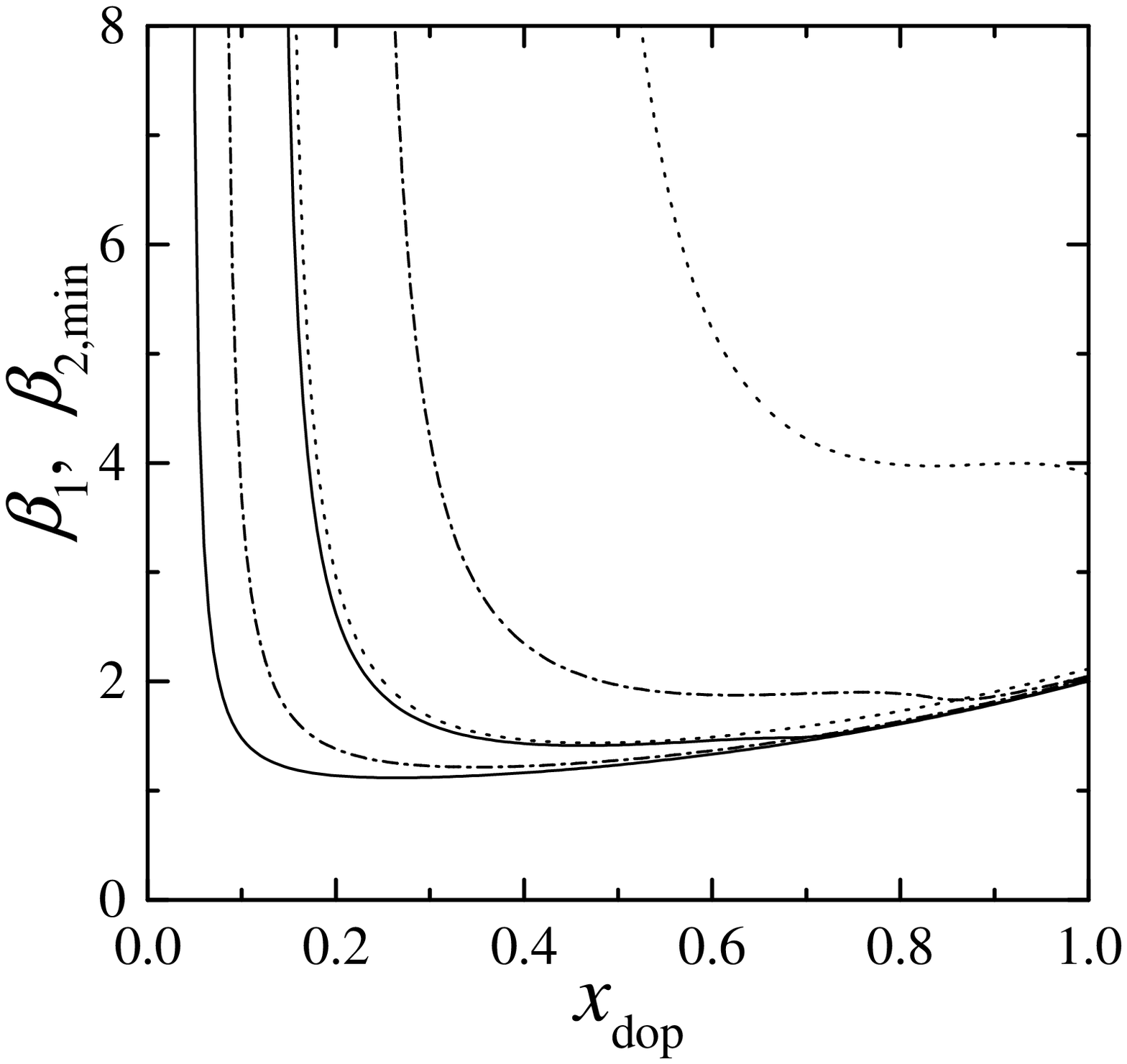}}
\end{center}
\caption{Lasing threshold $\protect\beta _{1}$ and minimum
instability threshold $\protect\beta _{2,min}$ as functions of the
scaled radius of the doped region $x_{\mathrm{dop}}$ for three
values of the gain parameter $G$: $G=600$ (solid lines), $G=200$
(dashed--dotted lines) and $G=60$ (dotted lines). For each value
of $G$ the singlemode solution exists and is stable in the region
between the two lines.} \label{fig:4}
\end{figure}
\begin{figure}[t]
\begin{center}
\scalebox{0.5}{\includegraphics{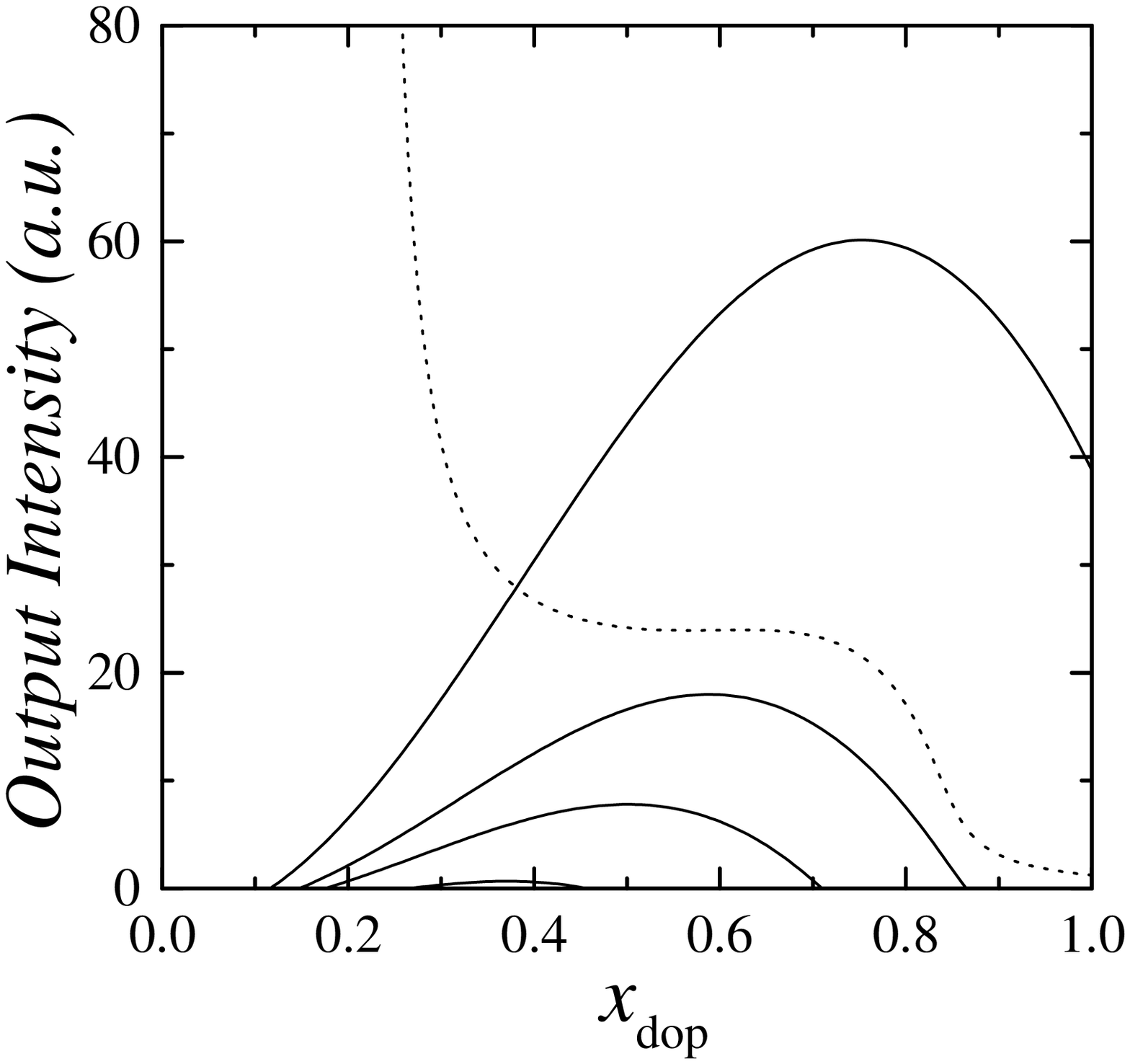}}
\end{center}
\caption{Solid lines: output intensity as a function of the scaled
radius of the doped region $x_{\mathrm{dop}}$ for $G=200$ and for
different values of the pump intensity, $\protect\beta
=1.25,\,1.5,\,1.75,\,2.5$ from bottom to top. Dashed line: output
intensity in correspondence of the minimum instability threshold
$\protect\beta _{2,\min }$.} \label{fig:5}
\end{figure}
The proper pump parameter now is the pump intensity $\beta $. We call $\beta
_{1}$ the lasing threshold and $\beta _{2}$ the threshold for the RNGHI. The
minimum instability threshold, which is the one corresponding to the
critical frequency space $\alpha _{\mathrm{c}}$ where the upper and lower
instability boundaries merge, is denoted by $\beta _{2,\min }$.

A first important result is shown in Fig. \ref{fig:4}, where the dependence
of $\beta _{1}$ and $\beta _{2,\min }$ on the normalized doped radius $x_{%
\mathrm{dop}}$ is shown for three experimentally accessible values of the
gain parameter $G$ . All curves display a vertical asymptote for small doped
regions, and it may be noted that, for each value of $G$, the asymptote for $%
\beta _{2,\min }$ corresponds to a value of $x_{\mathrm{dop}}$ which is
about three times larger than the one associated with the asymptote for $%
\beta _{1}$. The existence of these asymptotes can be explained making
reference to the plane--wave limit, to which the model reduces in the limit $%
x_{\mathrm{dop}}\rightarrow 0$ (which implies $u_{\mathrm{m}}\rightarrow 0$)
\cite{Urchueguia00}. It can be shown that the two asymptotes are well
approximated by $x_{\mathrm{dop}}=1/\sqrt{0.879G}$ and $x_{\mathrm{dop}}=3/%
\sqrt{0.879G}$, where the factor $0.879$ comes from $u_{\mathrm{m}%
}=0.879\,x_{\mathrm{dop}}^{2}$.

Thus, there exists a band of values of $x_{\mathrm{dop}}$ between the two
asymptotes, where singlemode lasing is stable for all pump strengths. One is
tempted to conclude that to stabilize the laser it suffices to dope only a
small fraction of the fibre core. But Fig. \ref{fig:4} also tells us that
the lasing threshold $\beta _{1}$ rapidly increases for small $x_{\mathrm{dop%
}}$, which means that, for a given pump $\beta $, the output intensity
decreases. A compromise between stability and not too low output intensity
must be found.

Fig. \ref{fig:5} shows how the output intensity $I$ depends on $x_{\mathrm{%
dop}}$ for four increasing values of the pump intensity $\beta $, and $G=200$%
. All curves are similar and display a maximum that shifts to
larger doped area radii as $\beta $ increases. Also displayed, as
a dashed line, is the critical intensity $I_{2,\min }$ associated
with the minimum instability threshold $\beta _{2,\min }$. All
stationary intensities in Fig. \ref{fig:5} which are above
$I_{2,\min }$ are unstable (multimode emission takes place). We
show only the case $G=200$ for the sake of clarity. The vertical
asymptote of $I_{2,\min }$ corresponds to that of $\beta _{2,\min
}$ in Fig. \ref{fig:4} for $G=200$. Interestingly, $I_{2,\min }$
shows a relatively large plateau for $0.4\leq x_{\mathrm{dop}}\leq
0.7$. In this region the maximum stable singlemode intensity
available is almost insensitive to $x_{\mathrm{dop}}$, what
constitutes a genuine transverse effect, not explainable in the
plane--wave limit.

The best operating conditions are given by an interplay between plane--wave
limit and transverse effects. In the particular case considered ($G=200$),
the plane--wave limit indicates that stable singlemode emission is
guaranteed if $0.1\leq x_{\mathrm{dop}}\leq 0.25$, but in this region very
large values of the pump intensity are required in order to obtain a large
output intensity. However, the particular dependence of the output intensity
on $x_{\mathrm{dop}}$, which is a consequence of the transverse profile of
the laser and pump beams, shows that stable emission with sufficiently high
intensity can be achieved for smaller pump intensities when the radius of
the doped area is larger ($x_{\mathrm{dop}}\simeq 0.6$).

A similar analysis was performed in Ref. \cite{Roldan03a} for a ring Nd:YAG
(four--level) laser using Eqs. (\ref{tra-fufl}--\ref{tra-dufl}) with $\delta
_{N,3}=0$. At difference with the erbium--doped fiber laser, in the case of
a bulk Nd:YAG laser the dopant distribution can be assumed to be
homogeneous, and to some approximation the same holds for the pump (at least
in side-pumped designs). Then, the only transverse effects that have to be
taken into account are the Gaussian profile of the laser field and the
finite dimension of the active medium (the laser rod). In \cite{Roldan03a}
it was demonstrated that the instability occurs whenever the beam waist $%
w_{0}$ is slightly larger, or at least equal to $r_{\mathrm{m}}$,
the radius of the active medium's rod. To accomplish this is not a
trivial task, as it is discussed in detail in \cite{Roldan03a},
but it seems feasible with present day technology. As Nd is a
four--level medium, in this case the instability to lasing
threshold ratio is slightly larger than 9 in the optimal
conditions (see Sect. \ref{model-3lev4lev}), \textit{i.e.}, in
this case the instability threshold is well separated from the
lasing one. We refer the reader to \cite{Roldan03a} for full
details.

\subsection{Outside the uniform field limit}

\label{homo-nufl}

The study of lasers instabilities outside the uniform field limit is in
general a difficult task, because even the stationary state is spatially
dependent \cite{Lugiato86b}. But the best candidates for the observation of
the RNGHI are fibre lasers, for which the UFL usually does not apply,
because the effective reflectivity parameter $\mathcal{R}$ is far from
unity. On the other hand, these lasers satisfy very well the class--B limit,
because the parameter $\gamma $ is as small as $10^{-5}$ \cite%
{P97,P99,Voigt01,Voigt04}. This is a fortunate circumstance because, as we
shall show in the present section, at least in the study of the RNGHI, the
class--B limit is as powerful as the UFL, in the sense that the analytical
results obtained in the UFL can be generalized to arbitrary $\mathcal{R}$ in
class-B lasers.

Specifically, in this section we will show that for class--B
lasers outside the UFL the following results hold: (i) the
boundaries of the stability domain of the homogeneous solution can
be determined analytically, (ii) the simple self--pulsing
solutions can be studied in a completely analytic way, (iii) the
multimode dynamics arising from the instability can be described
in terms of a limited number of Fourier modes. In all cases cavity
detuning will be ignored after the discussion in Sect.
\ref{homo-ufl-detuning}.

This section is organized as follows: In Sect.
\ref{homo-nufl-rnghi} we derive the RNGHI outside the UFL. Then we
generalize the result by taking into account distributed losses
(Sect. \ref{homo-nufl-loss}). In Sect. \ref{homo-nufl-rate} we
derive generalized rate equations for class--B lasers, and in
Sect. \ref{homo-nufl-pulse} we use those equations to analyze
self--pulsing. In Sect. \ref{pulse-numres} we discuss the issue of
the numerical integration of the model and present some numerical
results in order to compare the different integration methods.
Finally, in Sect. \ref{homo-nufl-sub} we address the issue of the
super-- or sub--criticality of the bifurcation.

\subsubsection{RNGH outside the UFL}

\label{homo-nufl-rnghi}

First we consider the case when no distributed losses exist in the
laser cavity ($\chi =0$) and recall the Maxwell--Bloch equations
(\ref{eqF3})--(\ref{eqD3}) valid in the resonant case
\begin{eqnarray}
\left( \partial _{\tau }+\partial _{\zeta }\right) F &=&\sigma
AP\,, \\
\partial _{\tau }P &=&\gamma ^{-1}\left( FD-P\right) \,, \\
\partial _{\tau }D &=&\gamma \left( 1-D-FP\right) \,,
\end{eqnarray}%
together with the boundary condition
\begin{equation}
F(0,\tau )=\mathcal{R}F(\zeta_{\mathrm{m}},\tau
)\,,\qquad\zeta_{\mathrm{m}}=2\pi/\tilde\alpha\,,
\label{bouGnonUFL}
\end{equation}%
where $\zeta =0$ and $\zeta =\zeta_{\mathrm{m}}$ denote the
location of the entry and exit faces of the active medium, and
$\tilde{\alpha}$ is the scaled cavity free spectral range, Eq.
(\ref{scaledFSR}).

In these equations we have assumed that the fields $F$ and $P$ are
real because we showed in Sect. \ref{homo-ufl-rnghi} that in the
resonant case and in the UFL, the RNGHI is an amplitude
instability. Lifting the UFL does not change the nature of the
bifurcation, as can be checked following the lines of our study in
Sect. \ref{homo-ufl-rnghi}.

Unlike in the UFL, now all variables at steady state are in
general functions of the coordinate $\zeta $, as we showed in
Section \ref{model-ss} . According to Eq. (\ref{ischizero}) the
steady value of the field intensity $I_\mathrm{s}=\left\vert
F_\mathrm{ s}\right\vert ^{2}$ at the medium exit face is given by
\begin{equation}
I_\mathrm{s}(\zeta_{\mathrm{m}})=\frac{|\ln
\mathcal{R}^{2}|}{1-\mathcal{R}^{2}}(A-1) \,,  \label{Fs2}
\end{equation}
and $A=1$ corresponds to the first laser threshold.

As in previous sections, the linear stability analysis is performed by
writing $F\left( \zeta ,\tau \right) =F_\mathrm{s}\left( \zeta \right)
+\delta F\left( \zeta \right) \exp \left( \lambda \tau \right) $, and
similar expressions for $P$ and $D$, and linearizing the resulting dynamical
equations. As now the problem depends explicitly on the longitudinal
coordinate $\zeta $ and the boundary condition is not periodic, one cannot
make the ansatz $\delta F\sim \exp \left( i\alpha \zeta \right) $ as in the
UFL. In fact, now the spatial dependence of $\delta F$ is unknown a priori
and an equation for it must be sought. After solving for the material
perturbations we set $\delta F(\zeta)=F_\mathrm{s}(\zeta)\delta f(\zeta)$
and obtain the following equation for $\delta f$
\begin{eqnarray*}
\frac{d}{d\zeta }\delta f &=&-\lambda \delta f-H\left( I_\mathrm{s}\right)
\delta f\,, \\
H\left(I_\mathrm{s}\right) &=&\frac{\sigma A}{1+I_\mathrm{s}}\frac{%
\gamma\lambda(\lambda+\gamma)+2\gamma I_\mathrm{s} }{\left( \gamma +\lambda
\right) \left( 1+\gamma \lambda \right) +\gamma I_\mathrm{s}},
\end{eqnarray*}
whose formal integration yields
\begin{eqnarray*}
\delta f\left( \zeta \right) &=&\delta f\left( 0\right) \exp \left[ -\lambda
\zeta -\psi\left( \zeta \right) \right] \, , \\
\psi\left( \zeta \right) &=&\int_{0}^{\zeta }d\zeta ^{\prime }H\left(I_%
\mathrm{s}\left( \zeta ^{\prime }\right) \right) \,.
\end{eqnarray*}%
The evaluation of the integral $\psi\left( \zeta \right) $ requires a change
of variable from $\zeta $ to $I_\mathrm{s}$ \cite{Lugiato85a} through the
steady state equation $dI_\mathrm{s}/d\zeta =2\sigma AI_\mathrm{s}/\left(
1+I_\mathrm{s}\right) $, after which the following result is obtained \cite%
{deValcarcel99}
\begin{equation}
\psi \left( \zeta \right)=\frac{\gamma \lambda }{2\left( 1+\gamma \lambda
\right) }\ln \frac{I_\mathrm{s}\left( \zeta \right) }{I_\mathrm{s}\left(
0\right) } +\frac{2+\gamma \lambda }{2\left( 1+\gamma \lambda \right) } \ln
\frac{\left( 1+\gamma \lambda \right) \left( \lambda +\gamma \right) +\gamma
I_\mathrm{s}\left( \zeta \right)}{\left( 1+\gamma \lambda \right) \left(
\lambda +\gamma \right) +\gamma I_\mathrm{s}\left(0 \right) }\,.
\end{equation}
Finally we return to the initial perturbation $\delta F(\zeta)$ and write
\begin{equation}
\delta F\left( \zeta \right)=\delta F\left( 0\right) \frac{F_\mathrm{s}%
\left( \zeta \right) }{F_\mathrm{s}\left( 0\right) }\exp \left[ -\lambda
\zeta -\psi \left( \zeta \right) \right]\,.  \label{dFnondef}
\end{equation}
The boundary condition (\ref{bouGnonUFL}) applied to the
perturbation $\delta F\left( \zeta \right)$ gives the following
characteristic equation
\begin{equation}
\lambda =-in\tilde{\alpha}-\frac{\psi \left(
\zeta_{\mathrm{m}}\right) }{\zeta_{\mathrm{m}}},
\label{chareqnonUFL}
\end{equation}%
where Eq. (\ref{Setam}) has been used. Recalling (\ref{bouGnonUFL}) and
making use of the steady intensity Eq. (\ref{Fs2}), the function $\psi
\left( \zeta_{\mathrm{m}}\right) $ reads%
\begin{eqnarray}
\psi \left( \zeta_{\mathrm{m}}\right) &=&\frac{\gamma \lambda }{2\left(
1+\gamma \lambda \right) }\left\vert \ln \mathcal{R}^{2}\right\vert +\frac{%
2+\gamma \lambda }{2\left( 1+\gamma \lambda \right) }  \notag \\
&&\times \ln \left[ 1+\frac{\gamma |\ln \mathcal{R}^{2}|(A-1)}{\left(
1+\gamma \lambda \right) \left( \lambda +\gamma \right) +\gamma \frac{%
\mathcal{R}^{2}|\ln \mathcal{R}^{2}|}{1-\mathcal{R}^{2}}(A-1)}\right] .
\label{PSI}
\end{eqnarray}
Equation (\ref{chareqnonUFL}) is the characteristic equation
governing the eigenvalues $\lambda $. It is a transcendental
equation in $\lambda $ and hence no explicit expressions can be
obtained. However in the class B limit $ \gamma \rightarrow 0$,
the use of an asymptotic expansion of (\ref{chareqnonUFL}) in
powers of $\gamma $ similar to that introduced in Sect.
\ref{homo-ufl-expansion} allows to completely unfold the problem.
After writing $\lambda =\lambda _{0}+\gamma \lambda _{1}+\gamma
^{2}\lambda _{2}+\cdots $ one obtains \cite{deValcarcel99}
\begin{eqnarray*}
\lambda _{0} &=&-i\alpha\,, \\
\lambda _{1} &=&i\sigma \alpha \left[ 1-\frac{2}{\alpha ^{2}}\left(
A-1\right) \right]\,, \\
\mathrm{Re}\lambda _{2}
&=&-\frac{\sigma}{\alpha^2}\left\{\alpha^4-3\alpha^2\left(
A-1\right)+2\left( A-1\right) \left[ 1+\frac{A-1}{2}
\frac{1+\mathcal{R}^{2}}{1-\mathcal{R} ^{2}}|\ln
\mathcal{R}^{2}|\right] \right\}\,,
\\
\mathrm{Im}\lambda _{2} &=&-\sigma ^{2}\alpha \left[ 1-\frac{4}{\alpha ^{4}}%
\left( A-1\right) ^{2}\right]\,,
\end{eqnarray*}%
where $\alpha =n\tilde{\alpha}$. The RNGHI occurs at $\mathrm{Re}
\lambda _{2}=0$, which defines the following boundaries
\begin{eqnarray}
\alpha _{\pm }^{2} &=&\tfrac{1}{2}\left[ 3\left( A-1\right) \pm \sqrt{\left(
A-1\right) ^{2}\mathcal{D}-8\left( A-1\right) }\right] ,
\label{tongue-nonUFL} \\
\mathcal{D} &=&9-4\frac{1+\mathcal{R}^{2}}{1-\mathcal{R}^{2}}|\ln \mathcal{R}%
^{2}|.  \label{D-nonUFL}
\end{eqnarray}
Notice the formal similarity with Eq. (\ref{tongueclassB})
corresponding to the UFL (for $\mathcal{R}\rightarrow 1$,
$\mathcal{D}\rightarrow 1$ and then Eq. (\ref{tongueclassB}) is
recovered). Now the two curves join at the
critical point%
\begin{eqnarray}
A_\mathrm{c} &=&1+\frac{8}{\mathcal{D}},  \label{Ac-nonUFL} \\
\alpha _\mathrm{c}^{2} &=&\frac{12}{\mathcal{D}}.  \label{alphac-nonUFL}
\end{eqnarray}%
The boundaries of the unstable domain depend on the reflectivity $\mathcal{R}
$\ through the function $\mathcal{D}$. In the UFL $\mathcal{T}\equiv 1-%
\mathcal{R}^{2}\rightarrow 0$ ($\mathcal{T}$ denotes the fraction of
intracavity power lost per roundtrip), $\mathcal{D}$\ can be approximated as
$\mathcal{D}=1-2\left( \mathcal{T}^{2}+\mathcal{T}^{3}\right) /3+\mathcal{O}%
\left( \mathcal{T}^{4}\right) $ and therefore the expressions for $A_\mathrm{%
c}$ and $\alpha _\mathrm{c}$ become
\begin{eqnarray*}
A_\mathrm{c} &=&9+\frac{16}{3}\mathcal{T}^{2}\left(
1+\mathcal{T}\right) +
\mathcal{O}\left( \mathcal{T}^{4}\right) , \\
\alpha _\mathrm{c}^{2} &=&12+8\mathcal{T}^{2}\left(
1+\mathcal{T}\right) + \mathcal{O}\left( \mathcal{T}^{4}\right) .
\end{eqnarray*}%
Both quantities reduce for $\mathcal{T}=0$ to the ones found in the UFL,
namely $A_\mathrm{c}=9$ and $\alpha _\mathrm{c}^{2}=12$.
\begin{figure}[t]
\begin{center}
\scalebox{0.5}{\includegraphics{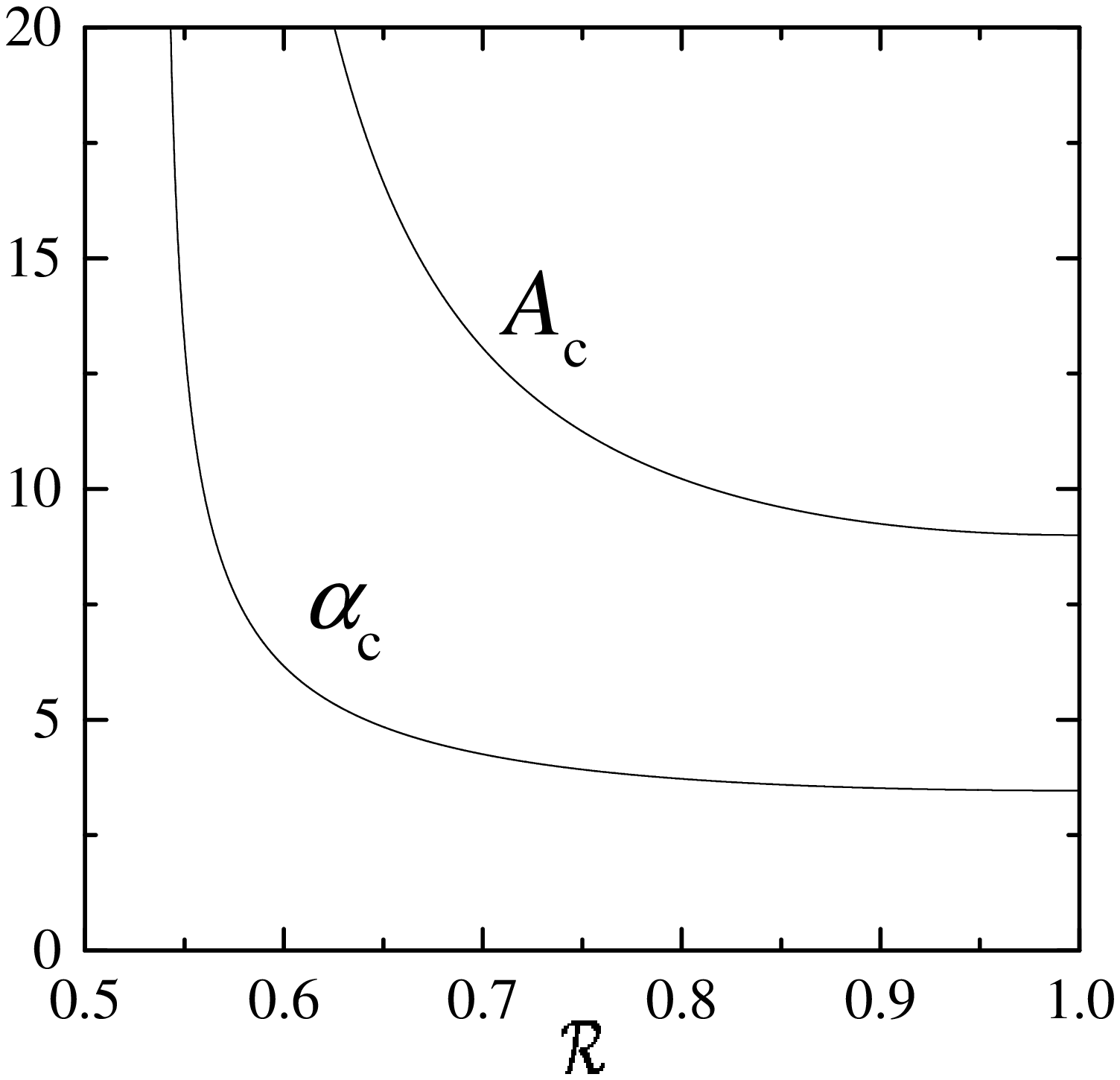}}
\end{center}
\caption{{}Behavior of the critical values $A_\mathrm{c}$ and
$\alpha_\mathrm{c}$ as functions of the reflectivity ${\cal R}$. }
\label{fig:nufl1}
\end{figure}
\begin{figure}[t]
\begin{center}
\scalebox{0.5}{\includegraphics{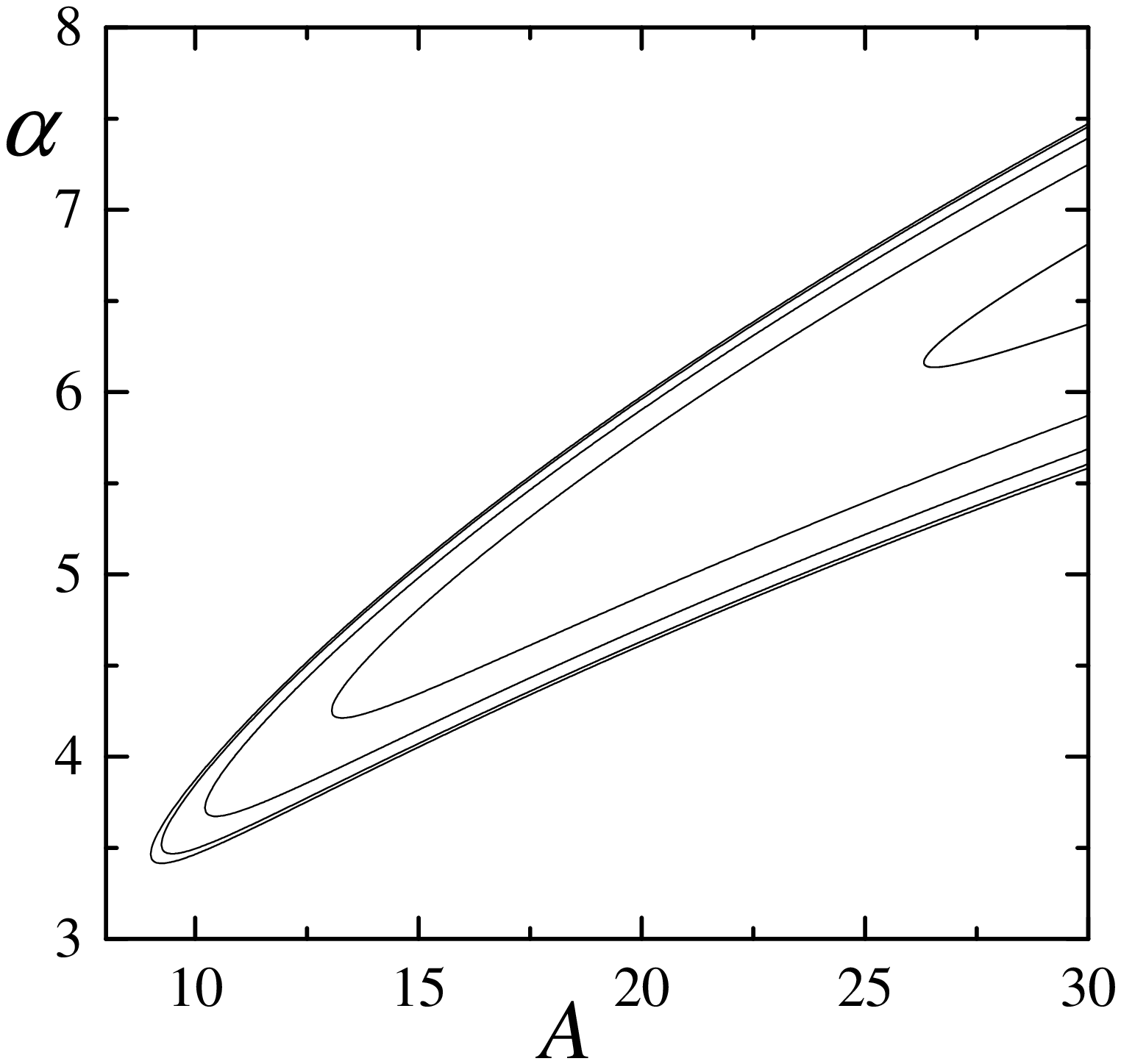}}
\end{center}
\caption{{}Unstable domains for five different values of the
reflectivity ${\cal R}$. From left to right ${\cal R}=0.99$ (which
corresponds to the UFL), 0.9, 0.8, 0.7, 0.6.} \label{fig:nufl2}
\end{figure}
Clearly Eq. (\ref{tongue-nonUFL}) is meaningful only if the argument of the
square root is non-negative, and this requires $\mathcal{D}>0$ or,
equivalently, that the reflectivity must be greater than%
\begin{equation}
\mathcal{R}_{\mathrm{\min }}\simeq 0.5379.  \label{Rmin}
\end{equation}%
This is an outstanding result as it sets an absolute minimum value
to $\mathcal{R}$ in order to observe the RNGHI
\cite{deValcarcel99} (see also \cite{Milovsky70}). In other words,
no matter how large the pump or how long the cavity be, the laser
will not display multimode instabilities if
$\mathcal{R}<\mathcal{R}_{\mathrm{\min }}$. Above this
reflectivity value, the effect of lowering $\mathcal{R}$ from the
UFL ($ \mathcal{R}\rightarrow 1$) is monotonous: Fig.
\ref{fig:nufl1} shows that the critical values $\left(
A_\mathrm{c},\alpha _\mathrm{c}\right) $ grow by decreasing
$\mathcal{R}$ and diverge by approaching
$\mathcal{R}_{\mathrm{\min }}$. Accordingly, as shown in Fig.
\ref{fig:nufl2}, the instability tongue (\ref{tongue-nonUFL})
shifts towards larger pumps and larger wavenumbers.

To conclude, let us remark that unlike in the UFL now the value of
cavity loss influences the second-to-first threshold ratio through
$\mathcal{R}$. In the UFL cavity loss affects the pump values
necessary for lasing and for instability, but not their ratio.
This fact marks a way for separating experimentally the two
thresholds in erbium--doped fiber lasers (remember that for these
lasers $W_{\mathrm{c}}/W_{\mathrm{on}}=1+\epsilon $ in the UFL,
i.e. for small cavity losses, rendering the observation of the
instability difficult). This is the strategy followed in the
experiments \cite{Voigt01,Voigt04}, which are treated in detail in
Sect. \ref{experimental}.

\subsubsection{Role of distributed losses}

\label{homo-nufl-loss}

Another factor present in almost any experiment and not considered by the
standard RNGH theory is the existence of distributed losses along the
amplifying medium, apart from those localized outside it. The two types of
losses should have different consequences as the role of distributed loss
should be to decrease gain, while localized loss determines the effective
reflectivity value $\mathcal{R}$, whose influence has been studied above. An
analysis of the combined effects of distributed and localized losses can be
found in Ref. \cite{Roldan03b}, which we summarize here. The proper model
reads now, Eqs. (\ref{eqF3})--(\ref{eqD3}),
\begin{eqnarray}
\left( \partial _{\tau }+\partial _{\zeta }\right) F &=&\sigma \left(
AP-\chi F\right) , \\
\partial _{\tau }P &=&\gamma ^{-1}\left( FD-P\right) \,, \\
\partial _{\tau }D &=&\gamma \left[ 1-D-\mathrm{\mathrm{Re}}\left( F^{\ast
}P\right) \right] \,,
\end{eqnarray}
where we have ignored again detuning, with the boundary condition
\begin{equation}
F(0,\tau )=\mathcal{R}F(\zeta_{\mathrm{m}},\tau
)\,,\qquad\zeta_{\mathrm{m}}=2\pi/\tilde\alpha\,.
\end{equation}%
The parameter $\chi =\alpha _{ \mathrm{m}}L_{\mathrm{m}}/\left(
|\ln \mathcal{R}^{2}|+\alpha _{\mathrm{m} }L_{\mathrm{m}}\right)
$, Eq. (\ref{ji}), verifies $0\leq \chi \leq 1$, and $\alpha
_{\mathrm{m}}$ is an intensity loss coefficient per unit length
\footnote{ We note that in Ref. \cite{Roldan03b} the parameter
$\chi $ was denoted by $ \eta $. Another relevant parameter used
there, $\gamma _{\mathrm{d}}$, can be written as $\gamma
_{\mathrm{d}}=\frac{1}{2}|\ln \mathcal{R}^{2}|\frac{ \chi }{1-\chi
}=\frac{1}{2}\alpha _{\mathrm{m}}L_{\mathrm{m}}$. The pump
parameter $A$ keeps the same definition as in \cite{Roldan03b}.}.

The intensity of the singlemode solution along $\zeta$, and its value at the
medium exit plane are given, respectively, by Eq. (\ref{ecIs}) and Eq. (\ref%
{ecIsetam}) with $\Delta =0$, and the lasing threshold is $A=1$.
\begin{figure}[t]
\begin{center}
\scalebox{0.5}{\includegraphics{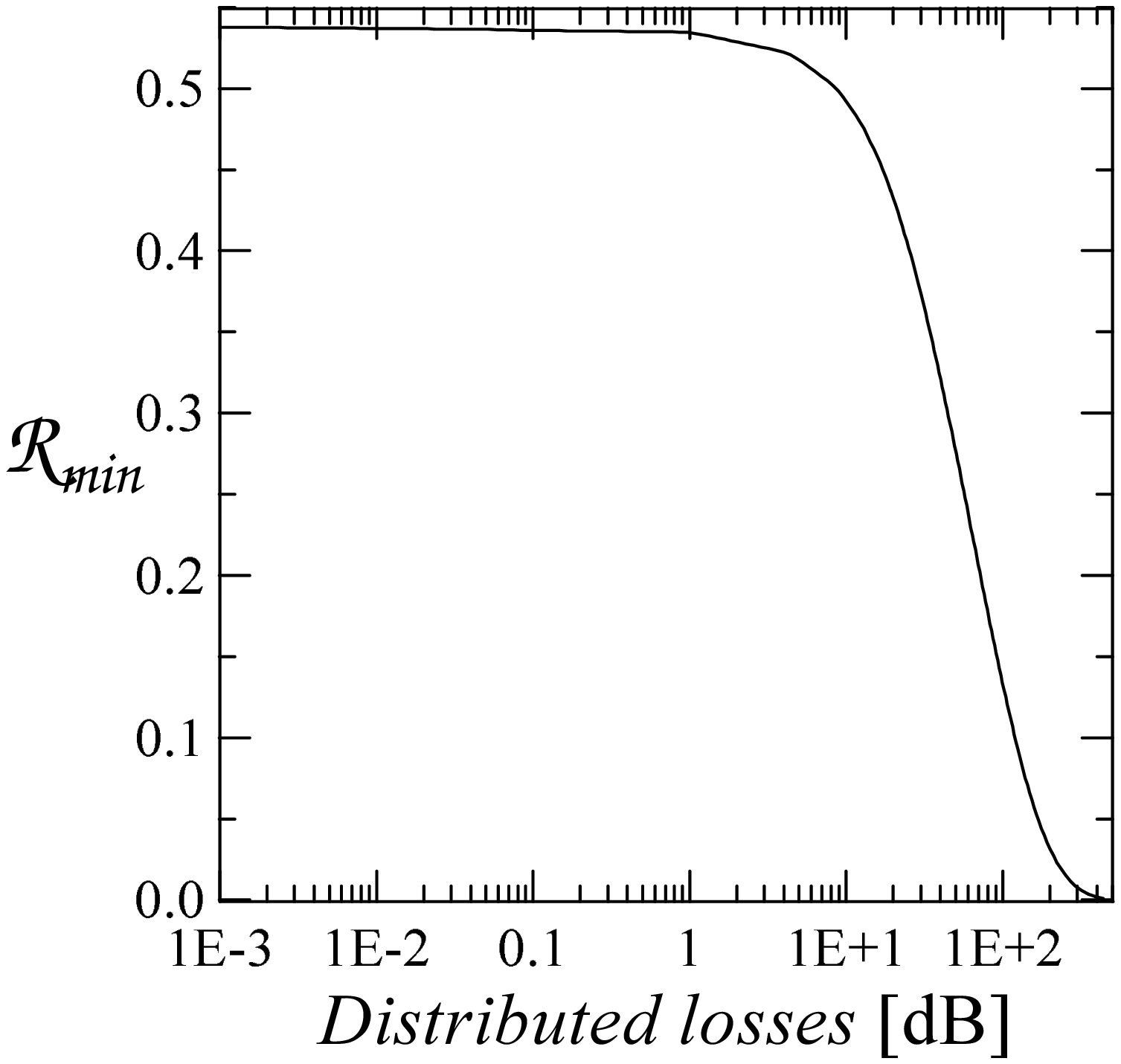}}
\end{center}
\caption{{}Dependence of the minimum reflectivity necessary for
the appearance of the RNGHI as a function of distributed losses
expressed in dB. } \label{fig:db}
\end{figure}
The linear stability analysis of the singlemode solution follows
the lines of the previous sections. As shown in \cite{Roldan03b}
the main effect of distributed loss is to favor the multimode
instability, i.e., to reduce the instability to lasing threshold
ratio (of course, the larger the losses the larger the pump needed
for reaching both the lasing and instability thresholds). For
instance, the minimum reflectivity value allowing the existence of
the RNGHI, $\mathcal{R}_{\min }$, is determined by
the equation%
\begin{equation}
\frac{8\left( 1-\chi \right) }{\chi \left( 8-9\chi \right) }\frac{1-\mathcal{%
R}^{2}}{\left\vert \ln \mathcal{R}^{2}\right\vert }\frac{1-\exp \left( -%
\frac{\chi }{1-\chi }\left\vert \ln \mathcal{R}^{2}\right\vert \right) }{1-%
\mathcal{R}^{2}\exp \left( -\frac{\chi }{1-\chi }\left\vert \ln \mathcal{R}%
^{2}\right\vert \right) }=1,
\end{equation}%
which can be simplified to
\begin{equation}
\frac{8}{8-9\chi }\frac{1-\mathcal{R}^{2}}{\alpha _{\mathrm{m}}L_{\mathrm{m}}%
}\frac{1-e^{-\alpha _{\mathrm{m}}L_{\mathrm{m}}}}{1-\mathcal{R}%
^{2}e^{-\alpha _{\mathrm{m}}L_{\mathrm{m}}}}=1.  \label{Rcrit}
\end{equation}%
This equation yields $\mathcal{R}_{\mathrm{\min }}\simeq 0.5379$
for $\chi =0 $, Eq. (\ref{Rmin}), as it should. For nonzero
distributed loss, $ \mathcal{R}_{\mathrm{\min }}$ is found to
decrease with increasing $\alpha _{ \mathrm{m}}L_{\mathrm{m}}$ so
that for $\alpha _{\mathrm{m}}L_{\mathrm{m} }\approx 13.6$,
$\mathcal{R}_{\mathrm{\min }}\simeq 0.1$, and for larger values of
$\alpha _{\mathrm{m}}L_{\mathrm{m}}$, $\mathcal{R}_{\mathrm{\min
}}\rightarrow 0$ exponentially, as shown in Fig. \ref{fig:db}. In
that figure distributed losses are expressed in decibels
($\mathrm{dB}$) through the relation
\begin{equation}
DL=8.646\,\alpha _{\mathrm{m}}L_{\mathrm{m}}.
\end{equation}%
We can conclude that for $DL<1$ dB (which is the usual situation),
$\mathcal{R}_{\mathrm{\min }}$ changes very little, and only for
$DL>10$ dB $\mathcal{R}_{\mathrm{\min }}$ decreases very quickly.
However, $10$ dB is an enormous amount of distributed loss, hardly
realistic for a usual laser. So it can safely be said that
distributed losses for realistic laser parameters do not
significantly affect the multimode emission threshold.
\begin{figure}[t]
\begin{center}
\scalebox{0.5}{\includegraphics{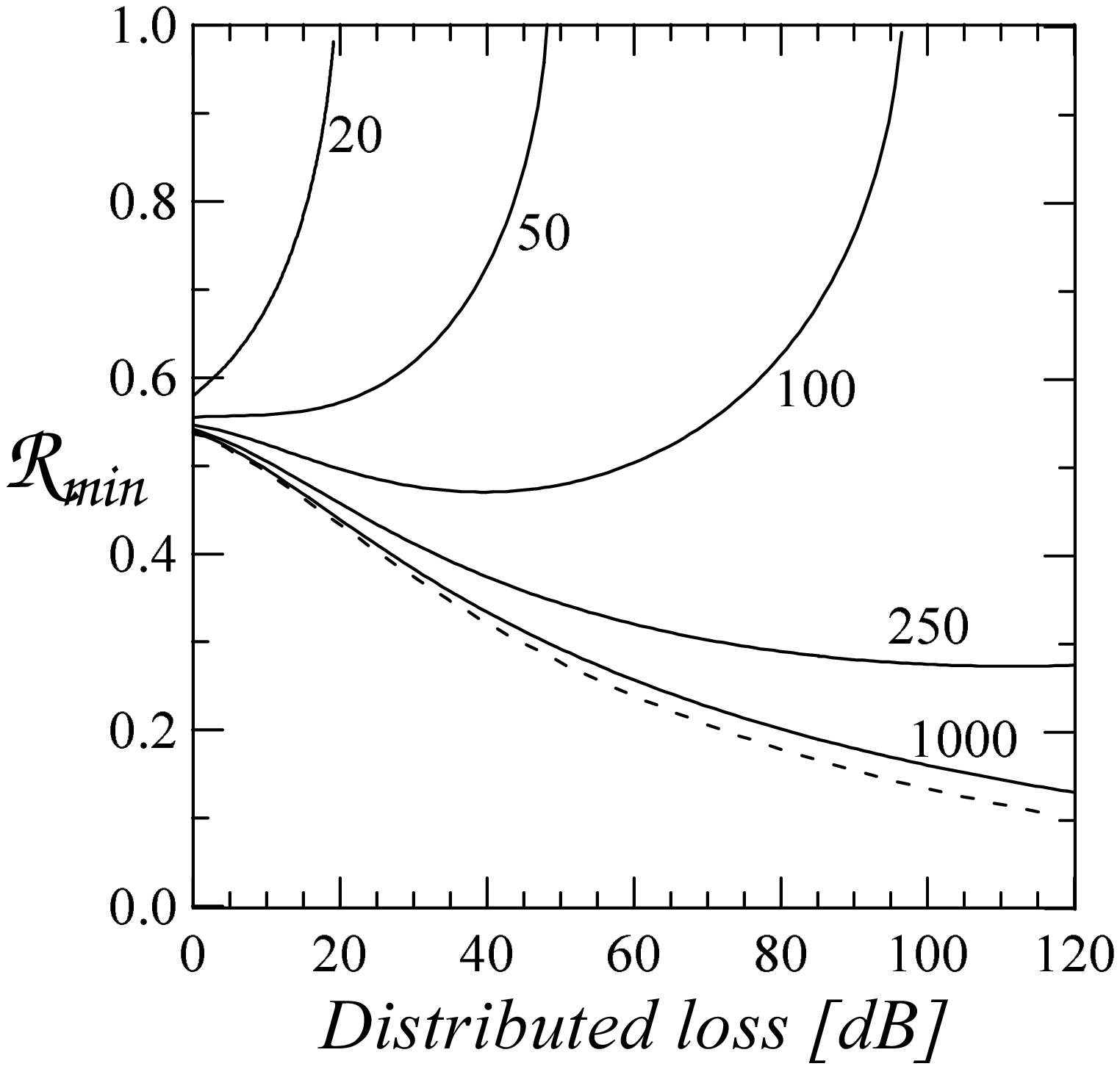}}
\end{center}
\caption{{}Dependence on distributed losses of the minimum
reflectivity necessary for the RNGHI to exist for three--level
lasers. The numbers
indicate the value of $G_{0}$ and the dashed line corresponds to $%
G_{0}\rightarrow \infty $. }
\label{fig:6}
\end{figure}
An important question is whether this conclusion can be extended
to three-- and four--level lasers: The answer is that it cannot.
For example, Eq. (\ref{Rcrit}) is only valid for two--level
lasers, i.e., it cannot be directly translated for three-- and
four--level lasers (this is due to the way it is derived, see the
discussion in \cite{Roldan03b}). When $\mathcal{R}_{\mathrm{ \min
}}$ is calculated for three--level lasers, the result is that of
Fig. \ref{fig:6}, which tends to the two--level laser result only
for very large $G$.

In general, although the effect of distributed losses can be
neglected for two--level lasers, as discussed above, this will not
be so for three-- and four--level lasers. The origin of this
different behavior lays in the fact that distributed losses enter
into the definition of the pump parameter, Eq. (\ref{A}), through
$G$, Eq. (\ref{G}). Nevertheless, the analysis we carried out for
two--level lasers leads to a simplified way of treating
distributed loss, as long as it is not too large (roughly, smaller
than $10$ dB): Perform the linear stability analysis without
taking distributed loss into account (which is accurate for
two--level lasers), and then take distributed loss into account
when translating the results derived for two--level lasers into
three-- and four--level lasers through $G$, Eq. (\ref{G}). We have
checked that this is a very accurate way of deriving the stability
properties for three-- and four--level lasers for not too large
distributed loss. We shall make use of this simplified way of
treating distributed losses in Sect. \ref{inhomo-nufl} when
inhomogeneous broadening be included

\subsubsection{Rate equations}

\label{homo-nufl-rate}

Class--B lasers are usually treated with rate equations, which are
derived by equating to zero the time derivative of the medium
polarization. If this is done in Eqs. (\ref{eqF3}--\ref{eqD3}),
then the RNGHI is lost. This means that in order to capture the
RNGH instability, coherent effects (\textit{i.e.}, those due to
the atomic polarization) must be preserved in the equations: The
standard adiabatic elimination technique fails in this case.

Nevertheless, the above circumstance does not prevent the derivation of
generalized rate equations that maintain enough information on the medium
polarization dynamics as for correctly describing the RNGHI. This is what we
do in this section: We derive a reduced set of rate equations, in which the
polarization variable is adiabatically eliminated in a generalized sense
\cite{deValcarcel03b}. The derivation of these equations stems from the
observation that the RNGHI instability can be captured only if in the
stability analysis terms up to order $\gamma ^{2}$ are kept. Therefore, the
instability certainly disappears if the rate equations are derived by means
of a standard adiabatic elimination of the polarization, which amounts to
truncate the polarization to the term of order $\gamma^0$. But with a more
refined adiabatic elimination, where terms up to order $\gamma ^{2}$ are
kept in the polarization, we can derive a set of generalized rate equations
that describe perfectly the RNGHI.

To derive the rate equations we start from Eqs.
(\ref{numint-f1}--\ref{numint-d1}) introduced in Sect.
\ref{model-ufl} for the primed variables $F'$, $P'$ and $D'$. The
most noticeable property of those equations is that, although they
are valid outside the UFL, they admit as a stationary solution the
singlemode solution of the UFL (\ref{ssufl}). In the following we
will consider power expansions of the dynamical variables in terms
of the small parameter $\gamma$. To this aim it is convenient to
work with variables which, at the leading order, are of order 1.
Therefore, we introduce the new dynamical variables $f$, $p$ and
$d$ defined as
\begin{equation}
f\left(\zeta,\tau\right)=\frac{F'\left(\zeta
,\tau\right)}{\sqrt{I_\mathrm{s}^\mathrm{UFL}}}\,,\quad
p\left(\zeta,\tau\right)=\frac{AP'\left(\zeta
,\tau\right)}{\sqrt{I_\mathrm{s}^\mathrm{UFL}}}\,,\quad
d\left(\zeta,\tau\right)=A\,D'\left(\zeta,\tau\right)\,.\label{fpdrate}
\end{equation}
The Maxwell--Bloch equations for $f$, $p$ and $d$ obtained from
Eqs. (\ref{numint-f1})--(\ref{numint-d1}) are
\begin{eqnarray}
\left( \partial _{\tau }+\partial _{\zeta }\right) f &=&\sigma
\eta \left( \zeta \right) \left[ p-\left( 1-i\Delta
\right)f\right] \,,  \label{numint-f}
\\
\partial _{\tau }p &=&\gamma ^{-1} \left[ fd-\left( 1+i\Delta \right)p\right]
\,,  \label{numint-p} \\
\partial _{\tau }d &=&\gamma \left\{ 1+\Delta^2-d+I_\mathrm{s}\left( \zeta
\right) \left[ 1-\mathrm{Re}\left(f^{\ast }p\right) \right]
\right\} \,, \label{numint-d}
\end{eqnarray}
The new electric field $f$ obeys the same periodic boundary
condition as $F'$
\begin{equation}
f\left( 0,\tau \right) =f\left( \zeta_{\mathrm{m}},\tau \right)
\,. \label{bounorm}
\end{equation}
and in the stationary state it is not only uniform but equal to 1.
In fact, Eqs. (\ref{numint-f})--(\ref{numint-d}) admit the
stationary solution $f_s=1,\,p_s=1-i\Delta,\,d_s=1+\Delta^2$. For
a perfect resonant laser ($\Delta=0$) they reduce to
\begin{eqnarray}
(\partial _{\tau }+\partial _{\zeta })f &=&\sigma \eta (\zeta )(p-f)\,,
\label{re-f} \\
\partial _{\tau }p &=&\gamma ^{-1}(fd-p)\,,  \label{re-p} \\
\partial _{\tau }d &=&\gamma \left\{ 1-d+I_\mathrm{s}(\zeta )\left[ 1-%
\mathrm{Re} \left( f^{\ast }p\right) \right] \right\} \,,  \label{re-d}
\end{eqnarray}
and the stationary solution is $f_s=p_s=d_s=1$. To obtain the
correct rate equations we first observe that, by setting $\gamma
=0$ in Eqs. (\ref{re-f}--\ref{re-d}), we obtain: (i) $\partial
_{\tau }d=0$, hence $d\left( \zeta ,\tau \right) =d_{0}\left(
\zeta \right) $, (ii) $p=d_{0}\left( \zeta \right) f$, and (iii)
$(\partial _{\tau }+\partial _{\zeta })f=\sigma \eta (\zeta
)\left[ d_{0}\left( \zeta \right) -1\right] f$. Note that these
equations do not fix the value of $d_{0}\left( \zeta \right) $.
However, since it is independent of time, it must be compatible,
in particular, with the stationary solution $d=1$, which implies
$d_{0}\left( \zeta \right) =1$. Hence, at order 0 in $\gamma $ we
have $d=1$ and $f=p$, from which it follows that $\partial _{\tau
}f=-\partial _{\zeta }f$. For $\gamma \neq 0$ we make the
following ansatz
\begin{equation}
p=f+\mathcal{O}(\gamma )\,,\quad d=1+\mathcal{O}(\gamma )\,,\quad \partial
_{\tau }f=-\partial _{\zeta }f+\mathcal{O}(\gamma )\,.  \label{approx}
\end{equation}%
We now observe that Eq. (\ref{re-p}) can be written as $p=fd-\gamma \partial
_{\tau }p$, which, iterated twice, yields
\begin{equation}
p=fd-\gamma \partial _{\tau }(fd)+\gamma ^{2}\partial _{\tau \tau }^{2}(fd)+%
\mathcal{O}(\gamma ^{3})\,.  \label{p1}
\end{equation}%
Taking into account Eqs. (\ref{re-f}), (\ref{re-d}) and (\ref{approx}) we
have
\begin{eqnarray}
\partial _{\tau }(fd) &=&-d\partial _{\zeta }f+\sigma \eta (p-f)-\gamma I_%
\mathrm{s}(|f|^{2}-1)f+\mathcal{O}(\gamma ^{2})\,, \\
\partial _{\tau \tau }^{2}(fd) &=&\partial _{\zeta \zeta }^{2}f+\mathcal{O}%
(\gamma )\,,
\end{eqnarray}%
which, inserted into Eq. (\ref{p1}), yield
\begin{equation}
p=fd+\gamma \left[ \sigma \eta (1-d)f+(d-\gamma \sigma \eta )\partial
_{\zeta }f\right] +\gamma ^{2}\left[ I_\mathrm{s}(|f|^{2}-1)f+\partial
_{\zeta \zeta }^{2}f\right] \,+\mathcal{O}(\gamma ^{3}).
\end{equation}%
Substituting this into Eqs. (\ref{re-f}) and (\ref{re-d}) we finally get the
generalized complex rate equations
\begin{eqnarray}
\left( \partial _{\tau }+\partial _{\zeta }\right) f &=&\sigma \eta \left[
f(d-1)(1-\gamma \sigma \eta )+\gamma (d-\gamma \sigma \eta )\partial _{\zeta
}f\right]  \notag \\
&&+\gamma ^{2}\sigma \eta \left[ I_\mathrm{s}\left( |f|^{2}-1\right)
f+\partial _{\zeta \zeta }^{2}f\right] \,,  \label{re-f1} \\
\partial _{\tau }d &=&\gamma \left[ 1-d+I_\mathrm{s}\left( 1-|f|^{2}d\right) %
\right] -\frac{1}{2}\gamma ^{2}I_\mathrm{s}\,\partial _{\zeta }|f|^{2}\,,
\label{re-d1}
\end{eqnarray}%
which are exact up to order $\gamma ^{2}$. In the UFL we can set $\eta =1$
and $I_\mathrm{s}=A-1$, and all the parameters in the right hand sides of
the rate equations become constant.

As a test of this reduced model we have determined the linear
stability of the steady lasing solution $\left( f=1,d=1\right) $
versus perturbations of the type $\delta x(\zeta ,\tau )=\delta
x_{0}(\zeta )\exp (\lambda \tau )$, with $x=f,\,d$, and $\delta
f_{0}(0)=\delta f_{0}(\zeta _{\mathrm{m}})$, Eq. (\ref{bounorm}).
Solving the characteristic polynomial perturbatively up to order
$\gamma ^{2}$ one recovers Eq. (\ref{tongue-nonUFL}) in Sect. \ref
{homo-nufl-rnghi}, as it must be. Notice finally that by making
the limit $\gamma \rightarrow 0$ in Eqs. (\ref{re-f1},\ref{re-d1})
(i.e., removing the terms multiplied by $\gamma $ in the field
equation and the terms multiplied by $\gamma ^{2}$ in the
inversion equation), the standard rate equations are recovered.

\subsubsection{The pulse equation}

\label{homo-nufl-pulse}

The simplest multimode solution which arises from the RNGHI
consists in regular pulses travelling along the cavity. If the
homogeneous stationary solution is first destabilized by the
$N$--th sidemode, $N$ identical pulses are present simultaneously
in the cavity. A semi--analytical study of the pulse solution can
be already found in \cite{RN68(b)}. In 1989 Fu analyzed in detail
the pulses and their stability in class--B lasers \cite{Fu89}. The
results of Fu were obtained in the UFL. Using the rate equations
(\ref{re-f1}) and (\ref{re-d1}) derived in the previous section we
can not only quickly recover them, but extend them to a class--B
laser outside the UFL.

We look for a solution of Eqs. (\ref{re-f1}) and (\ref{re-d1})
which travels at velocity $v$ along the cavity. To this aim we
introduce the new variable $\tau ^{\prime }=\tau -\zeta /v$, and
we consider the following expansions
\begin{eqnarray}
&&f=f_{0}(\tau ^{\prime })+\gamma f_{1}(\tau ^{\prime })+\mathcal{O}(\gamma
^{2})\,,\qquad d=1+\gamma I_\mathrm{s}(\zeta )d_{1}(\tau ^{\prime })+%
\mathcal{O}(\gamma ^{2})\,,  \notag \\
&&\frac{1}{v}=1-\gamma \sigma \beta +\mathcal{O}(\gamma ^{2})\,.
\end{eqnarray}%
We also assume that $f_{0}$ and $f_{1}$ are real and that the periodicity
condition
\begin{equation}
x(\tau ^{\prime }+NT)=x(\tau ^{\prime })\,,
\end{equation}%
holds for $x=f_{0},\,f_{1},\,d_{1}$, where $T$ is the duration of the single
pulse, and $N$ is the number of pulses circulating in the cavity. Denoting
differentiation with respect to $\tau ^{\prime }$ by a dot, the leading
order pulse equations are
\begin{eqnarray}
(\eta +\beta )\dot{f}_{0}-I_\mathrm{s}\eta f_{0}d_{1} &=&0\,,
\label{pulse-f1} \\
\dot{d}_{1}-1+f_{0}^{2} &=&0\,.  \label{pulse-d1}
\end{eqnarray}%
We can get rid of the dependence on $\zeta $ of $I_\mathrm{s}$ and $\eta
=A/(1+I_\mathrm{s})$ by averaging both sides of Eq. (\ref{pulse-f1}) on $%
\zeta $ from $0$ to $\zeta_{\mathrm{m}}$. We denote this average by $%
\left\langle \ldots \right\rangle $. Taking into account that the
variation of $I_\mathrm{s}$ along $\zeta $ is ruled by Eq.
(\ref{dFdz}), with $\chi =\Delta =0$, we find that
\begin{equation}
\left\langle \eta \right\rangle =1\,,\qquad \left\langle I_\mathrm{s}\eta
\right\rangle =A-1\,.
\end{equation}%
If we define $I_{0}=f_{0}^{2}$, the pulse equations read
\begin{eqnarray}
\dot{I}_{0} &=&\frac{2(A-1)}{1+\beta }I_{0}d_{1}\,,  \label{pulse-f2} \\
\dot{d}_{1} &=&1-I_{0}\,.  \label{pulse-d2}
\end{eqnarray}%
The parameter $\beta $ can be determined by means of the solvability
condition of the next order problem, which imposes that $\beta $ must take
one of the two values
\begin{equation}
\beta _{\pm }=\frac{2(A-1)}{\alpha _{\pm }^{2}}-1\,,
\end{equation}%
where $\alpha _{\pm }(A,\mathcal{R})$ are the boundaries of the
stability domain of the homogeneous solution given by Eqs.
(\ref{tongue-nonUFL}--\ref{D-nonUFL}) in Sect.
\ref{homo-nufl-rnghi}. The final form of the pulse equations is
then
\begin{eqnarray}
\dot{I}_{0} &=&\alpha _{\pm }^{2}\,I_{0}d_{1}\,,  \label{pulse-f3} \\
\dot{d}_{1} &=&1-I_{0}\,.  \label{pulse-d3}
\end{eqnarray}%
These equations allow to calculate with a very good approximation the pulse
shape in the limit $\gamma \rightarrow 0$, by--passing the problem of the
very long transients associated with the full set of Maxwell--Bloch
equations, and also with the generalized rate equations, in that limit. We
have just to determine, for any pump $A$, reflectivity $\mathcal{R}$, and
cavity length $\zeta_\mathrm{m}=2\pi /\tilde{\alpha}$, the correct initial
conditions which give a periodic solution.

To this aim we observe that Eqs. (\ref{pulse-f3}--\ref{pulse-d3}) have a
Hamiltonian structure, which can be better put in evidence with the change
of variables $x=\ln I_{0}$, $p=d_{1}$\cite{RN68(b)}. The equations for $x$
and $p$ read
\begin{eqnarray}
\dot{x} &=&\alpha _{\pm }^{2}\,p\,,  \label{pulse-x} \\
\dot{p} &=&1-\mathrm{e}^{x}\,,  \label{pulse-p}
\end{eqnarray}%
which are the equations of an anharmonic oscillator of mass $\alpha _{\pm
}^{-2}$ subjected to the potential $V(x)=\mathrm{e}^{x}-x$. A constant of
motion for this system is the energy
\begin{equation}
H(x,p)=\alpha _{\pm }^{2}\frac{p^{2}}{2}+V(x)\,,  \label{H}
\end{equation}%
from which Eqs. (\ref{pulse-x}--\ref{pulse-p}) can be derived through
Hamilton's equations. The anharmonic oscillator bounces between the point $%
x_{\min }$ and $x_{\max }$ where $p=0$. Hence, we have
\begin{equation}
H=V\left( x_{\min }\right) =V\left( x_{\max }\right) \,,  \label{xminxmax}
\end{equation}%
which establishes a relation between $x_{\min }$ and $x_{\max }$.
From Eq. (\ref{H}) we obtain
\begin{equation}
p=\frac{\sqrt{2}}{\alpha _{\pm }}\sqrt{H-V(x)}\,,
\end{equation}%
which can be inserted in Eq. (\ref{pulse-x}) to obtain the time spent by the
oscillator to move from point $x_{1}$ to point $x_{2}$
\begin{equation}
\tau _{12}=\frac{1}{\sqrt{2}\alpha _{\pm }}\int_{x_{1}}^{x_{2}}\frac{dx}{%
\sqrt{H-V(x)}}\,.
\end{equation}%
The period of a complete pulse is twice the time spent to go from $x_{\min }$
to $x_{\max }$. For the $N$--pulse solution it must be equal to $2\pi
/\alpha _{N}$, which means
\begin{equation}
\alpha _{\pm }(A,\mathcal{R})=\frac{\alpha _{N}}{\sqrt{2}\pi }\int_{x_{\min
}}^{x_{\max }}\frac{dx}{\sqrt{H-V(x)}}\,.  \label{pulse}
\end{equation}%
Since $x_{\min }$ and $x_{\max }$ are related by Eq. (\ref{xminxmax}), for
any given $\alpha _{N}$ and $\mathcal{R}$ Eq. (\ref{pulse}) allows to
determine $A$ as a function of $x_{\min }=\ln I_{\min }$ (or $x_{\max }=\ln
I_{\max }$) and then $I_{\min }$ (or $I_{\max }$) as a function of $A$.
\begin{figure}[t]
\begin{center}
\scalebox{0.5}{\includegraphics{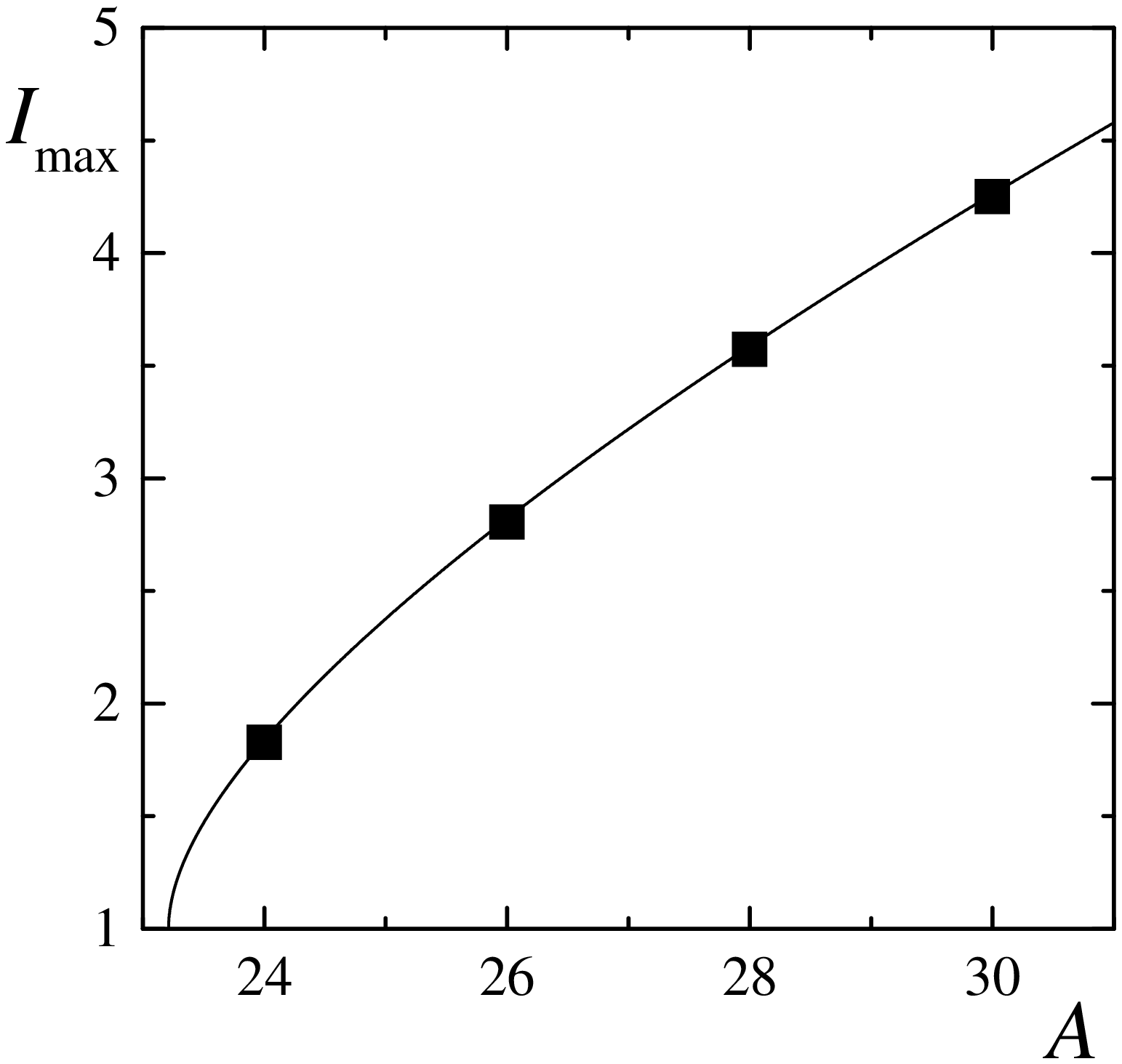}}
\end{center}
\caption{Maximum intensity of the 1--pulse solution according with Eqs. (%
\protect\ref{xminxmax}) and (\protect\ref{pulse}) for $\mathcal{R}=0.7$ and $%
\tilde{\protect\alpha}=2\protect\pi $. The bifurcation is
supercritical. The squares represent the value obtained by
numerical integration of the laser equations.} \label{fig:7}
\end{figure}
\begin{figure}[t]
\begin{center}
\scalebox{0.5}{\includegraphics{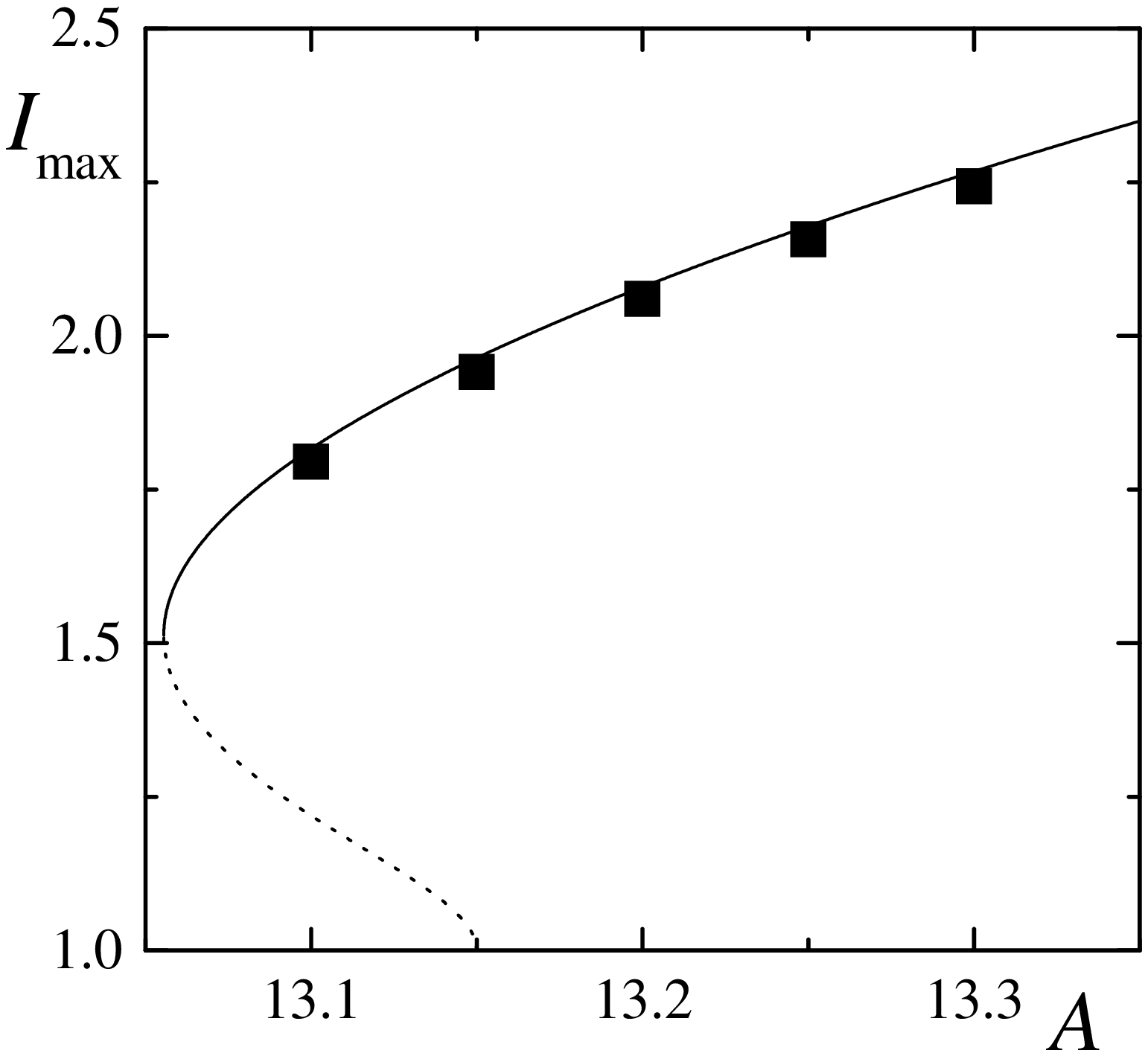}}
\end{center}
\caption{Same as Fig. \protect\ref{fig:7} for
$\tilde{\protect\alpha}=4.22$. Here the bifurcation is
subcritical.} \label{fig:8}
\end{figure}
Eq. (\ref{pulse}) has two solutions corresponding to the two signs
of $\alpha _{\pm }$. It can be easily demonstrated that the
integral in Eq. (\ref {pulse}) tends to $\sqrt{2}\pi $ in the
limit $x_{\min },x_{\max }\rightarrow 0$, \textit{i.e.} when the
pulse solution reduces to the homogeneous one $I=1$ \cite{Fu89}.
Hence, in that limit Eq. (\ref{pulse}) reduces to $\alpha _{\pm
}=\alpha _{N}$. This means that the pulse solutions associated
with $\alpha _{+}$ and $\alpha _{-}$ are the multimode solutions
emerging, respectively, from the upper ($\alpha _{+}$) and lower
($\alpha _{-}$) instability domain of the homogeneous solution.
General arguments of bifurcation theory corroborated by the
results of the numerical simulations indicate that only the former
is stable.

The plots of $I_{\max }=I_{\max }(A)$ show that the bifurcation
from the homogeneous solution to the pulse solution is
supercritical or subcritical depending on whether $\alpha _{N}$ is
larger or smaller than $\alpha _{\mathrm{c}}$. In Figs.
\ref{fig:7} and \ref{fig:8} we considered a rather small value of
the reflectivity, $\mathcal{R}=0.7$. This choice of $\mathcal{R}$
corresponds to a very bad resonator, in which the fraction of
power lost per roundtrip is $(1-\mathcal{\ R}^{2})=0.51$. Thus, we
are very far from the UFL. For this value of $\mathcal{R}$ the
critical and minimum values of $\alpha $ are, respectively,
$\alpha _{\mathrm{c}}=4.252$ and $\alpha _{\min }=4.212$, and the
critical pump is $A_{\mathrm{c}}=13.055$. In Fig. \ref {fig:7} we
consider the value $\alpha _{N}=2\pi $, larger than $\alpha _{
\mathrm{c}}$, while in Fig. \ref{fig:8} we set $\alpha _{N}=4.22$,
which is intermediate between $\alpha _{\min }$ and $\alpha
_{\mathrm{c}}$. The solid (dotted) lines correspond to the stable
(unstable) pulse solution associated with $\alpha _{+}$ ($\alpha
_{-}$). When the bifurcation is subcritical, the pulse solution is
stable also below the instability threshold up to $A_{
\mathrm{c}}$, and in this interval it coexists with the stable
homogeneous solution, see Sect. \ref{homo-nufl-sub}. However, we
notice that this happens only in the narrow range of values of
$\alpha _{N}$ between $\alpha _{\min }$ and $\alpha
_{\mathrm{c}}$.

In Figs. \ref{fig:7} and \ref{fig:8} the symbols indicate the peak
intensity of the pulse obtained by the numerical integration of
the dynamical equations. The agreement with the curve
$I_{\max}(A)$ calculated using Eqs. (\ref{xminxmax}) and
(\ref{pulse}) is very good.

The stability of the $N$--pulse with $N\geq 2$ was analyzed in \cite{Fu89}
in the UFL using the Floquet method. It was found that the $N$--pulse
solution becomes unstable beyond a certain value of $I_{\max }$ which
depends only on $N$. For instance $I_{\max}=4.914$ for $N=2$ and $I_{\max
}=2.339$ for $N=3$. For $N=1$ the stability analysis is more difficult and,
to our knowledge, it remains an unresolved problem.

\subsubsection{Numerical results}

\label{pulse-numres}

In the previous section we have illustrated a method to calculate
the $N$--pulse multimode solution developed by the laser beyond
the RNGHI threshold. In this way we can determine at once the
long-term regime of the laser, skipping all the transient
evolution. This is an enormous advantage from the computational
point of view, because we know that the unstable eigenvalue is of
order $\gamma^2$, which means that the time scale of the transient
is of order $\gamma^{-2}$.

But when the pulse solution itself destabilizes, the only way to
study the laser dynamics is the numerical integration of the
dynamical equations (\ref {eqF3}--\ref{eqD3}), with the boundary
condition (\ref{bou3})\footnote{ The pulsing regime has been
studied numerically by few authors \cite
{RN68(b),Ikeda89,Mayr81,Zorell81,Lugiato85a,Casini97} and these
studies have been quite superficial in most cases.}. These are
Partial Differential Equations (PDEs) that can be integrated using
the finite difference method originally proposed by Risken and
Nummedal \cite{RN68(b)}. However, the efficiency of this method is
strongly limited by the the fact that it imposes the constraint of
equal time and space step. Since the time step is proportional to
$\gamma $, in the limit of very small $\gamma $ the computational
time diverges, because the above mentioned constraint implies that
the number of points of the spatial grid is proportional to
$\gamma ^{-1}$.

An alternative method consists in a modal expansion of the electric field in
Fourier modes, which allows to convert the PDEs into a set of Ordinary
Differential Equations (ODEs). The big advantage of ODEs is that they can be
integrated using standard Runge-Kutta methods, which are easier to implement
and generally run faster than the finite difference methods used for
integrating PDEs.

It is commonly believed that modal expansion is convenient only in the UFL,
because only in that limit the electric field can be described properly by a
limited number of modes. Outside that limit, even the stationary solution
requires a high number modes, because it has a nontrivial spatial dependence
along the longitudinal direction \cite{Lugiato86a}, and the situation
certainly worsens when the RNGHI sets on.

However, it was demonstrated that in class B lasers a modal
expansion of the electric field is fully justified for the study
of the RNGHI, even outside the UFL, \textit{i.e.} for any mirror
reflectivity \cite{deValcarcel03a}. This result is a simple
consequence of the fact that, according to Eq. (\ref{dFnondef})
derived in Section \ref{homo-nufl-rnghi}, the unstable modes in
the RNGHI depend on $\zeta $ as
\begin{equation}
F_\mathrm{s}(\zeta )\exp \left[-\lambda_{n}\zeta -\psi _{n}(\zeta
)\right] \,,  \label{dF}
\end{equation}%
where $\psi _{n}(\zeta )$ is given by Eq. (\ref{PSI}) with
$\lambda=\lambda_n$, and $\lambda_n$ is a solution of the
characteristic equation (\ref{chareqnonUFL}). The important point
is that $\lambda_n\approx-i\alpha_n$, where $\alpha_n$ is the
spatial frequency of the $n$--th mode, and $\psi _{n}(\zeta
)\approx 0$ in two limiting cases: (i) in the UFL, and (ii) in the
class--B limit $\gamma \rightarrow 0$. Then, in both limits at the
leading order the unstable modes can be approximated by
\begin{equation}
F_\mathrm{s}(\zeta )\exp \left(i\alpha _{n}\zeta \right) \,.
\end{equation}%
In the UFL, where $F_\mathrm{s}=\sqrt{A-1}$ can be assumed
independent of $\zeta$, the unstable modes immediately coincide
with the empty cavity modes $\exp \left(i\alpha _{n}\zeta \right)
$. But this is true also for class--B lasers beyond the UFL,
provided the scaled electric field $f$ defined by Eq.
(\ref{fpdrate}), which is equal to 1 in the stationary state, is
considered in place of $F$.
\begin{figure}[t]
\begin{center}
\scalebox{0.5}{\includegraphics{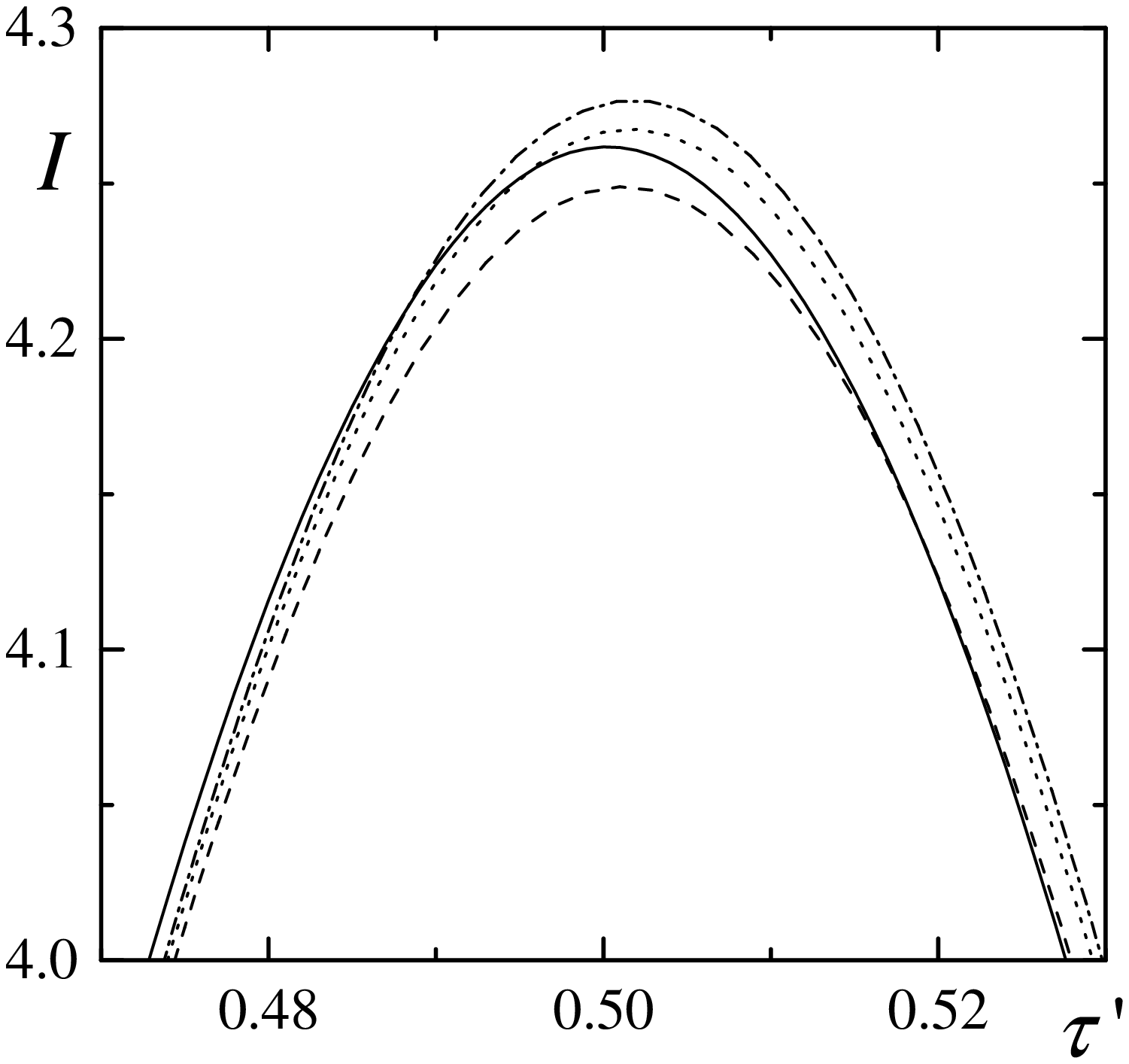}}
\end{center}
\caption{Enlargement of the steady pulse around the maximum for
$\protect\gamma=0.01$. The four outputs obtained with the four
different methods described in the text are compared.}
\label{fig:9}
\end{figure}
\begin{figure}[t]
\begin{center}
\scalebox{0.5}{\includegraphics{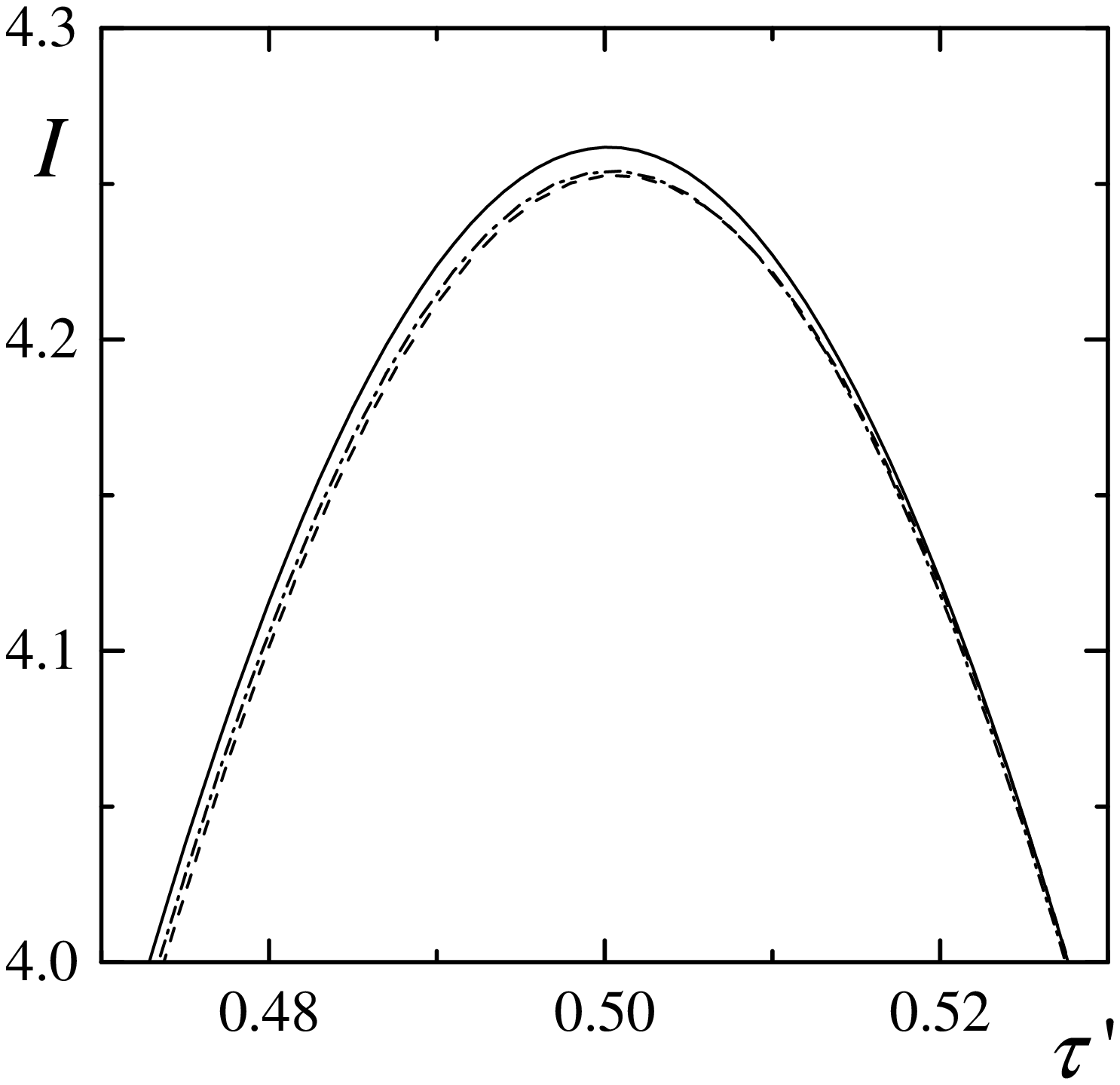}}
\end{center}
\caption{Same as Fig. \protect\ref{fig:9} for
$\protect\gamma=0.001$. Here only three curves are shown because
the integration of the PDEs was too time consuming.}
\label{fig:10}
\end{figure}
Therefore, the modal expansion can be applied to the Maxwell-Bloch
equations (\ref{numint-f}--\ref{numint-d}) written in Sect.
\ref{homo-nufl-rate} for the scaled variables $f$, $p$ and $d$ and
to the generalized rate equations that we derived from them. In
both cases the periodic boundary condition
\begin{equation}
f\left( 0,\tau \right) =f\left( \zeta_{\mathrm{m}}\,,\tau
\right)\,,\qquad\zeta_{\mathrm{m}}=2\pi/\tilde\alpha \,,
\end{equation}%
allows for a modal expansion of the electric field $f$
\begin{equation}
f(\zeta ,\tau )=\sum_{n=-\infty }^{+\infty }e^{i\alpha _{n}\zeta }f_{n}(\tau
)\,,
\end{equation}%
and the ODEs are simply obtained by inserting this expansion in the equation
for the electric field. For the complete Maxwell-Bloch equations the
equation for the generic mode amplitude $f_{n}$ is
\begin{equation}
\frac{df_{n}}{d\tau }=-i\alpha _{n}f_{n}+\sigma \int_{0}^{\zeta_{\mathrm{m}%
}}\!\!\!d\zeta \eta (\zeta )\exp \left( -i\alpha _{n}\zeta \right) (p-f)\,.
\label{fn}
\end{equation}%
The integral can be evaluated over a grid of points $\zeta _{l},\,l=1,\ldots
L$. To do that we need to know only the values of $p$ and $d$ in those
points. Thus, we have to consider a set of ODEs for the mode amplitudes $%
f_{n}(\tau )$ and the variables $p_{l}(\tau )=p(\zeta _{l},\tau )$ and $%
d_{l}(\tau )=d(\zeta _{l},\tau )$.

For the rate equations we proceed similarly, but with two big advantages:
the number of ODEs is smaller because the polarization has been
adiabatically eliminated, and the stiffness of the equations has been
reduced because the fast time scale associated with the polarization has
disappeared.

Summarizing, we can then use three different methods to study numerically
the laser equations: (i) the finite difference method applied to the full
Maxwell-Bloch equations, (ii) the modal expansion technique applied to the
Maxwell-Bloch equations, and (iii) the modal expansion technique applied to
the rate equations. Method (iii) is expected to be more efficient than
method (ii), and method (ii) more efficient than method (i). On the other
hand, only method (i) is applied to the full set of Maxwell-Bloch equations,
while with methods (ii) and (iii) some approximations are introduced.

In Figs. \ref{fig:7} and \ref{fig:8} the symbols refer to the calculations
made with method (ii), but on that scale the results of the three methods
are hardly distinguishable. A better check of the correctness of the three
methods and of their respective efficiency is the comparison among the shape
of the calculated pulses and that of the semi--analytic pulse solution found
in the previous section, which is the result to which any dynamical
simulation must tend in the limit $\gamma \rightarrow 0$.

We chose to make the comparison for $\mathcal{R}=0.7$ and $A=30$, and we
consider the $N=1$ pulse, setting $\alpha _{N}=\alpha _{1}=\tilde{\alpha}
=2\pi $, which implies $\sigma =0.357$. The results of the numerical
simulations are shown in Fig. \ref{fig:9} for $\gamma =10^{-2}$ and in Fig. %
\ref{fig:10} for $\gamma =10^{-3}$. In both figures we show an enlargement
of the portion of the pulse around the maximum to emphasize the differences
among the various integration methods. The solid lines represent the
semi--analytic solution, the dotted lines the integration of the PDEs, the
dashed lines the integration of the ODEs derived from the complete model,
and the dotted--dashed lines the integration of the ODEs derived from the
rate equations.

As expected, for the smaller value of $\gamma $ the agreement between the
numerical simulations and the analytic result is better. As for the
computation times, for $\gamma =0.01$ the ratios among the three methods are
5.8:1.6:1, so their efficiency is still comparable. But for $\gamma =0.001$
the ODEs derived from the rate equations are about 7 times faster than those
derived from the complete model, and the PDEs are already too slow to
converge in a reasonable time. These results demonstrate that to simulate
the dynamical behavior of a fibre laser, for which $\gamma \sim 10^{-5}$,
the ODEs derived from the rate equations are the only practical tool.

\subsubsection{Subcriticality of the RNGH}

\label{homo-nufl-sub}An important issue of the RNGHI that we have
referred to only superficially in Sect. \ref{homo-nufl-pulse}, is
the subcritical or supercritical character of the bifurcation
leading from singlemode to multimode emission. This issue has been
considered in the past, as we detail below, within the UFL and may
be important for the correct understanding of the experimental
results, see Sec .\ref{experimental}. We will here present some
numerical results outside that limit, and therefore treat this
question in this section.

In their second paper of 1968 \cite{RN68(b)}, Risken and Nummedal already
treated this question. They commented on their numerical finding that
multimode emission persisted for a given pump even when the spatial
frequency $\alpha $ was decreased below $\alpha _{-}$, see Eq. (\ref%
{tongue-nonUFL}), where the linear stability analysis predicts stable
singlemode emission. They explicitly comment that for $\alpha <\alpha _{-}$
there is bistability between the singlemode and the multimode solutions.

Later on, Haken and Ohno \cite{Haken76,Ohno76} obtained an equation for the
critical mode and found this coexistence of solutions again. In fact they
derived a condition for supercritical or subcritical bifurcation but they
did not give the dependence on the wavenumber $\alpha $ of their result
because of the excessive complexity of their equations. Lugiato \textit{et
al.} \cite{Lugiato85a} derived an approximated multimode solution (from a
five--mode truncation of the Maxwell--Bloch equations) that explicitly
showed the bistable behavior. The question was taken up again by Fu \cite%
{Fu89} who derived the unambiguous condition in the limit of
class--B lasers that we have already found in Sect.
\ref{homo-nufl-pulse}: The bifurcation is supercritical
(subcritical) when $\alpha >\alpha _\mathrm{c}$ ($\alpha <\alpha_
\mathrm{c}$). Finally, Carr and Erneux \cite{Carr94a} analytically
obtained the same result in a slightly different limit for
class--B lasers ($ \gamma _{||}/\kappa \ll 1$, which in our
notation corresponds to small $ \gamma $ and large $\sigma $).

As stated, all these studies were carried out within the UFL. The meaning of
these results is that multimode emission could be found for parameter
settings for which the singlemode solution is still stable. Nevertheless, in
all previous studies, the minimum instability threshold pump $r_{\mathrm{c}%
}=9$ for class-B lasers was found to be a lower bound for multimode
emission.
\begin{figure}[t]
\begin{center}
\scalebox{0.5}{\includegraphics{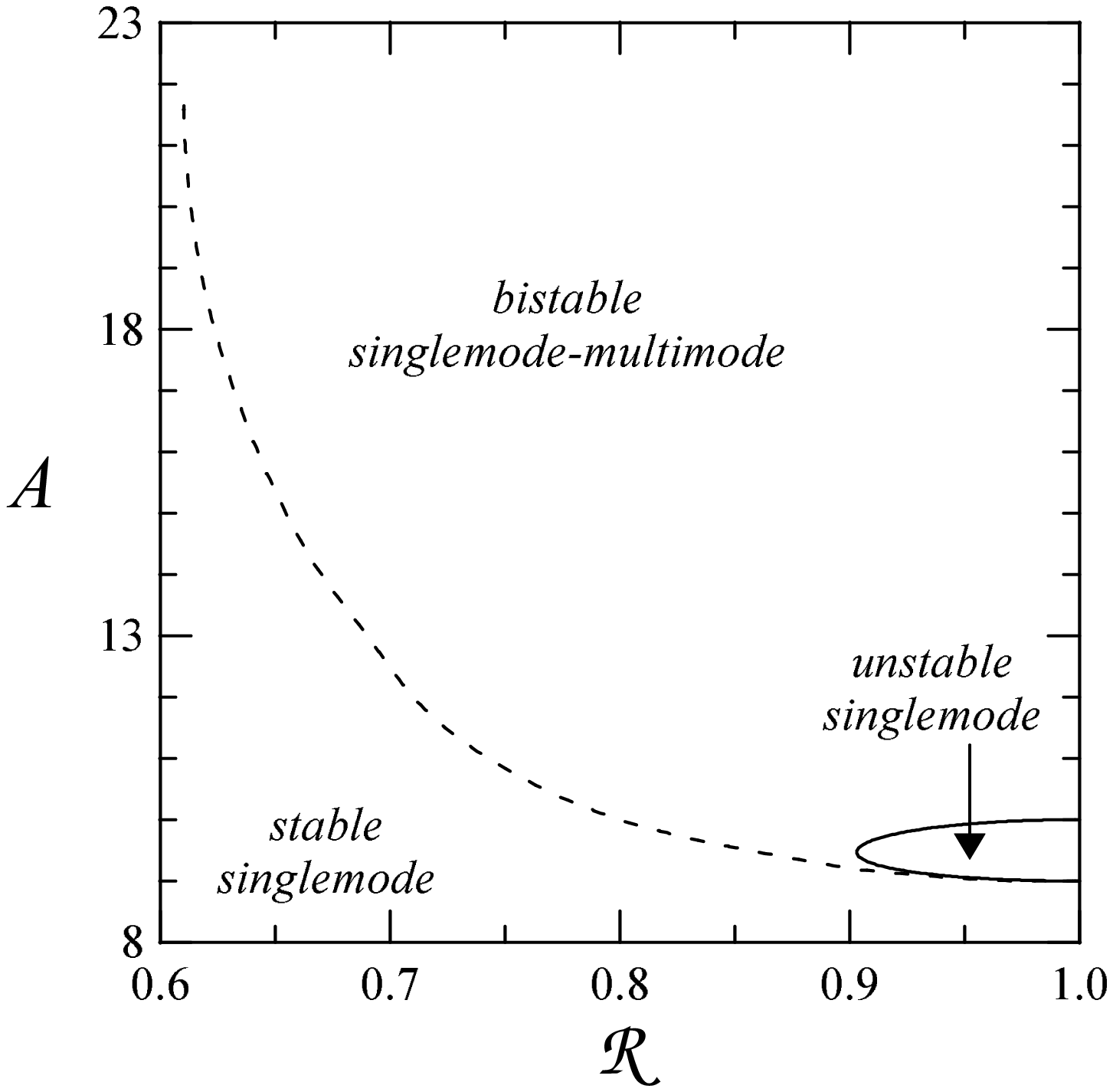}}
\end{center}
\caption{The dashed line indicates the limit of the domain of
existence of pulsing solutions in the $\left\langle
\mathcal{R},A\right\rangle $ plane for $\protect\alpha
=\protect\sqrt{12}$ and $\protect\gamma =0.01$, as
obtained by the numerical integration of the rate equations model, Eqs. (%
\protect\ref{re-f1},\protect\ref{re-d1}). The solid line is the
prediction of the linear stability analysis, Eq.
(\protect\ref{tongue-nonUFL}). } \label{fig:11}
\end{figure}
Outside the UFL this question has not been treated before, to the
best of our knowledge. The analytical treatment has in fact been
presented in Sect. \ref{homo-nufl-pulse}, where we have commented
on this. Here we present preliminary numerical results
\cite{Font04} obtained with the the modal expansion technique
applied to the rate equations (\ref{re-f1},\ref{re-d1}). In Fig.
\ref{fig:11} we represent both the linear stability multimode
emission threshold and the (numerically determined) domain of
existence of stable multimode emission as a function of the
reflectivity of the cavity
for $\alpha =\alpha _{\mathrm{c}}\left( \mathcal{R}\rightarrow 1\right) =%
\sqrt{12}$ and $\gamma =0.01$. The remarkable result is that
multimode emission extends well beyond the limits in reflectivity
marked by the linear stability analysis, although the pump
necessary is always larger than $A_{\mathrm{c}}=9$, the value
theoretically predicted in Sect. \ref {homo-nufl-pulse}. This
means that the domain of bistability between the singlemode and
the multimode solutions is very large and, consequently, multimode
pulsations could be experimentally observed for parameters for
which the theory predicts singlemode emission. We shall come back
to this later on when discussing the experimental observations.

\section{MULTILONGITUDINAL MODE EMISSION IN INHOMOGENEOUSLY BROADENED RING
LASERS}

\label{inhomo}

Up to now we have only considered ring lasers with a homogeneously broadened
active medium. Nevertheless most lasers present some kind of inhomogeneous
broadening in the gain line: In gas lasers the Maxwellian distribution of
atomic velocities introduces Doppler broadening, and in solid state and
fibre lasers differences in the sites of excited atoms introduce some spread
in their transition frequency \cite{Khanin,New83,Siegman,Svelto}.

From the point of view of multimode emission, inhomogeneous broadening in
ring cavity lasers is of primary importance as it is well known that these
lasers easily emit in several longitudinal modes because of spectral hole
burning \cite{Khanin,New83,Siegman,Svelto}. This mechanism for multimode
emission is intuitively understood by imagining the different cavity modes,
whose frequencies fall within the inhomogeneous linewidth, interacting
independently with different subsets of atoms whose frequencies are
quasiresonant with that of each longitudinal mode. If the different mode
frequencies are well separated, as compared with the homogeneous linewidth,
the above is a good picture of what actually happens in most inhomogeneously
broadened ring lasers. When the frequencies of adjacent modes are moved
closer together, or when their intensities grow and they undergo power
broadening, some interaction between the modes may appear, mainly through
saturation effects. All this has been well known for a long time and is not
particularly exciting from the viewpoint of laser instabilities.

Nevertheless this image may be too naive: In the previous sections
we have analyzed in detail how Rabi--splitting induced gain leads
to multimode emission in homogeneously broadened lasers. Then
natural questions arise immediately: How does this mechanism
affect multimode emission in inhomogeneously broadened lasers? Or
in other terms, how does inhomogeneous broadening affect the
RNGHI? Are there two different mechanisms, namely Rabi--splitting
induced gain and spectral hole burning, for multimode emission in
inhomogeneously broadened lasers?

The rest of this section is organized as follows. In Sect.
\ref{inhomo-mod} we present the Maxwell--Bloch and rate equations
models. Then we analyze the multimode emission threshold, first in
the UFL\ (Sect. \ref{inhomo-ufl}) and then outside this limit
(Sect. \ref{inhomo-nufl}). In Sect. \ref{inhomo-ufl} we pay
special attention to the comparison between Maxwell--Bloch
equations and rate equations.

\subsection{Modelling}

\label{inhomo-mod}

As in the previous sections, we consider an incoherently pumped
two--level active medium of length $L_{\mathrm{m}}$, contained in
a ring cavity of length $L_{\mathrm{c}}$, interacting with a
unidirectional plane wave laser field. However, now the field
interacts with an inhomogeneously broadened active medium. We
model the medium by considering a Lorentzian distribution of
atomic frequencies. We further assume, for the sake of simplicity,
that the cavity is resonant with the center of the atomic
transition frequency distribution.

\subsubsection{Maxwell--Bloch equations}

The Maxwell--Bloch equations describing such a laser can be easily
written by generalizing Eqs. (\ref{eqF3}--\ref{eqD3}). The
generalization consists in replacing the polarization appearing in
the field equation Eq. (\ref{eqF3}) by the weighted sum of the
polarizations corresponding to the different frequencies $\omega$.
The model equations read
\cite{Mandel97,Mandel85,Roldan01a,Roldan01b}
\begin{eqnarray}
\left( \partial _{\tau }+\partial _{\zeta }\right) F(\zeta ,\tau ) &=&\sigma
A\int_{-\infty }^{+\infty }\!\!\!\!d\omega \,\mathcal{L}(\omega )\,P\,,
\label{mod1ib} \\
\partial _{\tau }P(\omega ,\zeta ,\tau ) &=&\gamma ^{-1}\left[ -(1+i\omega
)P+FD\right] \,,  \label{mod2ib} \\
\partial _{\tau }D(\omega ,\zeta ,\tau ) &=&\gamma \left[ 1-D-\mathrm{Re}%
\left( F^{\ast }P\right) \right] ,  \label{mod3ib}
\end{eqnarray}%
supplemented by the boundary condition
\begin{equation}
F(0,\tau )=\mathcal{R}F\left( \zeta _{\mathrm{m}},\tau
\right)\,,\qquad\zeta_{\mathrm{m}}=2\pi/\tilde\alpha\,.
\label{boundary}
\end{equation}%
Notice that distributed losses have been neglected ($\chi =0$) and resonance
has been assumed between the cavity and the center of the atomic frequency
distribution ($\delta =0$).

In Eqs. (\ref{mod1ib})--(\ref{mod3ib}) $F(\zeta ,\tau )$ is the normalized
slowly varying envelope of the laser field, and $P(\omega ,\zeta ,\tau )$
and $D(\omega ,\zeta ,\tau )$ are the normalized slowly varying envelopes of
the medium polarization and the population inversion, respectively, for
molecules detuned by $\omega $ with respect to the cavity resonance.

Inhomogeneous broadening is accounted for by the function $\mathcal{L}%
(\omega )$. It represents the spectral distribution of atomic resonances,
which, in order to deal with analytical expressions, is taken to be a
Lorentzian distribution of width$\ $(HWHM) $u$
\begin{equation}
\mathcal{L}(\omega )=\frac{1}{\pi }\frac{u}{u^{2}+\omega ^{2}}\,,
\end{equation}%
where both $\omega $ and $u$ are frequencies scaled to the homogeneous
linewidth $\gamma _{\bot }$. With our notation the unscaled total gain
linewidth (HWHM) of the medium is $\gamma _{\bot }\left( 1+u\right) $.

\subsubsection{The uniform field limit}

As demonstrated in Sect. \ref{homo-ufl}, when the effective
amplitude reflectivity $\mathcal{R}$ is close to unity, one can
apply the uniform
field limit that consists in replacing the field Eq. (\ref{mod1ib}) by%
\begin{equation}
\left( \partial _{\tau }+\partial _{\zeta }\right) F(\zeta ,\tau )=\sigma
\left[ -F+A\int_{-\infty }^{+\infty }\!\!\!\!d\omega \,\mathcal{L}(\omega
)\,P\right] \,,  \label{mod4ib}
\end{equation}%
complemented with the new boundary condition $F(0,\tau
)=F\left(\zeta_\mathrm{m},\tau \right)$.

\subsubsection{Standard rate equations (uniform field limit)}

Class--B lasers (those for which $\gamma _{\bot }\gg \gamma
_{||},\kappa $, i.e., $\gamma ,\sigma \ll 1$), use to be described
with rate equations, which are obtained after the adiabatic
elimination of the medium polarization in the Maxwell--Bloch
equations. These rate equations are generally assumed to be
appropriate for describing multimode emission due to spectral hole
burning \cite{Khanin}. As we shall later compare the predictions
of the full set of Maxwell--Bloch equations with those of the
simpler rate equations, we give here this simpler model, whose
detailed derivation can be found in \cite{Khanin,Prati04}. After
expanding the field as
\begin{equation}
F\left( \zeta ,\tau \right) =\sum_{n=-N}^{+N}\sqrt{I_{n}}e^{i\phi
_{n}}\exp\left[ i\alpha _{n}\left(\zeta -\tau \right) \right]
\end{equation}%
with $\alpha _{n}=n\alpha $, expanding the medium polarization in
a similar way and adiabatically eliminating the polarization, one
is left with the standard rate equations
\begin{eqnarray}
\frac{dI_{n}}{d\tau } &=&2\sigma I_{n}\left[ A\int_{-\infty }^{\infty
}\!\!d\omega \frac{\mathcal{L}(\omega )D}{1+(\omega -\alpha _{n})^{2}}-1%
\right] \,,  \label{rate-a} \\
\frac{\partial D}{\partial \tau } &=&\gamma -\gamma D\left[ 1+\sum_{n}\frac{%
I_{n}}{1+(\omega -\alpha _{n})^{2}}\right] \,.  \label{rate-b}
\end{eqnarray}%
The phases can be calculated through%
\begin{equation}
\frac{d\phi _{n}}{d\tau }=-\sigma A\int_{-\infty }^{\infty }\!\!d\omega
\frac{\mathcal{L}(\omega )D(\omega -\alpha _{n})}{1+(\omega -\alpha _{n})^{2}%
}\,.  \label{rate-fas}
\end{equation}%
In these equations, the integer index $n$ denotes the
\textit{n}-th longitudinal mode with respect to the central
resonant mode ($n=0$). As we show below, the rate equations
description is not appropriate for too long cavities, where
\textit{long} may actually be quite short.

\subsection{Multimode emission in the uniform field limit}

\label{inhomo-ufl}

The stability properties of the uniform field limit were treated first by
Mandel \cite{Mandel85,Mandel86} (see also \cite{Mandel97}), who showed that
in the limit $u\rightarrow \infty $, the multimode emission threshold is
just above the lasing threshold. The general case has been treated at depth
in \cite{Roldan01a}, and the reader is referred to that publication for full
details.

Eqs. (\ref{mod2ib},\ref{mod3ib},\ref{mod4ib}) have two stationary
solutions: The laser--off solution ($F=0,\,P=0,\,D=1$), and the
monomode solution, for which the intensity of the lasing mode,
$I_{0}=F^{2}$, can be written as $A=\sqrt{1+I_{0}}\left(
\sqrt{1+I_{0}}+u\right) $. The threshold for lasing emission is
found by taking $I_{0}=0$ in this expression, and thus it occurs
at a pump $A=A_{0}\equiv 1+u$. Then, by defining the normalized
pump parameter $r=A/A_{0}$, the lasing solution reads
\begin{eqnarray}
r &=&\frac{R\left( R+u\right) }{1+u}\,,  \label{r} \\
R &=&\sqrt{1+I_{0}},  \label{R}
\end{eqnarray}%
or%
\begin{equation}
I_{0}=r\left( 1+u\right) -1+\frac{1}{2}u\left( u-\sqrt{u^{2}+4r\left(
1+u\right) }\right) ,
\end{equation}%
which for $u=0$ gives $I_{0}=r-1$. At threshold, $R=r=1$.

\subsubsection{Two estimates of the multimode emission threshold}

Let us assume momentarily the following na\"{\i}ve approach: Assume that the
threshold for amplification of a detuned mode is not affected by the already
existing resonant lasing mode. This assumption will be reasonably accurate
whenever (i) the inhomogeneous width $u$ is large, because in that case the
intensity $I_{0}$ of the resonant mode at threshold for multimode emission
will be small, and (ii) when the frequency $\Omega $ of the detuned mode is
(sufficiently) larger than the normalized homogeneous width $\gamma ^{-1}$
(which is a measure of the width of the spectral hole), \textit{i.e.}, for
large $\alpha $.

In \cite{Roldan01a} we showed that under these assumptions the emission
threshold for a sidemode of spatial frequency $\alpha $ can be estimated to
be given by
\begin{equation}
r_{\mathrm{thr}}(\alpha )=1+\left( \frac{\gamma \alpha }{\gamma \sigma +1+u}%
\right) ^{2}\,\longrightarrow\, r_{\mathrm{thr},B}(\alpha
)=1+\left( \frac{\gamma \alpha }{1+u}\right) ^{2},
\label{estimacion}
\end{equation}%
where the limit corresponds to class--B lasers ($\gamma \ll 1$).
Thus multimode emission occurs, in this simple approach, for
$r>r_{\mathrm{thr}}$. The term $\gamma \alpha /(1+u)$ represents
the sidemode frequency offset divided by the sum of the cavity
linewidth and the total gain linewidth. One can think of Eq.
(\ref{estimacion}) as the multimode emission threshold when
spectral hole burning is the only relevant mechanism for multimode
emission and there is not any interaction between modes.

A second, more accurate, estimate for the multimode emission
threshold for class--B lasers is obtained from the rate equations
model, Eqs. (\ref{rate-a}, \ref{rate-b}). The derivation can be
found in Chapter 4 of Khanin's book \cite{Khanin} (see also
\cite{Roldan01a}). It reads
\begin{eqnarray}
0 &=&p\left( R\right) +\gamma ^{2}\alpha ^{2},  \label{Khanin-thres} \\
p\left( R\right) &=&R^{2}(1-u)+R(1-u)(u+2)+(u+1)^{2}\,,  \label{p(R)}
\end{eqnarray}%
when the cross relaxation coefficient $\gamma _{\mathrm{CR}}$ is set equal
to zero in Khanin's equations \cite{Khanin}. By using Eq. (\ref{r}), the
above equation can be rewritten as
\begin{equation}
r_{\mathrm{rate}}=r_{\mathrm{thr},B}(\alpha
)+\frac{2R(R+1)}{\left( 1+u\right) ^{2}}\,,  \label{Khanin}
\end{equation}%
with $r_{\mathrm{thr},\mathrm{B}}\left( \alpha \right)$ given by
Eq. (\ref {estimacion}) and $R$ given by Eq. (\ref{R}), the
monomode solution being unstable for $r>r_{\mathrm{rate}}$. We see
that $r_{\mathrm{rate}}\geq r_{\mathrm{thr},\mathrm{B}}(\alpha )$,
i.e., the instability threshold given by Eq. (\ref{Khanin}) is
always larger than the one predicted by Eq. (\ref{estimacion}),
which was admittedly valid for $\alpha \gg 1$. Both
boundaries tend to coincide in the limits of validity of Eq. (\ref%
{estimacion}), \textit{i.e.}, $u\gg 1$. This is perfectly
reasonable as Eq. ( \ref{Khanin}) contains one more ingredient
with respect to Eq. (\ref{estimacion}), namely the saturation of
the sideband gain due to the resonant mode, which explains why it
gives a larger value for the instability threshold.

Notice that Eq. (\ref{Khanin}) predicts that there will be no
multimode emission unless $u>1$, \textit{i.e.}, it fails in
predicting multimode emission due to Rabi--splitting induced gain
(RNGHI). This is entirely normal as the rate equations do not
contain information about the atomic coherence, i.e., about the
medium polarization dynamics, which is the responsible for the
RNGHI as we have seen in the previous
section. On the other hand, the limitation $u>1$ does not appear in the na%
\"{\i}ve approach of Eq. (\ref{estimacion}). Then the physical picture is
the following: For $u<1$, even if the threshold for emission for more than
one mode is crossed, the nonlinear competition between the different cavity
modes enforces that only one of them lases. This is an example of ``the
winner takes all''competition and is the common picture of multimode
dynamics --- which we know to be false because of the RNGHI.

The existence of a minimum instability threshold in the limit
$\alpha \rightarrow 0$ predicted by the rate equations, Eq.
(\ref{Khanin}), is quite unphysical, because it would mean that
the closer are the sidemodes to the central mode, the lower is its
instability threshold. This erroneous result is a consequence of
the approximations introduced in the derivation of the rate
equations, and it disappears when the stability analysis of the
homogenous solution is performed using the full set of
Maxwell--Bloch equations. As will be shown below, the complete
model predicts that, for a given pump value, there is always a
minimum value of the frequency spacing $\alpha $ below which there
is no instability because the gain of the sidemodes is saturated
by the strong central mode.

\subsubsection{Multimode emission threshold}

The multimode emission threshold is rigorously obtained by performing a
linear stability analysis of the single--mode solution Eq. (\ref{r}). The
analytical expressions obtained are cumbersome and it is not trivial to
extract analytical information. We refer the interested reader to \cite%
{Roldan01a} where several limits of interest are treated explicitly. Here we
shall make a resume of the results concerning class--B lasers. Two limits of
interest can be treated explicitly for class--B lasers: The short and long
cavity limits, \textit{i.e.}, the limits $\alpha \gg 1$ and $\alpha =%
\mathcal{O}(1)$ or smaller.

For short cavities, $\alpha \gg 1$, the instability threshold
turns out to be given by Eq. (\ref{Khanin}). This amounts to
saying that for short cavities the rate equations approach is
valid for deriving the multimode emission threshold.

For long cavities, $\alpha =\mathcal{O}(1)$, an approximate expression for
the multimode emission threshold can be derived. It is given by%
\begin{eqnarray}
\mathcal{P} &=&0,  \label{MBthres} \\
\mathcal{P} &=&p\left( R\right) \alpha ^{4}-3(R^{2}-1)\left[
(R+u+1)^{2}-Ru \right] \alpha ^{2}  \notag \\
&&+R(R^{2}-1)(R+1)^{2}(R+u)(u+2),  \notag
\end{eqnarray}%
with $p\left( R\right) $ given by Eq. (\ref{p(R)}). The monomode solution is
unstable for $\mathcal{P}<0$.
\begin{figure}[t]
\begin{center}
\scalebox{0.45}{\includegraphics{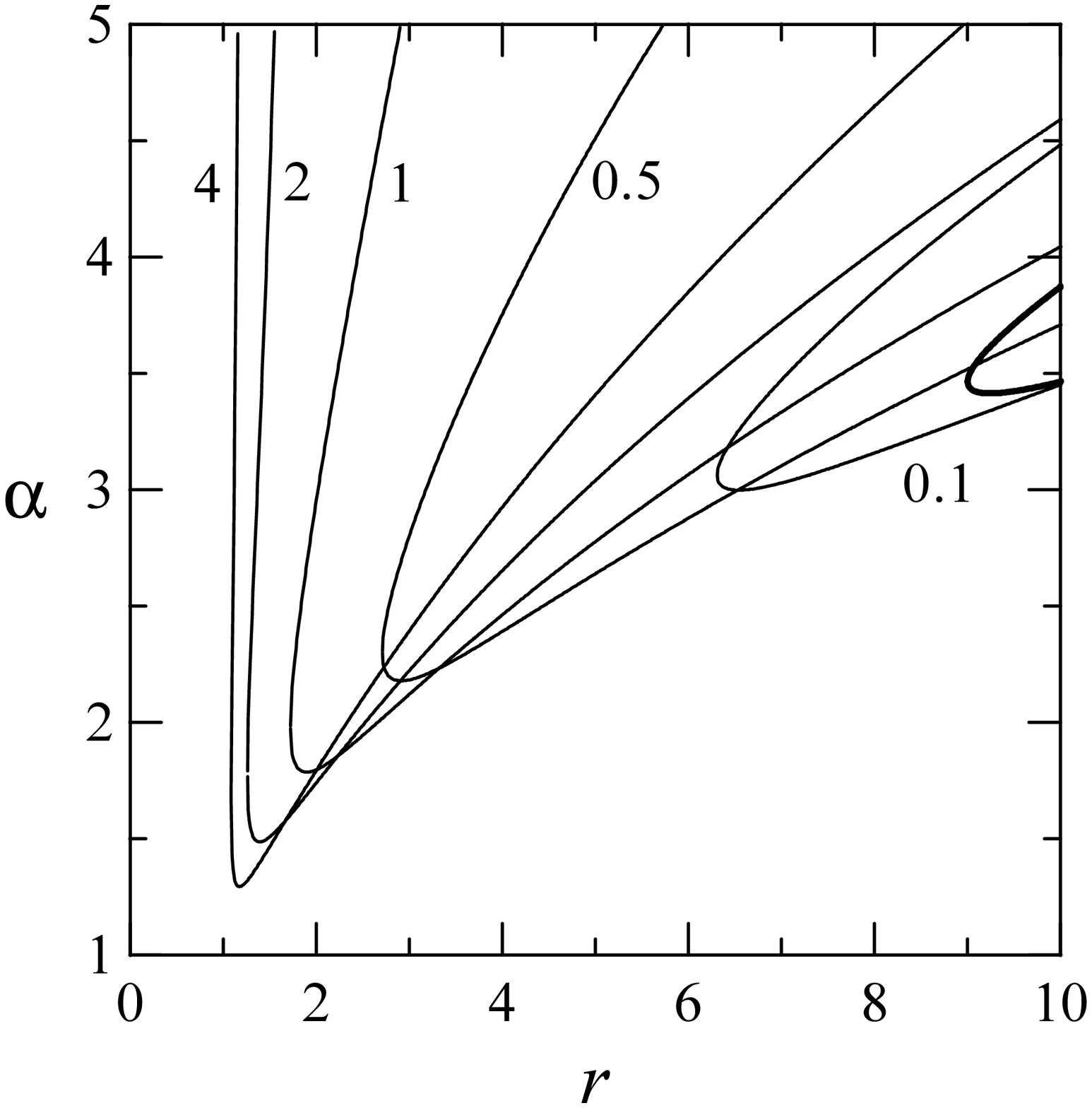}}
\end{center}
\caption{{}Multimode emission threshold in the $\left\langle
r,\protect\alpha \right\rangle $ plane for the values of the
inhomogeneous to homeogeneous linewidth ratio, $u$, marked in the
figure. The thick line corresponds to the homogeneous limit
$u=0$.} \label{fig:12}
\end{figure}
\begin{figure}[t]
\begin{center}
\scalebox{0.45}{\includegraphics{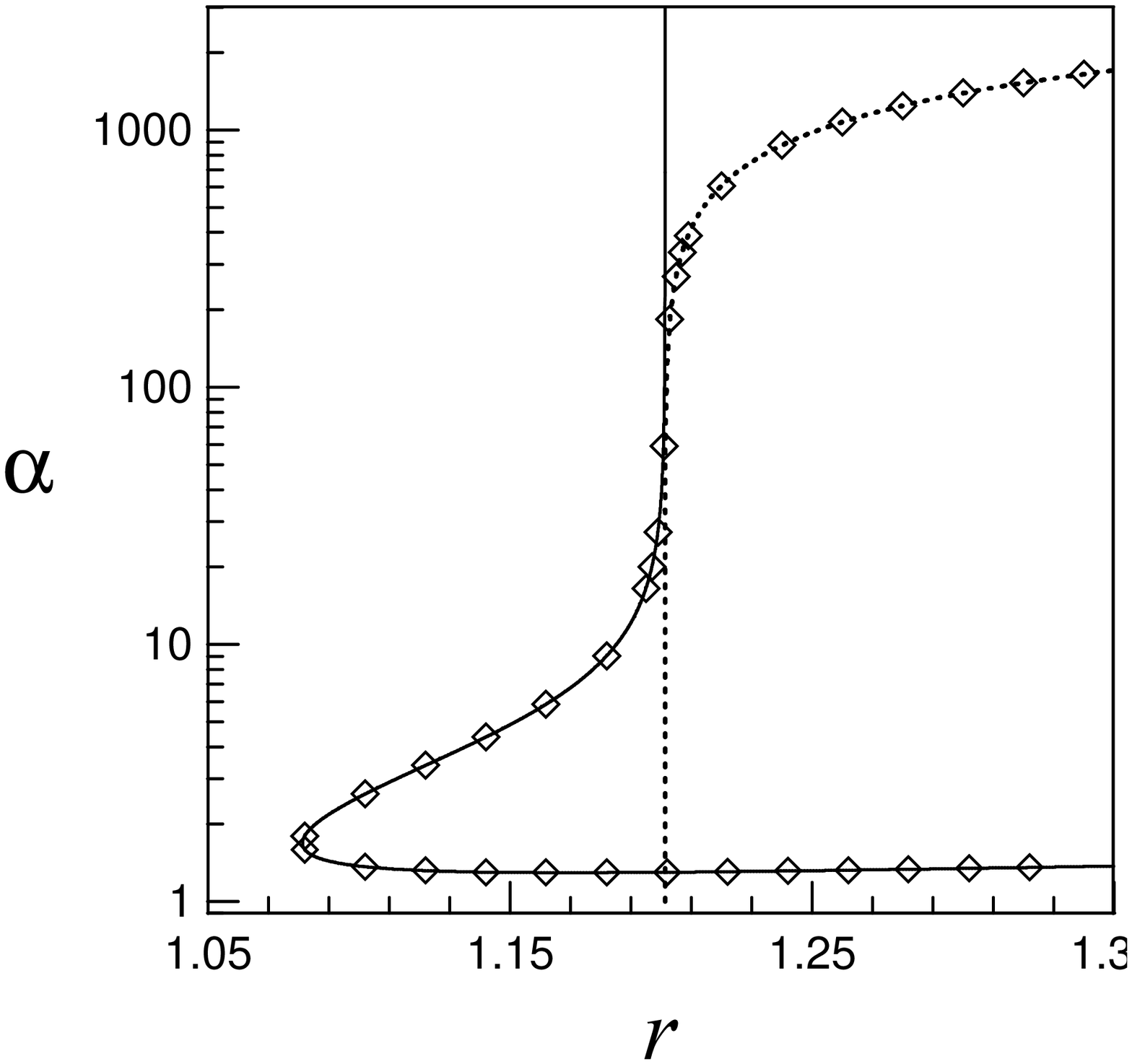}}
\end{center}
\caption{Multimode emission threshold for $u=4$. {}The solid
(dashed) lines have been obtained from Eqs.
(\protect\ref{MBthres}) and (\protect\ref{Khanin-thres}),
respectively. The diamonds correspond to the exact stability
analysis of the Maxwell--Bloch equations, Eq. (22) in Ref.
\protect\cite{Roldan01a}, and have been calculated for
$\protect\gamma=10^{-3}$ and $\protect\sigma =0.1$.}
\label{fig:13}
\end{figure}
Before comparing this multimode emission threshold with that predicted by
the rate equations, we show the influence of the inhomogeneous broadening on
the multimode instability threshold. In Fig. \ref{fig:12} we represent the
multimode emission threshold predicted by Eq. (\ref{MBthres}) for several
values of $u$ in the $\left\langle r,\alpha \right\rangle $ plane. Notice
the enormous quantitative effect that the inhomogeneous broadening has on
the instability threshold. If fact, it can be shown analytically \cite%
{Roldan01a} that the critical point $\left\langle r_{\mathrm{c}},\alpha _{%
\mathrm{c}}\right\rangle $, that for which $r$ is minimum at the instability
threshold, varies as
\begin{eqnarray}
r_{\mathrm{c}} &=&9-36u+\mathcal{O}(u^{2})\,, \\
\alpha _{\mathrm{c}}^{2} &=&12-33.75u+\mathcal{O}(u^{2})\,,
\end{eqnarray}%
for $u\ll 1$, and as
\begin{eqnarray}
r_{\mathrm{c}} &=&1+32/\left( 17u^{2}\right) +\mathcal{O}\left(
u^{-3}\right) \,, \\
\alpha _{\mathrm{c}}^{2} &=&8/3+80/\left( 17u^{2}\right) +\mathcal{O}\left(
u^{-3}\right) \,.
\end{eqnarray}%
for $u\gg 1$.

In Fig. \ref{fig:13} the thresholds predicted by Eq.
(\ref{Khanin-thres}) (dashed line) and Eq. (\ref{MBthres}) (full
line) are represented in the $ \left\langle r,\alpha \right\rangle
$ plane for $u=4$. The exact instability threshold (diamonds) as
obtained by numerically solving Eq. (22) in \cite {Roldan01a}, is
also shown. Several conclusions can be extracted from this figure.

(i) There is a single instability threshold, \textit{i.e.}, there are not
two mechanisms for multimode emission (spectral hole burning and RNGHI) but
a single mechanism;

(ii) The asymptotic expressions Eqs. (\ref{Khanin-thres}) and (\ref{MBthres}%
) compare perfectly well with the exact result in their respective domains
of validity: The threshold predicted by the rate equations becomes invalid
for small $\alpha $ (long cavity) while the approximate expression Eq. (\ref%
{MBthres}) becomes invalid for large $\alpha $ (short cavity);

(iii) The unphysical instability predicted for small $\alpha $ by the rate
equations result is removed, and there is a minimum value of $\alpha $ below
which there is no instability; and

(iv) there is a domain of intermediate cavity lengths where the two analytic
expressions connect.

The last item allows to determine which is the value of the cavity length
beyond which the rate equations approach is no more valid. By analyzing Eqs.
(\ref{Khanin-thres}) and (\ref{MBthres}) one obtains that the distance
between the two functions is minimum for $\alpha \sim 2\gamma ^{-1/2}$ when $%
u\geq 2$. This leads to a \lq\lq coherence length\rq\rq estimate
\begin{equation}
L_{\mathrm{coh}}=\frac{\pi c}{\gamma _{\bot }}\left( \frac{\gamma _{\bot }}{%
\gamma _{\Vert }}\right) ^{1/4}\,,  \label{Lcoh}
\end{equation}%
for lasers with a significant inhomogeneous broadening ($u\geq 2$).

Then, for cavities larger than $L_{\mathrm{coh}}$, atomic coherence effects (%
\textit{i.e.}, Rabi--splitting induced gain) are important for the
determination of the multimode emission threshold while they are irrelevant
for shorter cavities. For example, for CO$_{2}$ lasers $L_{\mathrm{coh}%
}\approx 100$ m, for $632.8$ nm HeNe lasers $L_{\mathrm{coh}}\approx 20$ m,
for Er$^{3+}$--doped fibre lasers $L_{\mathrm{coh}}\approx 5$ cm, and for
Nd--glass lasers $L_{\mathrm{coh}}\approx 3$ mm. Thus, the necessity of
using the full Maxwell--Bloch description of the laser in order to describe
multimode emission depends strongly on the particular laser system under
consideration: It is necessary for Nd--glass and Er$^{3+}$--doped fibre
lasers and unnecessary for CO$_{2}$ and HeNe lasers.

Let us remark that the expression for $L_{\mathrm{coh}}$ gives a good
estimate for the critical length for values of $u\geq 2$; for smaller values
of $u$, $L_{\mathrm{coh}}$ decrease, tending to $0$ for $u\rightarrow 1$, so
that the necessity of considering coherent effects is still more important
in this limit.

The above shows that Rabi--induced sidemode gain (RNGHI) is far
from being a negligible mechanism for multimode emission in
inhomogeneously broadened ring lasers. Quite to the contrary, it
is an essential ingredient for those lasers with cavity lengths
larger than $L_{\mathrm{coh}}$ or even shorter if the ratio of
inhomogeneous to homogeneous linewidth is small.

\subsubsection{Dynamics beyond the multimode emission threshold}

The numerical integration of the inhomogeneously broadened model is far from
being a simple task, especially for class--B lasers. In fact, this task has
been carried out very recently for the first time in \cite{Prati04}. For the
sake of illustration, let us comment that some of the calculations we
comment below implied a running time of more than one month on a R12000
Silicon Graphics processor. This illustrates the enormous difficulties of
making numerical simulations for class--B laser parameters, due to the
stiffness of the equations in this limit.

There were previous attempts, during the eighties, by Brunner et
al. \cite{Brunner1,Brunner2} of numerically studying multimode
emission in inhomogeneously broadened lasers. But the model they
used was that of a standing--wave cavity and the third--order Lamb
theory approximation was assumed. As this approximation does not
contain atomic coherence effects, their results did not give
information about the RNGHI mechanism in standing--wave lasers.

In \cite{Prati04}, the Maxwell--Bloch Eqs.
(\ref{mod2ib},\ref{mod3ib},\ref {mod4ib}) and the rate equations
model Eqs. (\ref{rate-a},\ref{rate-b}) were numerically integrated
for fixed pump ($r=5$), inhomogeneous broadening ($u=2$) and
spatial frequency (equal to the homogeneous linewidth), and three
different values of $\gamma$ ($10^{-1},10^{-2},$ and $10^{-3}$).
As we change the value of $\gamma $ over three orders of
magnitude, we change the Rabi frequency of the field. In doing so,
the behavior makes the transition from that predicted by the
rate--equations model (for smaller $\gamma $) to that in which
Rabi--induced gain effects are expected to be important (for
larger $\gamma $).
\begin{figure}[t]
\begin{center}
\scalebox{0.45}{\includegraphics{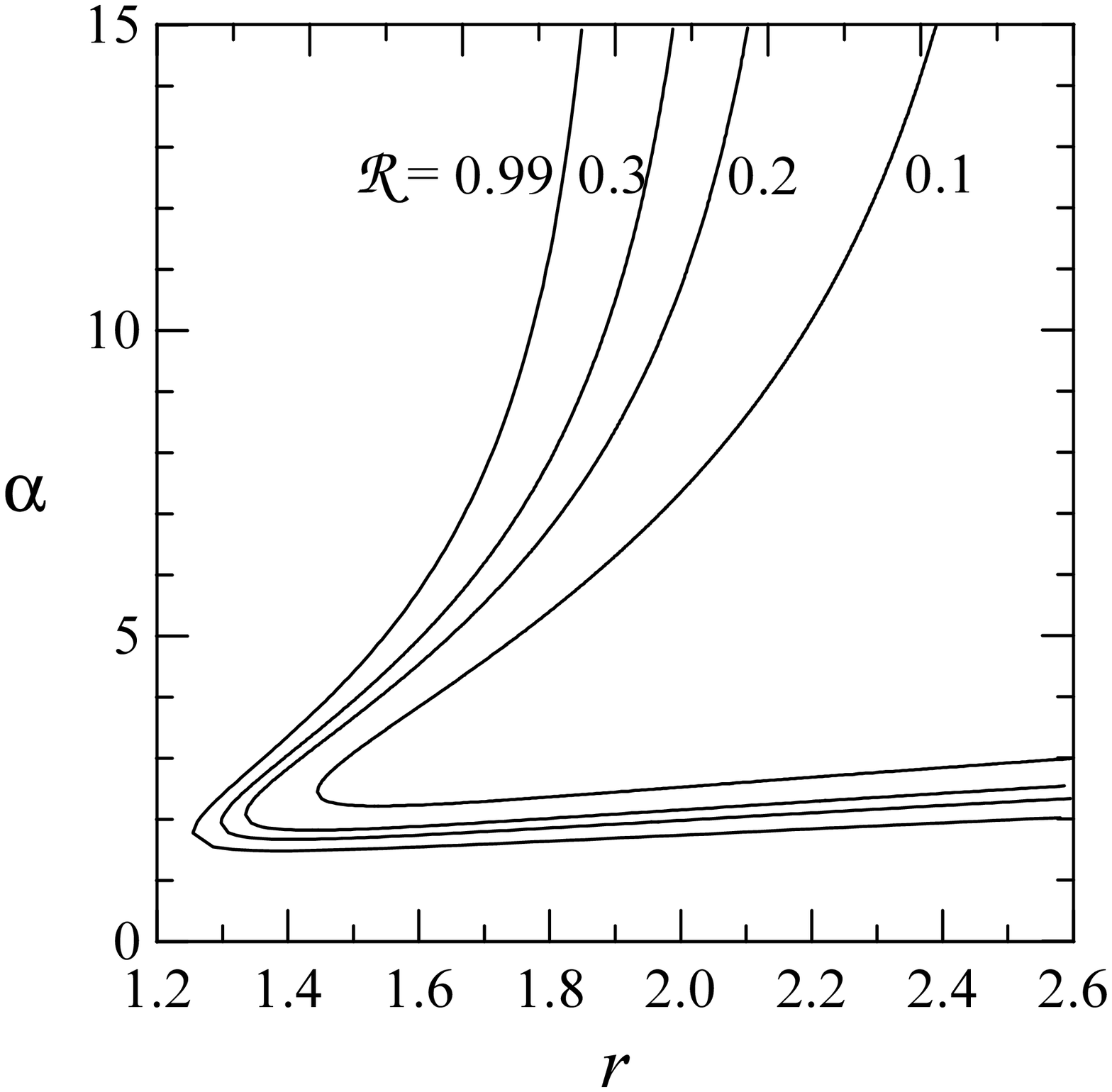}}
\end{center}
\caption{{}Multimode emission threshold in the $\left\langle
r,\protect\alpha \right\rangle $ plane for $u=2$ and the values of
the reflectivity marked in the figure as obtained from Eq. (36) in
\protect\cite{Roldan01b}.} \label{fig:14}
\end{figure}
\begin{figure}[t]
\begin{center}
\scalebox{0.5}{\includegraphics{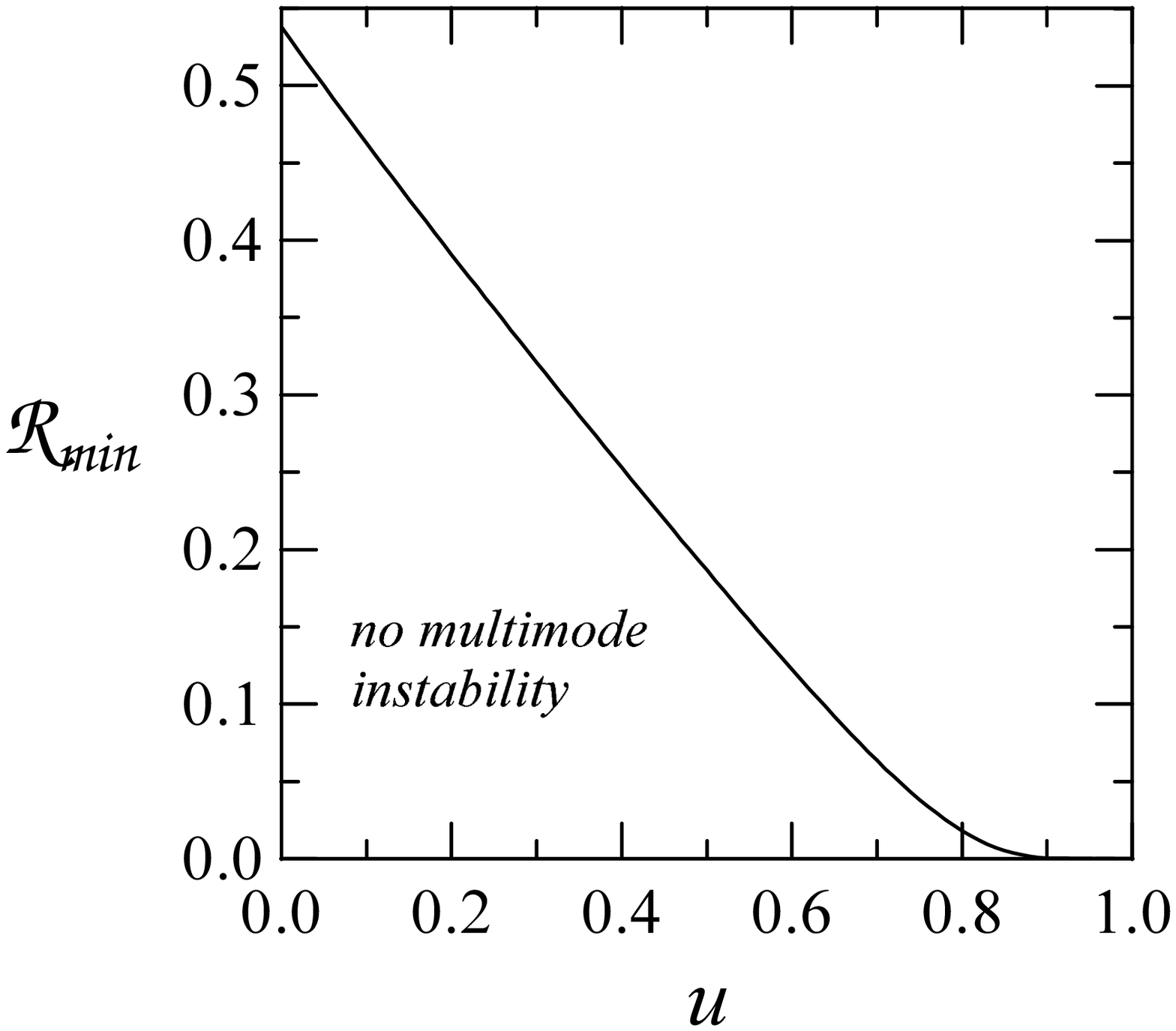}}
\end{center}
\caption{{}Minimum value of the reflectivity for the multimode
instability to exist, $\mathcal{R}_{\min }$, as a function of the
inhomogeneous to homogeneous linewidth ratio.} \label{fig:15}
\end{figure}
The main finding in \cite{Prati04} consists in the confirmation of
the conclusions extracted from the stability properties: Rate
equations describe properly the multimode dynamics only for short
cavities, that is, under conditions for which the free spectral
range is large as compared to the Rabi frequency of the
intracavity field. In this limit rate equations and Maxwell--Bloch
equations provide the same results for the mode intensities at
steady state, and no phase locking is found even in the
Maxwell--Bloch equation, where all relevant information about the
phase dynamics is preserved (remember that mode--locking cannot be
described by the rate equations model as modal phases do not enter
in the dynamics of mode intensities and population inversion, see
Eqs. (\ref{rate-a}--\ref{rate-fas})). Then, for short cavities the
use of rate equations is perfectly legitimate.

This is no longer true for longer cavities, where the inhomogeneously
broadened laser behaves like a homogeneously broadened one, with a multimode
dynamics which presents all the features of the classical RNGHI: The number
of excited modes is larger than expected from linear stability
considerations and phases spontaneously lock, giving rise to regular pulses
in the total intensity. We refer the reader to \cite{Prati04} for full
details.

\subsection{Multimode emission outside the uniform field limit}

\label{inhomo-nufl}

Outside the uniform field limit, \textit{i.e.}, for arbitrary values of the
amplitude reflectivity $\mathcal{R}$ the analytical expressions for the
instability threshold are much more involved than in the uniform field limit
previously analyzed. For class--B lasers the problem has been treated in
detail in \cite{Roldan01b}, and the interested reader is referred to that
paper for full details. Here we shall stress some important points.

As it happens in the homogeneously broadened model, a decrease in
the value of $\mathcal{R}$ leads to an increase of the multimode
emission threshold. This can be seen in Fig. \ref{fig:14} where
the multimode instability threshold is shown, as usual, in the
$\left\langle r,\alpha \right\rangle $ plane for $u=2$ and several
values of the amplitude reflectivity $\mathcal{R} $. For
$\mathcal{R}=0.99$, the instability threshold is very close to
that predicted by the homogeneously broadened model (Eq.
(\ref{MBthres})), and decreasing values of $\mathcal{R}$ increase
the instability threshold as expected. What is unexpected is that
multimode emission exists for values of $\mathcal{R}$ as low as
$0.1$. This is unexpected because we showed analytically in Sect.
\ref{homo-nufl-rnghi} that in homogeneously broadened lasers, the
decrease of $\mathcal{R}$ leads to the disappearance of the
multimode instability when $\mathcal{R}<\mathcal{R}_{\min }\simeq
0.538$.

Fortunately, $\mathcal{R}_{\min }$ can be analytically determined even
outside the uniform field limit (see \cite{Roldan01b}). It is given by $%
\mathcal{D}\left( \mathcal{R},u\right) =0$ with
\begin{equation}
\mathcal{D}\left( \mathcal{R},u\right) =9-4\left[ 1-\frac{u\left( u+1\right)
}{2}\right] \frac{1+\mathcal{R}^{2}}{1-\mathcal{R}^{2}}\left\vert \ln
\mathcal{R}^{2}\right\vert ,  \label{D(R)}
\end{equation}%
which reduces to Eq. (\ref{D-nonUFL}) for $u=0$ \footnote{%
Notice that Eq. (\ref{D(R)}) corrects a typographic error in Eq. (46) of
\cite{Roldan01b}.}.

In Fig. \ref{fig:15} $\mathcal{R}_{\min }$ is shown as a function of $u$ .
Notice that the limitation on the value of $\mathcal{R}$ for the existence
of multimode emission disappears at $u=1$ (although in practical terms it
disappears for $u>0.9$ as $\mathcal{R}_{\min }$ decreases exponentially as $%
u\rightarrow 1$). Thus the multimode instability does not have any
limitation in cavity losses when inhomogeneous broadening is about the same
as the inhomogeneous width or larger. This marks an important difference
with the case of homogeneous broadening. The fact that this qualitative
change appears at $u=1$ can be related to the fact that the rate equations
model predicts the multimode instability only for $u>1$. This result
suggests that for $u>1$ spectral hole burning makes multimode emission
possible when the Rabi--induced gain mechanism could not do it by itself (%
\textit{i.e.}, there is no more a $\mathcal{R}_{\min }$, which is a
characteristic of the RNGHI), while for $u<1$, $\mathcal{R}_{\min }$ exists
(although its values decreases for increasing $u$). This suggests that
spectral hole burning plays a less important role in this limit.

This result is particularly relevant from the experimental point of view: An
increase of the cavity losses eventually leads to the impossibility of
multimode emission when $u<0.9$ (\textit{i.e.}, for homogeneously broadened
lasers or lasers with small inhomogeneous broadening), but this does not
happen for lasers with enough inhomogeneous broadening at $u>0.9$.

\subsubsection{Effect of distributed losses in three--level and four--level
lasers}

In Sect. \ref{homo-nufl-loss}, where homogeneous broadening was considered,
it was shown that distributed losses have quantitative importance for the
determination of the multimode emission threshold for three-- and
four--level lasers, although it was not really important for two--level
lasers as far as distributed losses have reasonable values, roughly below $%
10 $\textrm{dB}. The reason lies in the way how losses enter in the
transformation of the pump parameter for applying the results of the
two--level theory to three-- and four--level lasers.

For inhomogeneously broadened lasers the same argument holds, and
distributed losses must be taken into account when calculating the multimode
instability threshold for three-- and four--level lasers. Performing the
linear stability analysis with inhomogeneous broadening and distributed
losses outside the uniform field limit is a very heavy task that does not
provide any analytical insight. Rather than going to that trouble, it
suffices at least in the case of not too large distributed losses to use the
results of the linear stability analysis outside the uniform field limit
without distributed losses, and introduce them when calculating the
transformation of the pump parameter, as already explained in Sect. \ref%
{homo-nufl-loss}. This is the way we calculated the instability
threshold presented in \cite{Voigt04} and reproduced in the
following section (cf. Fig. \ref{fig:18} below). For more details,
see Ref. 24 in \cite{Voigt04}.

\section{ON EXPERIMENTAL STUDIES OF THE MULTIMODE INSTABILITY IN EDFL's}

\label{experimental}

In this section we describe experimental observations that
contribute to the discussion of multimode laser instabilities. We
already discussed in Sect. \ref{introduction} that experimental
research on the RNGHI has been limited to the dye laser
\cite{Hillman1,Hillman2} and that the observed multimode emission
was explained as being caused by the band structure of the lasing
levels \cite{FuHaken1,FuHaken2,FuHaken3}, not a manifestation of
the RNGHI. Prior to Refs.~\cite{F95,P97} no other experimental
work on the subject has appeared. The reason is probably that the
constraints imposed on the cavity length by Eqs. (\ref{Ac}) and
(\ref{Lcmin}) have discouraged experimentalists: According to Eq.
(\ref{Lcmin}), cavity lengths need to be enormously long (in some
cases several orders of magnitude longer than standard values),
while at the same time there is the requirement of a very large
instability to lasing threshold ratio (the \textquotedblleft
factor--of--nine\textquotedblright) \cite{WeissVilaseca}. By now
it is known, however, that the \textquotedblleft
factor--of--nine\textquotedblright\ is irrelevant if the active
medium is better modelled as a three--level laser medium.
Moreover, for such lasers Eq.~(\ref{Lcmin}) does not constitute a
hard limit, as for cavity lengths well below this value
instabilities occur at accessible pump values. This makes it much
more likely that the RNGHI becomes observable in some lasers.

By this reasoning, Er-doped fiber lasers are rendered the most
promising candidate for a clear observation of the RNGHI: these
are three--level lasers, and their cavities can be made very long.
We will therefore concentrate on this laser type.

Many laser applications require monochromatic operation, and therefore
several researchers attempted to operate Er-doped fiber laser in single mode
operation. However, after many failures and a few partial successes, this is
now considered as notoriously difficult. Where it was attempted, researchers
chose one of these strategies: (i) increase the cavity's effective free
spectral range beyond the gain bandwidth (see, e.g., \cite{Zhang96}), or
(ii) reduce the gain bandwidth below the free spectral range by insertion of
filters with very narrow bandwidth (see, e.g. \cite{Guy95}). None of these
approaches led to a full success. For example, in a publication subsequent
to \cite{Guy95} it is reported that single mode operation became impossible
to maintain above a certain pump level \cite{Chang96}. The reason for this
behavior was not further investigated.

The pragmatic conclusion is that EDFL's obviously have a natural tendency to
operate on several modes simultaneously. The small free spectral range
caused by the necessarily long resonator contributes to the difficulty, but
by itself cannot explain the underlying reason.

\subsection{A first dedicated approach}

Probably the first experimental study of fundamental causes for the
instability of an EDFL was presented in \cite{F95}. The authors made a
distinction between relaxation oscillation and self mode locking, and
investigated the latter. They used a resonator consisting of a WDM coupler
to bring in the pump light, and an output coupler to steer fully 90\% of the
power out. A polarization-insensitive optical isolator enforced
unidirectional operation. An Er-doped active fiber provided gain;
alternatively either a 15~m length at low dopant concentration (300~ppm), or
a 0.8~m length at high concentration ($\approx$5000~ppm) was used. Allowing
for some extra length of component pigtails, this brought the resonator's
free spectral range to about 10~MHz or 70~MHz, respectively.

The laser output was directed through another isolator and then monitored
either by a fast photodiode hooked up to a fast oscilloscope or an RF
spectrum analyzer, or by a background-free autocorrelator. No stable cw
emission was observed. For the longer cavity it was reported that
self-modelocking produced a train of pulses with a repetition rate given by
the resonator's free spectral range. An occasional presence of satellite
pulses at intermediate times was mentioned. The pulses were reported to have
a temporal width (FWHM) of a few ns. Inspection with the autocorrelator was
performed to check for substructure. The autocorrelator could monitor a time
window of 50~ps, and within this window no substructure was found. For the
shorter cavity, a modulation of the output power at the free spectral range
was also observed. However, it did not take the form of pulses but rather
had a nearly sinusoidal shape. This is plausible because for a larger free
spectral range, fewer modes will fall into the bandwidth of the gain.

When the pump power was varied near lasing threshold, in the shorter cavity
the instability was seen whenever there was lasing, while in the longer
cavity a small interval of single-mode operation seemed to exist just above
lasing threshold.

\subsection{The follow-up}

In a subsequent study \cite{P97}, basically the same group of
authors replaced the polarization-insensitive isolator with a
polarizing isolator. They also added polarization controllers to
the cavity. Two output couplers branching out 95\% or 50\% were
used alternatively. The active fiber was again of the low Er
concentration type, and was 13~m long. The cavity free spectral
range was thus near 10~MHz. Again, self mode locking at the cavity
free spectral range was observed; pulses had a duration of
1.74\dots3.00~ns. The RF spectrum contained $\approx 250$ beat
notes, indicative of a similar number of oscillating modes.

In this experiment there was a combination of fiber birefringence
plus polarizing elements in the cavity. This raises the issue
whether mode locking due to Nonlinear Polarization Rotation (NPR)
might have occurred. The authors argue that NPR can be ruled out
for the following reasons: (i) In the previous setup, there
definitely was no picosecond structure, and when the polarizing
isolator was introduced, the pulse shape was not modified. This
suggests --- somewhat indirectly --- that the polarizing action is
not responsible. The authors further noted that (ii) at times
intermediate to the pulses there was a constant background of
random signal, possibly satellite pulses of some kind, and that
NPR would likely suppress such structure. However, random groups
of pulses are routinely seen in NPR lasers. Finally, (iii) power
levels in the fiber were deemed insufficient for NPR, in
particular since a wide core fiber was used. NPR requires
remarkably little power, however, and at the powers stated, NPR
cannot be ruled out entirely. However, the authors kindly inform
us that the instability also existed just above threshold, and in
that case NPR is indeed highly unlikely to occur.

There remains an unresolved discrepancy about the measured spectral shape of
the pulses which is reported to be Gaussian and, for 2~ns pulses, must have
been 0,001~nm wide. The optical spectrum analyzer reportedly used for this
measurement, however, has a specified spectral resolution of 0,05~nm.

One of the most prominent features of RNGHI, namely its threshold-like
onset, was not addressed in the experiment of \cite{P97}. The publication
provides a comment that the experiments took place far above lasing
thresholds, but that instability persisted down to at least a
less-than-tenfold threshold power. As the authors kindly inform us, the
instability was seen immediately above lasing threshold. From this
information one must conclude that the matter deserves more clarification
before anything is definitely proven.

\subsection{A systematic assessments of thresholds}

\begin{figure}[t]
\begin{center}
\scalebox{0.4}{\includegraphics{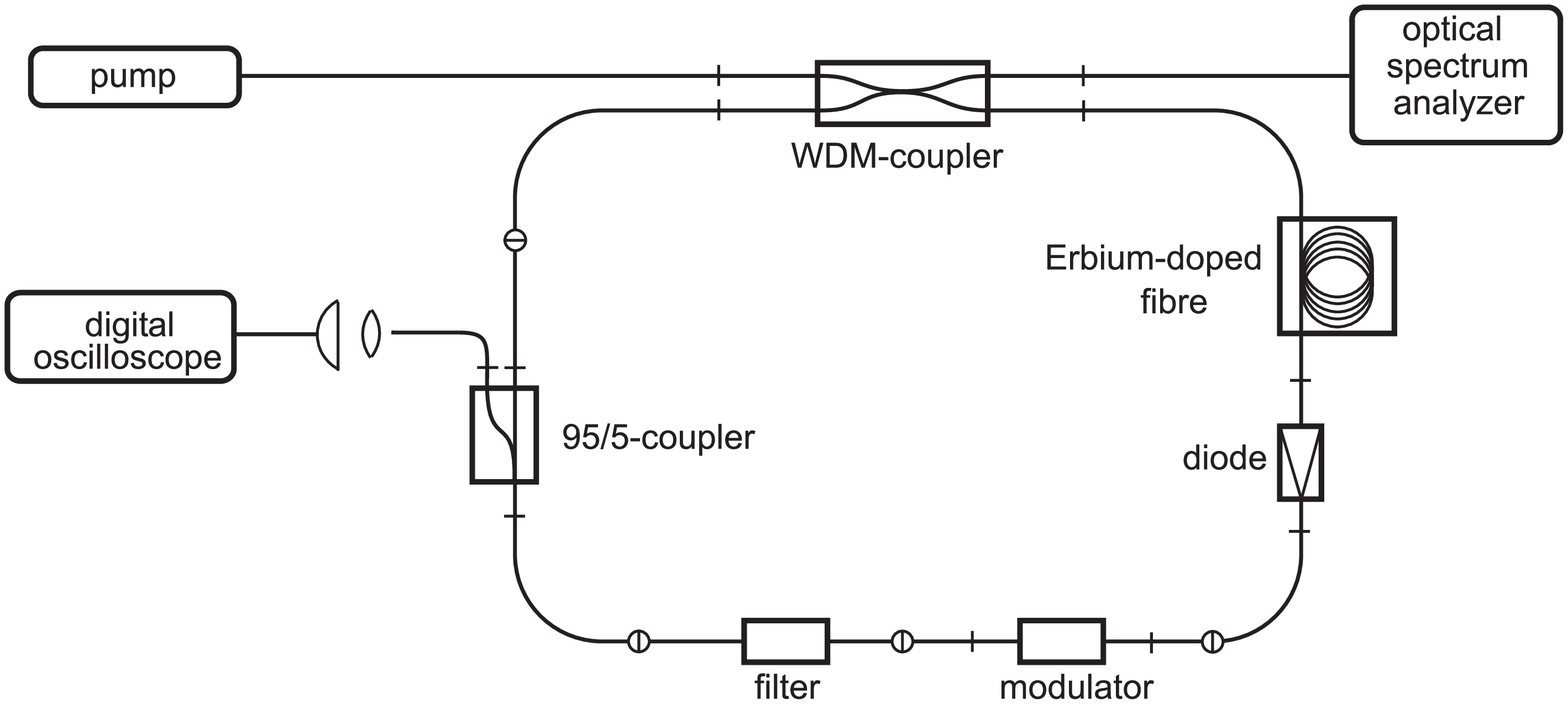}}
\end{center}
\caption{Experimental setup. The modulator serves to introduce
well--defined loss; it can take the form of either an
electro--optic modulator or a tightly wound fiber coil. For
further detail see text.} \label{fig:16}
\end{figure}

In subsequent work, an Er-doped fiber ring laser specifically
designed for observation of instabilities was set up by some of
the present authors. As a starting point a standard configuration
was chosen, see Fig. \ref{fig:16}. The cavity
contained $8.2\,\mathrm{m}$ of active fiber (585~ppm Er dopant level) in a $%
22\,\mathrm{m}$ long ring (the remainder consisted of standard single mode
fiber). Pump light came from a $100\,\mathrm{mW}$, $980\,\mathrm{nm}$ laser
diode; it was launched into the ring by a WDM coupler. Light coupled out by
this WDM coupler, and also from an additional 95/5 coupler, was used to
monitor the system simultaneously by a fast photodiode and an optical
spectrum analyzer. An optical isolator ensured unidirectional operation, and
a moderately narrow bandpass filter gently restricted the bandwidth
available for lasing to about $1\,\mathrm{nm}$. As a unique feature,
variable loss was inserted into the cavity. To this end either an amplitude
modulator or a tight fiber coil with well--defined radius was employed. High
variable loss served to bring out the onset of instability more clearly.
Great effort was made to characterize the exact amount of loss for each
setting: all components (localized losses) including splices etc. were
tested individually, and the distributed loss in the Er fiber was
determined. Finally, as a cross-check the total loss was calculated from the
operational laser's output power vs. pump power relation. For more detail
see \cite{Voigt04}.
\begin{figure}[t]
\begin{center}
\scalebox{0.7}{\includegraphics{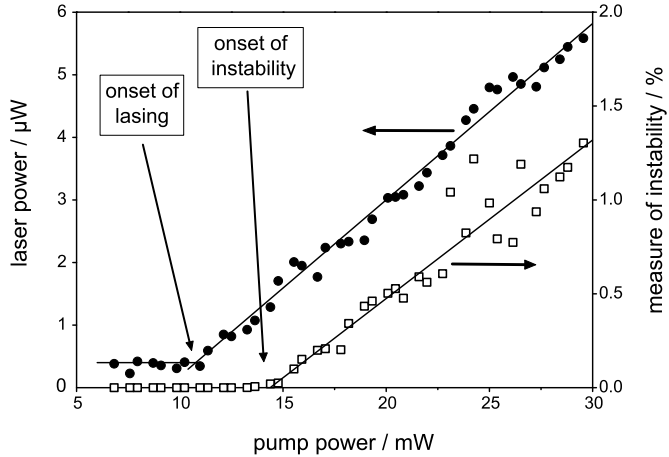}}
\end{center}
\caption{Laser power (filled circles) and measure of instability
(open squares) as a function of pump power.} \label{fig:17}
\end{figure}
\begin{figure}[t]
\begin{center}
\scalebox{0.5}{\includegraphics{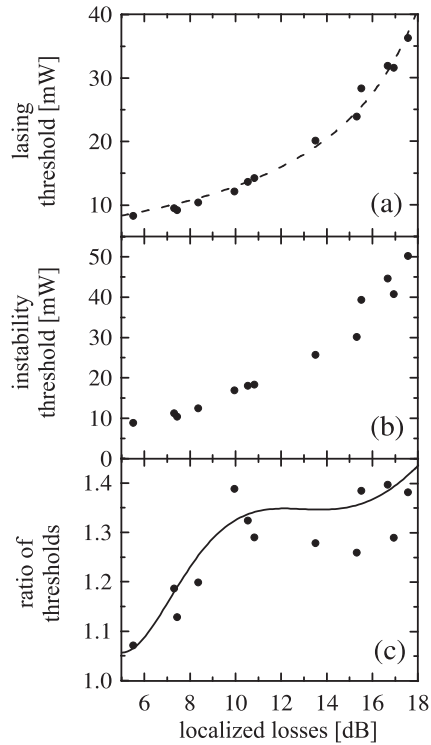}}
\end{center}
\caption{Influence of localized cavity loss. Shown are (a) the
laser threshold, (b) the instability onset, and (c) the ratio of
both. The dashed line in (a) is a fit with theory (see
\protect\cite{Voigt04} for more detail); while the line in (c)
only serves to guide the eye.} \label{fig:18}
\end{figure}
\begin{figure}[t]
\begin{center}
\scalebox{0.5}{\includegraphics{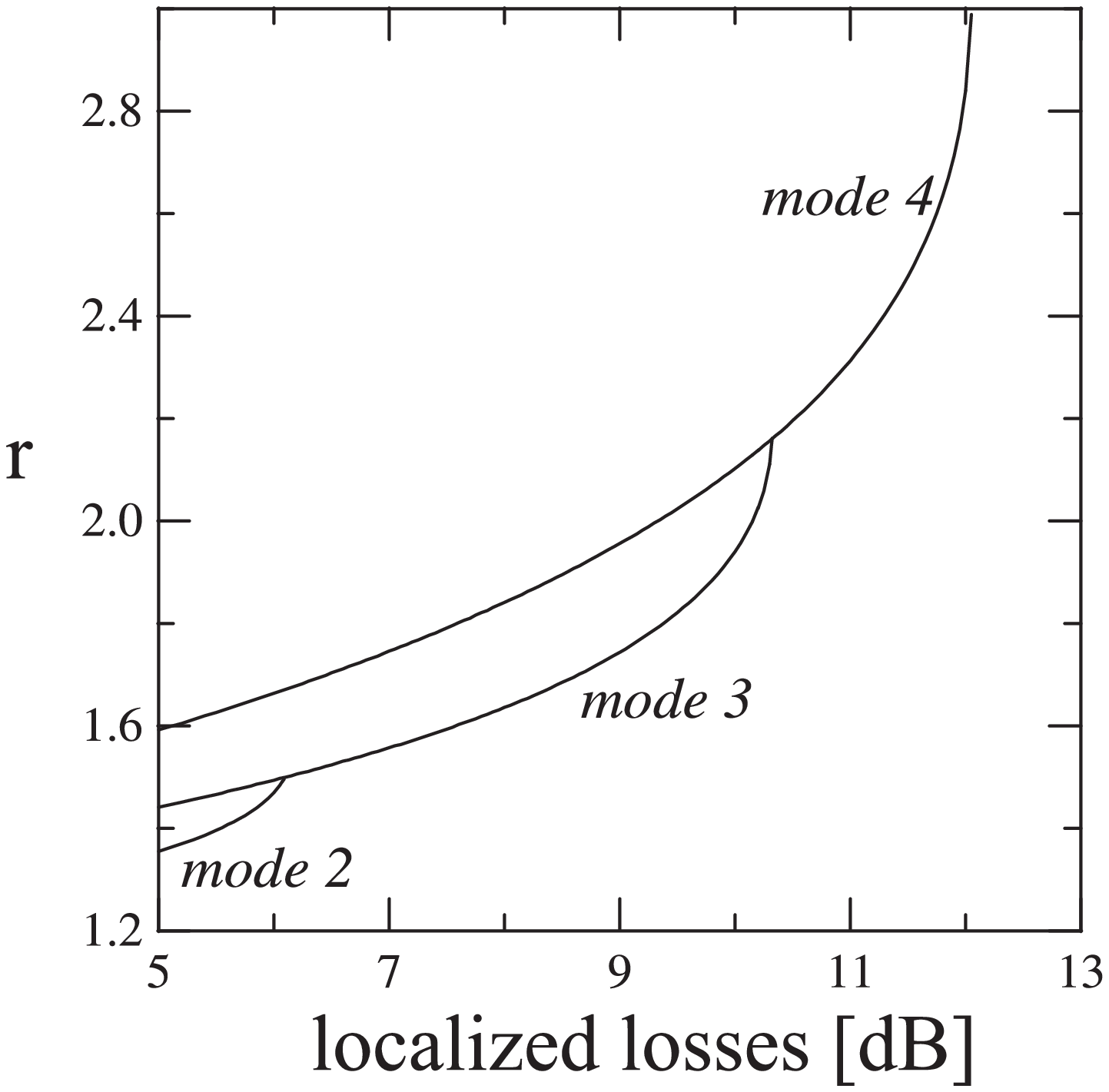}}
\end{center}
\caption{Ratio of instability onset to lasing threshold, as
predicted by a three--level laser model with inhomogeneous
broadening at $u=2$. Different modes become unstable at different
points; the labels refer them to the central lasing mode. For the
instability onset, always the lowest line counts.} \label{fig:19}
\end{figure}
Just above lasing threshold this laser indeed operated in a single
longitudinal mode. As the pump power was increased, modulations of
the output power appeared at frequencies which were integer
multiples of the cavity round trip frequency ($9\,\mathrm{MHz}$)
without exception. Obviously these were beat notes between
different longitudinal modes.

These beat notes, however, were neither steady in amplitude, nor
in frequency: The frequencies involved hopped rapidly and
apparently at random all across the range up to several GHz, and
during much of the time in between there was no beat at all. After
the insertion of the bandpass filter already mentioned above, the
frequency hopping range was limited to hundreds of MHz and thus
more manageable. Still, the beats remained unsteady. A typical
beat note episode lasted on the order of tens of milliseconds to a
few seconds. The temporal profile of the beat note was almost
always very nearly sinusoidal, indicative of a beat between only
two resonator modes. In the presence of the high loss
intentionally introduced here, the laser could not be pumped very
far above threshold, so that it may be not too surprising that
there was just dual--mode, but no multimode operation.

We wish to point out that for this experiment nonlinear polarization
rotation (NPR) \cite{NLPR} can be safely ruled out. Polarization-dependent
losses, on which NPR hinges, were carefully avoided.

Given a random phenomenon, statistical means were employed to
characterize it. Time series of the instantaneous power were
recorded which --- in view of the typical timescales --- were
chosen to be several seconds long. Unfortunately, adequate
sampling to correctly assess all up to the highest frequencies
would have required data rates of GB/s and file sizes of about 10
GB per shot; that is just not feasible. To keep data files at
manageable size, undersampling at 20~000 samples per second was
chosen. While this way information about the actual beat frequency
is lost, all episodes of mode beating longer than $100\,\mu
\mathrm{s}$ still can be detected from the time series. Shorter
episodes seemed not to occur anyway. Occasional occurrences of
relaxation oscillation were easily identified in the file by their
very different amplitude, and were discarded. The fraction of time
during which valid mode beats were detected served as the measure
of instability $M$.

The data-taking procedure consisted in setting a particular loss
value, then incrementing the pump power in small steps, and
determining $M$ at each step. (At the same time, as described
above, the total power was recorded to help assess the loss
value). Next, the loss was incremented, and the procedure was
repeated until the accessible range of loss values was exhausted.
After an evaluation of these extensive data, the following
conclusions are reached: The laser power data in Fig. \ref{fig:17}
show the universally known threshold behavior: the power is close
to zero below threshold, but not exactly so due to fluorescence.
Above threshold the power makes a good fit to a straight line by
which the slope efficiency is defined. The measure of instability,
on the other hand, strictly remains at zero up to some point above
the laser threshold. Beyond that point, $M$ sharply sets on to
nonzero values, and continues to rise as the pump power is raised
further. This is clear evidence that the instability has a
well-defined, sharp onset, a fact which had not previously been
demonstrated experimentally.

In the next step both the lasing threshold and the instability
onset are taken from data as in Fig. \ref{fig:17}. Fig.
\ref{fig:18} reveals that the ratio of both values
$P_{\mathrm{inst}}/P_{\mathrm{0}}$ (where $P_{\mathrm{inst}}\ $and
$P_{\mathrm{0}}$ denote the pump power at the instability and
lasing thresholds, respectively) does indeed scale with loss. At
low loss, $P_{\mathrm{inst} }/P_{\mathrm{0}}\approx 1$. Under
typical operating conditions of Er fiber lasers, losses would be
even lower, and experimenters would be unable to tell apart both
onsets. For larger loss, however, $P_{\mathrm{inst}}/P_{
\mathrm{0}}$ increases up to about $1.5$. The intentionally high
loss of this experiment pays off nicely here: Both onsets are
clearly distinguishable, and the interval of single mode operation
in between is clearly identifiable. This constitutes a
considerable progress over previous work.

However, we must emphasize an important fact regarding the range
of loss values used in that experiment. We showed in Sect.
\ref{homo-nufl-rnghi} that in a homogeneously broadened gain
medium $\mathcal{R}_{\mathrm{\min }}$ describes a maximum value of
loss beyond which there exists no instability at all.
$\mathcal{R}_{\mathrm{\min }}$ as defined in Eq. \ref{Rmin}
corresponds to 4.5~dB of localized loss in the experiment
\cite{Voigt04}, but Fig. \ref{fig:18} clearly shows that
instability persists at much higher loss.

To resolve this discrepancy, we reconsider the structure of the Er
gain line. In Sect. \ref{inhomo-nufl} we showed that in the case
of an inhomogeneous contribution to the line as expressed by
$u\neq 0$, $\mathcal{R}_{\mathrm{\min }}$ goes to zero (see the
discussion of Eq. \ref{D(R)} and Fig. \ref{fig:15}); hence the
corresponding maximum loss diverges. It is difficult to make
precise statements about the value of $u$ for the fiber used in
the experiment in \cite{Voigt04}, but a choice of $u=2$ is
reasonable. For this value, Fig. \ref{fig:19} shows the
theoretically expected threshold ratio as a function of localized
cavity loss (the calculation takes distributed losses into
account, see \cite{Voigt04} for details). Evidently, the
disagreement with experimental data is reduced dramatically. While
quantitatively $r$ is systematically predicted too high, in
particular at the highest loss values, at least the discrepancy
about the \emph{existence} of an instability onset is resolved.

Finally we need to address a caveat about the interpretation of
the experimentally observed instability onset. As was discussed in
Secs.~\ref {homo-nufl-pulse},\ref{homo-nufl-sub}, the possibility
exists that the instability threshold is either a supercritical or
a subcritical bifurcation. Based on experimental data alone, a
decision between these possibilities cannot be made. Consider a
subcritical bifurcation, which would in all likelihood imply a
range of bistability between the single--mode and the multimode
solution (at least this is the case in the homogeneously broadened
case, see Fig. \ref{fig:15}). The observed intermittent behavior
would make some sense in that case: The experimentally determined
instability onset would then be the lower limit point of the
unstable branch, and the bifurcation proper was never reached due
to limited available pump power. In fact, not even the point of
\textquotedblright Maxwell's construction\textquotedblright\
(where both branches are occupied for equal amounts of time on
average) was reached. Even if unlimited pump power had been
available, it is not at all clear whether a final conclusion about
the nature of the bifurcation could have been reached, because a
sizeable increase of the pump power eventually brings on other
processes like Brillouin scattering, thermal effects, etc., which
further complicate the issue. We must therefore leave this
question open for now.

Let as finally remark that, in contrast to \cite{P97}, \emph{multi}mode
operation and pulsing was never observed in \cite{Voigt04}: there was only
dual--mode operation and sinusoidal modulation. Also, there were only
intermittent, not steady, beat notes between modes. Surely, this must have
to do with the fact that the laser was intentionally made lossy so that it
could never be pumped very much above its first threshold. Whether there are
still other factors involved (codopants of the fiber, etc.) must remain
unresolved at this point.

\section{CONCLUSION AND OUTLOOK}

\label{conclus}

We have introduced the different models required for the study of
the RNGHI. In particular, we have treated the applicability of the
two--level laser model to three-- and four--level lasers and,
importantly, the rigorous derivation of the uniform field limit.
We have then revised the basics of the RNGHI, and we have reviewed
our continued work on the subject over recent years. This research
was motivated by the suggestion by Lugiato and coworkers in 1997
\cite{P97} that the pulsations exhibited by a unidirectional EDFL
could be a manifestation of this elusive phenomenon.

At long last, the threshold-like onset of the multimode
instability in an EDFL was demonstrated in \cite{Voigt04}. Data
show a qualitative and, with appropriate corrections, even
semi-quantitative agreement to theoretical expectations.
Nevertheless, data do not represent a clear-cut textbook rendition
of the RNGHI, but only an approximation. Therefore the bottom line
of our combined theoretical and experimental research is this: The
processes observed in EDFLs very likely constitute a manifestation
of the RNGHI, but it is a manifestation in a 'dressed' way.
Inhomogeneous broadening, distributed losses, and the three--level
structure of erbium ions take their imprint on the instability.
Moreover, it is very likely that noise plays a central role in the
'intermittent' appearance of multimode emission.

Theories are very often neater and much more elegant than
real--world experiments. The closest thing to Lorenz-type laser
chaos that was ever found experimentally suffered from
complications that are absent from the model, as we commented in
Sect. \ref{introduction}. In a similar way, the experiment in
\cite{Voigt04} presents the closest thing to RNGHI that has been
found to date. In any case, we do not have any doubt that the
experimental observations are a clear manifestation of the
resonant Rabi instability, i.e., that Rabi sideband gain is the
responsible for the observed instability.

Another remarkable result of our research is that the RNGH
mechanism is essential for understanding the multimode emission
threshold in inhomogeneously broadened lasers, even for relatively
short values of the laser cavity depending on the active medium.
This fact suggests that RNGHI could be important for correctly
understanding mode--locking in lasers with an accessible coherence
length (see Eq. (\ref{Lcoh}) and the subsequent discussion).

Finally we would like to remark that there are open questions from
both the theoretical and the experimental sides. On the one hand,
the observed intermittent pulsations needs to be theoretically
explained, and it must be determined up to what extent
subcriticality and noise could explain them, or wether other
phenomena we have not yet considered (such as dispersion or fiber
nonlinearity \cite{Tartwijk97,Tartwijk98}) need to be taken into
account. On the other hand, experimental research in other laser
types, as NdDFLs or Nd:YAG lasers, would help to understand how
the RNGH mechanism affects real lasers.

We gratefully acknowledge J.L. Font, F. Fontana, L.A. Lugiato, M.
Lenz, E.M. Pessina, J. Redondo, F. Silva, and J.F. Urchuegu\'{\i}a
for continued discussions on the subject along the recent years.
We thank J.L. Font for carrying out the numerical calculations
represented in Fig. \ref{fig:11} \cite{Font04}. This work has been
supported by the Spanish Ministerio de Ciencia y Tecnolog\'{\i}a
and European Union FEDER (Fonds Europ\'{e}en de D\'{e} velopement
R\'{e}gional) through Project PB2002-04369-C04-01, and by Deutsche
Forschungsgemeinschaft

\end{document}